\documentclass[sigconf]{acmart}

\usepackage{symbols}
\usepackage{svg}
\usepackage[off]{svg-extract}
\svgsetup{clean=true}
\usepackage{enumitem}
\usepackage{makecell}
\setlist[enumerate]{leftmargin=5mm,topsep=1mm}
\setlist[itemize]{leftmargin=5mm,topsep=1mm}

\DeclareUnicodeCharacter{03BC}{$\mu$}

\newcommand{\lmdb}{\textsc{LMDB}\xspace}
\newcommand{\rmi}{\textsc{RMI}\xspace}
\newcommand{\cdfshop}{\textsc{CDFShop}\xspace}
\newcommand{\pgm}{\textsc{PGM-index}\xspace}
\newcommand{\alex}{\textsc{ALEX/APEX}\xspace}
\newcommand{\plex}{\textsc{PLEX}\xspace}
\newcommand{\dcal}{\textsc{Data Calculator}\xspace}
\newcommand{\name}{AirIndex}
\newcommand{\btree}{\textsc{B-tree}\xspace}
\newcommand{\system}{\textsc{\name}\xspace}
\newcommand{\ourmodel}{\textsc{\name-Model}\xspace}
\newcommand{\oursearch}{\textsc{AirTune}\xspace}
\newcommand{\githuborg}{illinoisdata}

\newcommand{\supawit}[1]{{\color{blue} Supawit: #1}}
\newcommand{\tofix}[1]{{\color{red} #1}}
\newcommand{\ignore}[1]{}
\newcommand{\todo}[1]{{\color{red} #1}}
\newcommand{\fixed}[1]{{\color{purple} #1}}

\definecolor{Acolor}{HTML}{264653}
\definecolor{Bcolor}{HTML}{2A9D8F}
\definecolor{Ccolor}{HTML}{E9C46A}
\definecolor{Dcolor}{HTML}{F4A261}
\definecolor{Ecolor}{HTML}{E76F51}

\definecolor{lmdbcolor}{HTML}{944003}
\definecolor{rmicolor}{HTML}{7678ED}
\definecolor{pgmcolor}{HTML}{213F80}
\definecolor{alexcolor}{HTML}{8ECAE6}
\definecolor{plexcolor}{HTML}{2A9D8F}
\definecolor{dcalcolor}{HTML}{E9C46A}
\definecolor{btreecolor}{HTML}{F4A261}
\definecolor{airindexcolor}{HTML}{E76F51}

\definecolor{rootcolor}{HTML}{E76F51}
\definecolor{layeronecolor}{HTML}{F4A261}
\definecolor{layertwocolor}{HTML}{E9C46A}
\definecolor{datalayercolor}{HTML}{2A9D8F}

\definecolor{iocolor}{HTML}{FFFFFF}
\definecolor{cachecolor}{HTML}{E76F51}
\definecolor{deserializecolor}{HTML}{F4A261}
\definecolor{predictcolor}{HTML}{E9C46A}
\definecolor{findcolor}{HTML}{2A9D8F}
\definecolor{othercolor}{HTML}{264653}  % TODO

\definecolor{nfscolor}{HTML}{2A9D8F}
\definecolor{ssdcolor}{HTML}{E76F51}

\definecolor{buildcolor}{HTML}{2A9D8F}
\definecolor{costcolor}{HTML}{E76F51}

\definecolor{barOrange}{HTML}{bb5555}
\definecolor{BlueColor}{HTML}{bb5555}
\definecolor{RedColor}{HTML}{bb5555}
\definecolor{OrangeColor}{HTML}{bb5555}
\definecolor{barYellow}{HTML}{ee9944}
\definecolor{barLightGreen}{HTML}{A3A847}
\definecolor{barGreen}{HTML}{5588bb}
\definecolor{GreenColor}{HTML}{5588bb}

\definecolor{vvlowcolor}{HTML}{7678ED}
\definecolor{vlowcolor}{HTML}{7678ED}
\definecolor{lowcolor}{HTML}{7678ED}
\definecolor{normalcolor}{HTML}{000000}
\definecolor{highcolor}{HTML}{E76F51}
\definecolor{vhighcolor}{HTML}{E76F51}
\definecolor{vvhighcolor}{HTML}{E76F51}

\makeatletter
\g@addto@macro\normalsize{%
\setlength\belowdisplayskip{1pt}
\setlength\belowdisplayshortskip{1pt}
}
\makeatother

\AtBeginDocument{%
  \providecommand\BibTeX{{%
    \normalfont B\kern-0.5em{\scshape i\kern-0.25em b}\kern-0.8em\TeX}}}

\setcopyright{acmcopyright}
\copyrightyear{2024}
\acmYear{2024}
\acmDOI{XXXXXXX.XXXXXXX}

\acmConference[SIGMOD '24]{ACM SIGMOD/PODS International Conference on Management of Data}{June 11--16,
  2024}{Santiago, Chile (Accepted 23 May 2023)}
\acmPrice{15.00}
\acmISBN{978-1-4503-XXXX-X/18/06}

\begin{document}
% \title{\system: Holistic Learned Index Tuning on Data and Storage}
% \title{\system: Towards Highly Adaptive Index on Storage}
% \title{\system: Circumvent Data Complexity and Storage Diversity}
% \title{\system: Indefinite Data Volume and Storage Shape}
\title{\system: Versatile Index Tuning Through Data and Storage}
% adapt, tune
% flexible, adaptive, versatile
% data
% gas: compressibility, expandability, diffusibility 

%%
%% The "author" command and its associated commands are used to define the authors and their affiliations.
\author{
Supawit Chockchowwat,
Wenjie Liu,
Yongjoo Park}
\email{{supawit2,wenjie3,yongjoo}@illinois.edu}
\affiliation{%
  \institution{CreateLab, University of Illinois at Urbana-Champaign}
}

\begin{abstract}

The end-to-end lookup latency of a hierarchical index---such as a B-tree or a learned index---is determined by its structure such as the number of layers, the kinds of branching functions appearing in each layer,  the amount of data we must fetch from layers, etc. Our primary observation is that by optimizing those structural parameters (or \emph{designs}) specifically to a target system’s I/O characteristics (e.g., latency, bandwidth), we can offer a faster lookup compared to the ones that are not optimized. Can we develop a systematic method for finding those optimal design parameters? Ideally, the method must have the potential to generate almost any existing index or a novel combination of them for the fastest possible lookup.
    
In this work, we present a new data-and-I/O-aware index builder (called \system) that can find high-speed hierarchical index designs
in a principled way. Specifically, \system minimizes an objective expressing the end-to-end latency in terms of various \emph{designs}---the number of layers, types of layers, and more---for given data and a \emph{storage profile}, using a graph-based optimization method purpose-built to address the computational challenges rising from the inter-dependencies among index layers and the exponentially many candidate parameters in a large search space. Our evaluations confirm that \system can find optimal index designs, build them within the times comparable to existing methods, and deliver up to $4.1\times$ faster lookup than a lightweight B-tree library (\lmdb), $3.3\times$--$46.3\times$ faster than state-of-the-art learned indexes (\rmi/\cdfshop, \pgm, \alex, \plex), and $2.0\times$ faster than \dcal's suggestion on various dataset and storage settings.

\end{abstract}

%    <concept>
%        <concept_id>10002951.10002952.10002971</concept_id>
%        <concept_desc>Information systems~Data structures</concept_desc>
%        <concept_significance>500</concept_significance>
%        </concept>
%    <concept>
%        <concept_id>10002951.10002952.10003212.10003216</concept_id>
%        <concept_desc>Information systems~Autonomous database administration</concept_desc>
%        <concept_significance>500</concept_significance>
%        </concept>
%  </ccs2012>

% \ccsdesc[500]{Information systems~Data structures}
% \ccsdesc[500]{Information systems~Autonomous database administration}

% \keywords{tuning, indexing, learned indexes, storage, external memory}

\received{15 January 2023}
\received[revised]{20 April 2023}
\received[accepted]{23 May 2023}

\maketitle

%%%%%%%%%%%%%%%%%%%%%%%%%%%%%%%%%%%%%%%%%%%%%%%%%%%%%%%%%%%%%
%%%%%%%%%%%%%%%%%%%%%%%%%%%%%%%%%%%%%%%%%%%%%%%%%%%%%%%%%%%%%

\section{Introduction}
\label{sec:intro}

Hierarchical indexes (e.g., binary search tree, B-tree)
    allow us to
        quickly locate a relevant item
    by fetching (only) a small fraction of data inside each index layer.
The B-tree~\cite{Bayer1972,Comer1979,Zhou2014, Bin2014} has been a conventional choice
        for many data systems such as PostgreSQL~\cite{postgresql}, 
        MySQL~\cite{mysql}, ZLog~\cite{zlog}, BTRFS~\cite{rodeh2013btrfs}, etc.
More recently,
    it has been shown that by using \emph{learned models} (i.e., regression functions offering approximate pointers)
        in place of the (exact) child pointers inside a B-tree's internal nodes,
    we can reduce the amount of data we need to fetch for each layer,
        lowering the overall lookup latency~\cite{kipf2018learned,DBLP:conf/sigmod/KipfMRSKK020RadixSpline,DBLP:journals/pvldb/FerraginaV20PGMIndex,DingMYWDLZCGKLK2020ALEX,lu2021apex}.
In general, the amount of data indicated by those exact/approximate branching
    pointers---along with the number of layers---determine 
        the overall lookup latency of an index.

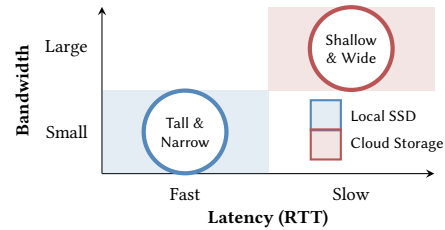
\begin{figure}[t]
\pgfplotsset{
mystyle/.style={
    axis lines = left,
    width=60mm,
    height=38mm,
    legend style={
        at={(-0.2,1.1)},anchor=south west,column sep=2pt,
        draw=black,fill=none,line width=.5pt,
        /tikz/every even column/.append style={column sep=5pt},
        font=\footnotesize,
    },
    xlabel near ticks,
    xlabel shift=-1mm,
    tick label style={font=\footnotesize},
    label style={font=\footnotesize\bf},
    xtick style={draw=none},
    ytick style={draw=none},
    xmin=1,
    xmax=5,
    xtick={1,2,3,4,5},
    xticklabels={,Fast,,Slow,},
    xlabel=Latency (RTT),
    ymin=1,
    ymax=5,
    ytick={1,2,3,4,5},
    yticklabels={,Small,,Large,},
    ylabel=Bandwidth,
    clip=false,
}
}

\tikzset{
mynode/.style={
    circle,draw=gray!50!white,minimum width=11mm,ultra thick,
    align=center,font=\scriptsize\sf,fill=white,
},
myshade/.style={
    fill opacity=0.15,inner xsep=0,inner ysep=0
},
}

\centering
\begin{tikzpicture}

\begin{axis}[
    mystyle,
]

\node[fit={(axis cs: 1,1) (axis cs: 3,3)},myshade,fill=GreenColor] {};
\node[fit={(axis cs: 3,3) (axis cs: 5,5)},myshade,fill=BlueColor] {};

\node[mynode,draw=GreenColor] at (axis cs: 2,2) {Tall \& \\Narrow};
\node[mynode,draw=BlueColor] at (axis cs: 4,4) {Shallow \\ \& Wide};

\def\x{3.5}
\node[myshade,fill=GreenColor,anchor=west,minimum width=4mm,minimum height=4mm,
    draw=GreenColor,thick] 
    (L1) at (axis cs: \x, 2.4) {};
% \node[anchor=west,font=\scriptsize\sf] at ($(L1.east)+(0.1,0)$) {Local SSD};
\node[anchor=west,font=\scriptsize\sf] at ($(L1.east)+(1.0,1.0)$) {Local SSD};
    
\node[myshade,fill=BlueColor,anchor=west,minimum width=4mm,minimum height=4mm,
    draw=BlueColor,thick] 
    (L2) at (axis cs: \x, 1.7) {};
% \node[anchor=west,font=\scriptsize\sf] at ($(L2.east)+(0.1,0)$) {Cloud Storage};
\node[anchor=west,font=\scriptsize\sf] at ($(L2.east)+(1.0,1.0)$) {Cloud Storage};

\end{axis}

\end{tikzpicture}
\vspace{-4mm}
\caption{
Expected optimal structures by I/O characteristics. 
\system finds the optimal structure in a large design space.}
\label{fig:intro}
% \vspace{-2mm}
\end{figure}

Despite their effectiveness,
    (most) existing indexes have a common limitation,
making them often \emph{suboptimal} for novel system environments with different I/O characteristics.
That is,
    \emph{due to their nearly fixed index structures (e.g., the number of layers, the types of layers),
     existing indexes cannot deliver as fast performance as the
        ones specifically designed
     in consideration of the data access cost of a target system environment.}
For instance, in \cref{fig:intro},
    if an index is maintained directly on remote storage with relatively high I/O latency 
        (e.g., in cloud systems 
        like Amazon Aurora~\cite{amazon-aurora} and Delos~\cite{delos}),
    we might be able to achieve faster lookup speeds
    by creating a wider/shallower index (e.g., larger fanout in B-trees, more accurate models in learned indexes),
    mitigating the impact of high I/O latency
        with fewer data fetches required for index traversal.
In contrast,
    if an index is maintained on a local SSD 
    (e.g., SQLite~\cite{sqlite}) with fast I/O latency
        (or relatively smaller bandwidth),
    we can create
        a narrower/taller index (e.g., smaller fanout in B-trees, less accurate models in learned indexes)
            for faster lookup.
In other words,
    depending on a different system environment,
        an optimal index structure may vary,
    causing significant performance gaps between fixed index structures
        and the ones adapting to target environments (\cref{sec:motivation}).
This observation closely resembles the one made by Gray and Graefe~\cite{gray1997FiveMinuteRules}
    for determining an optimal B-tree page size in relation to page access cost.
However, its utility-based approach is specific to conventional indexes with exact child pointers,
    meaning it cannot be generalized to a wider class of indexes with learned models.

\begin{table*}[t]
\begin{center}
\caption{Summary of existing work. \tmark/\xmark indicates limited/no support.}
\label{tab:existing}
\vspace{-2mm}
\small
\begin{tabular}{ l c c l l }
\toprule
 \textbf{Method} & \textbf{Approx.~Pointer?} 
    & \textbf{I/O Aware?} & \textbf{Novelty} 
    & \textbf{Weakness/Difference} \\ 
\midrule
 B-trees~\cite{Bayer1972,Comer1979} & \xmark & \xmark & First B-tree proposal & No search for optimal fanout \\
 \rmi~\cite{Kraska2018} & \cmark & \xmark 
    & First approximate branching 
    & No hyperparameter tuning, fixed two-layer structure \\
 \alex~\cite{Ding2020}  & \cmark & \xmark 
    & Updatable with learned models
    & No optimization for end-to-end latency \\  
 5minRules~\cite{gray1997FiveMinuteRules} & \xmark & \cmark 
    & Heuristic for B-tree page size
    & Restricted tuning, fixed branching across layers \\
 \cdfshop~\cite{DBLP:conf/sigmod/MarcusZK20CDFShop}
    & \cmark & \xmark
    & Searches Pareto efficient RMIs
    & Inconclusive tuning, not optimize for latency \\
 \dcal~\cite{DBLP:conf/sigmod/IdreosZHKG2018DataCalc}
    & \tmark & \cmark & Evaluates end-to-end performance
    & Inefficient tuning, restricted branching functions \\
 \textbf{Ours (\system)} & \cmark & \cmark
    & I/O-aware exact/approx.~layers
    & Optimization focuses on lookup than updates  \\
\bottomrule
\end{tabular}
\end{center}
\end{table*}

In this work, 
we tackle this limitation with \emph{\system,
    a general index-building framework that can efficiently find an (optimal) low-latency index structure
        in consideration of profiled I/O characteristics as well as data distribution.}
The core difference of \system is that
    % with unidirectional layer construction (from top to bottom; or bottom to top),
it can balance the properties of all the layers simultaneously (including their total count) 
    to optimize the expected (cache-aware) end-to-end latency
        by solving a mathematical optimization problem.
During the optimization, \system
    considers many different design choices (e.g., the number of layers,
        fanout, model types, model size/accuracy),
    often producing a heterogeneous index 
        consisting of different types of layers:
    an index may have
        a B-tree-like design for one layer
            and a model-based design for another layer.
While the primary contribution of \system
    is demonstrating significant improvements in lookup speeds
        enabled by its principled search technique,
its approach can easily integrate with existing orthogonal work (e.g., ALEX~\cite{DingMYWDLZCGKLK2020ALEX})
    to allow insert/delete without complete reconstruction.
To our knowledge,
    \emph{no previous work has formally studied an efficient search technique
        for optimizing the entire index structure 
            in consideration of I/O performance
    and has evaluated its practical performance benefits
        with comprehensive experiments.}

\paragraph{Challenge}

The core challenge in finding an optimal index structure
    is the lack of an efficient mechanism that
        can compare the quality of exponentially many candidate structures.
For example, we need to consider indexes with different heights (i.e., 1, 2, $\ldots$);
    for an index of height $L$,
        there are $L$ layers to build;
    and for each layer,
        we need to compare various branching functions (e.g., B-tree-like child pointers,
    regression-based learned models). 
Moreover, the design decision we make for layers are dependent on one another;
    that is, if we alter the design of a layer, it affects the other components that depend on this layer
        (i.e., all the other layers on top of it)
    due to the inherent nature of hierarchical indexes,
        complicating the search process.
Finally,
    we must be able to evaluate the \emph{goodness} of each index
        for (arbitrary) target storage media.

\paragraph{Our Approach}

\system can be understood as a search process (\oursearch) 
    inside a high-dimensional design space consisting of exponentially many feasible candidate structures.
\;\; \textit{\textbf{Each (entire) candidate structure is a model:}}
We mathematically represent each candidate structure using a \emph{unified index model} (called \ourmodel),
    a high-level abstraction that expresses diverse hierarchical indexes
parametrically $\calL(\bfTheta)$,
    where $\bfTheta = (\Theta_1, \ldots, \Theta_L)$ are layer-specific design parameters (such as fanout, model accuracy/size),
        and $\calL(\bfTheta)$ is the (cache-aware) end-to-end lookup latency 
        (i.e., cumulative time for traversing the index).
That is, given a parameter set $\bfTheta$, there exists an associated physical index; and
    the index takes $\calL(\bfTheta)$ time for finding a relevant item via traversal (thus, the lower the better).
\;\; \textit{\textbf{Search:}}
    Our key contribution is an efficient search algorithm for finding an optimal design $\bfTheta^*$
    at which $\calL(\bfTheta)$ has the smallest value.
Since the design includes both numeric and categorical values, gradient-descent-like optimization algorithms are inapplicable.
We devise a novel algorithm by translating our problem into a graph traversal;
    that is, each design represents a vertex (or state) in the search space
        and an edge is drawn from $\bfTheta_A$ to $\bfTheta_B$ if $\bfTheta_B$ is immediately reachable from $\bfTheta_A$
            by stacking another layer on top.
While our algorithm involves actual index construction---because for some types of layers,
    their quality depends on data distribution requiring actual construction---our 
        algorithm can quickly find $\bfTheta^*$ with bounded time
by avoiding visits to the (low-quality) states unlikely to reach an optimal one.

\paragraph{Orthogonal Work}

Database research has a long history in index design
    such as (1) leveraging the properties of emerging hardware,
        (2) employing machine learning for compact layers, and
    (3) targetting for different data layouts.
We summarize more closely related work in \cref{tab:existing} while providing comprehensive discussions in \cref{sec:related}.
Our work is largely orthogonal to those existing work
    because we do not propose a new type of index per se;
instead, we study how to combine existing techniques under a unified framework
    to build an index specifically tuned for target I/O characteristics.

\paragraph{Contribution}
% To our knowledge,
%     \emph{no known work has formally investigated the problem of optimizing indexes 
%         from both data distributions and storage performance,
%     and conducted extensive empirical studies to show its practical performance impact.}
% \cref{fig:intro:quadrant} compares \system 
%     in the context of existing methods.
% Traditional indexes (e.g., B-tree~\cite{Bayer1972,Comer1979, Graefe2001, Chen2001, Chen2002, Wang2020, Wu2010, Zhou2014, Bin2014, bender2000CacheObliviousBtree}, skip lists~\cite{Dick2017, Sprenger2017, Daly2018, He2018})
%     are agnostic to data distributions, producing essentially the same indexes
%         if the numbers of indexed keys are the same.
% \emph{Learned indexes}~\cite{kipf2018learned,DBLP:journals/pvldb/FerraginaV20PGMIndex, DBLP:conf/sigmod/KipfMRSKK020RadixSpline,DingMYWDLZCGKLK2020ALEX} can adaptively 
%     fit to data distributions; however, 
%         they do not consider storage performance in their learning.
% In contrast,
%     \system learns indexes from both data distribution and storage performance,
% which serves as the core reason
%     for its up to 2.6$\times$--11.7$\times$ faster lookup speed 
% than state-of-the-art methods (\cref{sec:exp}).
In this work, we make the following contributions:
\begin{enumerate}
    \item We illustrate the significance of optimizing indexes with 
        both data and storage (\cref{sec:bg}).
    % \item We propose an index building algorithm (called \system) on top of generic storage model (\cref{sec:overview}).
    \item We formulate the index optimization as a search
        for the optimal design parameters of
            a \emph{unified index model} (\cref{sec:system-overall}).
    \item We design an efficient graph-based search method (\cref{sec:system-building}).
    % \item To learn optimal indexes from both data and storage, we formulate an optimization problem () and design an optimization method to solve the problem (\cref{sec:system-building}).
    \item We empirically study \system with various datasets and storage and show that \system can offer up to 2.0$\times$--46.3$\times$ faster lookup speed compared to state-of-the-art methods (\cref{sec:exp}).
\end{enumerate}

\noindent
Finally, \cref{sec:related} discusses related work, and \cref{sec:con} concludes this work.

\begin{figure*}[t]

\tikzset{
mynode/.style={
    draw=black,thick,font=\scriptsize\sf,
    minimum height=3mm,minimum width=4mm,
},
mylarge/.style={
    mynode,minimum width=16mm,
},
mylayer/.style={
    draw=gray!50!white,ultra thick,rounded corners=1mm,
    inner ysep=0.5mm,inner xsep=1mm
}
}

\pgfplotsset{mystyle/.style={
    ybar,
    % bar shift=0pt,
    xtick=data,
    width=55mm,
    height=32mm,
    bar width=4mm,
    ymin=0.5,
    ymax=2.0,
    axis lines=left,
    % axis y line*=none,
    % axis x line*=none,
    xmin = 0.5,
    xmax = 2.5,
    xtick={1,2},
    xticklabels = {\enva, \envb},
    xtick style={draw=none},
    ytick={0, 0.5, 1.0, 1.5, 2.0},
    tick label style={font=\scriptsize},
    legend style={
        at={(1.0,1.0)},anchor=north east,column sep=2pt,
        draw=black,fill=white,line width=.5pt,
        /tikz/every even column/.append style={column sep=5pt},
        font=\scriptsize,
    },
    legend cell align={left},
    legend columns=2,
    ylabel near ticks,
    xticklabel shift=-1mm,
    label style={font=\footnotesize},
    ylabel={Relative Latency},
    ymajorgrids,
    area legend,
    legend image code/.code={%
        \draw[#1, draw=none] (0cm,-0.1cm) rectangle (0.6cm,0.1cm);
    },
    clip=false,
}}

\begin{subfigure}[b]{0.32\textwidth}
\centering
\begin{tikzpicture}

\def\h{0.6}
\def\d{0.05}

\node[mynode] (L1) at (0,0) {};

\node[mynode] (L21) at ($(L1)+(-0.8,-\h)$) {};
\node[mynode,anchor=west] (L22) at ($(L21.east)+(\d,0)$) {};
\node[mynode,anchor=west,draw opacity=0,font=\Large\bf] (L23) at ($(L22.east)+(\d,0)$) {$\cdots$};
\node[mynode,anchor=west] (L24) at ($(L23.east)+(\d,0)$) {};

\node[mynode] (L31) at ($(L21)+(-0.65,-\h)$) {};
\node[mynode,anchor=west] (L32) at ($(L31.east)+(\d,0)$) {};
\node[mynode,anchor=west,draw opacity=0,font=\Large\bf] (L33) at ($(L32.east)+(\d,0)$) {$\cdots$};
\node[mynode,anchor=west,draw opacity=0,font=\Large\bf] (L34) at ($(L33.east)+(\d,0)$) {$\cdots$};
\node[mynode,anchor=west] (L35) at ($(L34.east)+(\d,0)$) {};
\node[mynode,anchor=west] (L36) at ($(L35.east)+(\d,0)$) {};

\node[mynode,minimum width=50mm] (D) at ($(L31)!0.5!(L36) + (0,-\h)$) {Data Layer (1M records)};

% Layers
\node[mylayer,fit={(L21.north west) (L24.south east)}] {};

\node[mylayer,fit={(L31.north west) (L36.south east)}] {};

% Arrows
\draw[-latex,draw=black,thick] (L1) -- (L22.north);
% \draw[-latex,draw=black,thick] (L1) -- (L22.north east);
\draw[-latex,draw=black,thick] (L22) -- (L32.north);
% \draw[-latex,draw=black,thick] (L22) -- (L32.north east);
\draw[-latex,draw=black,thick] (L32) -- ($(D.north)+(-1,0)$);

\end{tikzpicture}
\caption{\sta: B-tree with 200 child pointers}
\label{fig:motivation:a}
\end{subfigure}
\hfill
\begin{subfigure}[b]{0.32\textwidth}
\centering
\begin{tikzpicture}

\def\h{0.7}
\def\d{-0.2}

\node[mylarge] (L1) at (0,0) {};

\node[mylarge,draw opacity=0,font=\Large\bf] (L22) at ($(L1)+(0,-\h)$) {$\cdots$};
\node[mylarge,anchor=east] (L21) at ($(L22.west)+(-\d,0)$) {};
\node[mylarge,anchor=west] (L23) at ($(L22.east)+(\d,0)$) {};

\node[mynode,minimum width=50mm] (D) at ($(L21)!0.5!(L23) + (0,-\h)$) {Data Layer (1M records)};

% Layers
\node[mylayer,fit={(L21.north west) (L23.south east)}] {};

% Arrows
\draw[-latex,draw=black,ultra thick] (L1) -- (L21);
\draw[-latex,draw=black,ultra thick] (L21) -- ($(D.north)+(-1,0)$);

\end{tikzpicture}
\caption{\stb: B-tree with 5,000 child pointers}
\label{fig:motivation:b}
\end{subfigure}
\hfill
\begin{subfigure}[b]{0.32\textwidth}
\centering
\begin{tikzpicture}
\begin{axis}[
    mystyle
]

% 200 fanout
\addplot[fill=barGreen,draw=none]
table[x=x,y=y] {
x y
1 1
2 1.32
};

% 50,000 fanout
\addplot[fill=barOrange,draw=none]
table[x=x,y=y] {
x y
1 1.21
2 1.0
};

\addlegendentry{\sta}
\addlegendentry{\stb}

\end{axis}
\end{tikzpicture}
\vspace{-2mm}
\caption{Lookup Speed Comparison}
\label{fig:motivation:c}
\end{subfigure}

\vspace{-2mm}
\caption{Need for I/O-aware Optimization. 
Depending on system environments (\enva and \envb),
    employing different B-tree designs (\sta and \stb)
        achieve higher expected performance 
    in end-to-end lookup latency.}
% \vspace{-2mm}
\end{figure*}
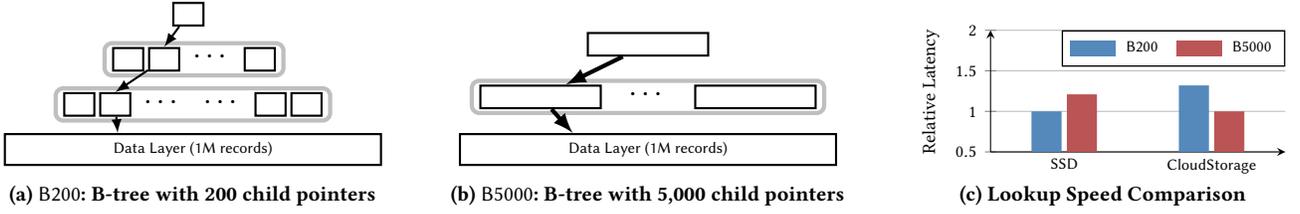

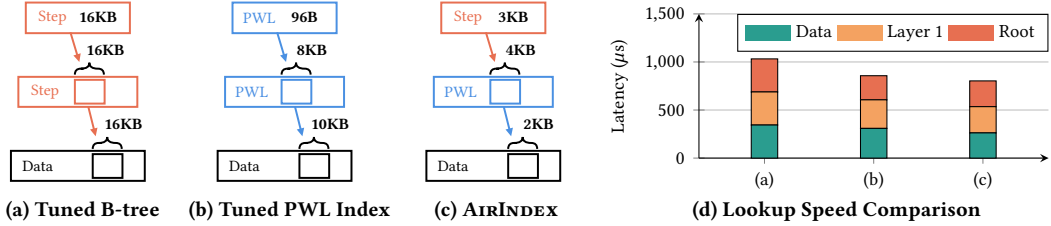
\begin{figure*}[t]
\def\subfigwidth{0.15\linewidth}
\def\subfigbarwidth{0.35\linewidth}
\pgfplotstableread{ % data 
label       root    layerone    data
tunedbtree  344     344         344
% band        258     344         344
tunedband   251     297         309
airindex    268     273         262
}\testdata
\begin{subfigure}[b]{\subfigwidth}
    \centering
    \tikzset{every picture/.style={line width=0.75pt}} %set default line width to 0.75pt        

\begin{tikzpicture}[x=0.75pt,y=0.75pt,yscale=-0.75,xscale=0.75]
%uncomment if require: \path (0,122); %set diagram left start at 0, and has height of 122

%Shape: Rectangle [id:dp7180312855569165] 
\draw  [color={rgb, 255:red, 231; green, 111; blue, 81 }  ,draw opacity=1 ] (10,0) -- (80,0) -- (80,20.6) -- (10,20.6) -- cycle ;
%Shape: Rectangle [id:dp0814412770186217] 
\draw  [color={rgb, 255:red, 231; green, 111; blue, 81 }  ,draw opacity=1 ] (5,50) -- (85,50) -- (85,70.2) -- (5,70.2) -- cycle ;
%Shape: Rectangle [id:dp010615945758501777] 
\draw  [color={rgb, 255:red, 0; green, 0; blue, 0 }  ,draw opacity=1 ] (0,99.8) -- (90,99.8) -- (90,120) -- (0,120) -- cycle ;
%Straight Lines [id:da9044861502978696] 
\draw [color={rgb, 255:red, 231; green, 111; blue, 81 }  ,draw opacity=1 ]   (40,20) -- (45.14,37.13) ;
\draw [shift={(46,40)}, rotate = 253.3] [fill={rgb, 255:red, 231; green, 111; blue, 81 }  ,fill opacity=1 ][line width=0.08]  [draw opacity=0] (5.36,-2.57) -- (0,0) -- (5.36,2.57) -- cycle    ;
%Straight Lines [id:da911892995804679] 
\draw [color={rgb, 255:red, 231; green, 111; blue, 81 }  ,draw opacity=1 ]   (52,70) -- (56.27,87.09) ;
\draw [shift={(57,90)}, rotate = 255.96] [fill={rgb, 255:red, 231; green, 111; blue, 81 }  ,fill opacity=1 ][line width=0.08]  [draw opacity=0] (5.36,-2.57) -- (0,0) -- (5.36,2.57) -- cycle    ;
%Shape: Rectangle [id:dp2127014847397226] 
\draw  [color={rgb, 255:red, 231; green, 111; blue, 81 }  ,draw opacity=1 ] (43,52) -- (63,52) -- (63,68) -- (43,68) -- cycle ;
%Shape: Rectangle [id:dp5875760120983763] 
\draw  [color={rgb, 255:red, 0; green, 0; blue, 0 }  ,draw opacity=1 ] (55,102) -- (75,102) -- (75,118) -- (55,118) -- cycle ;
%Shape: Brace [id:dp3015215805356195] 
\draw   (62.5,48.5) .. controls (62.5,45.83) and (61.16,44.49) .. (58.49,44.49) -- (58.49,44.49) .. controls (54.66,44.49) and (52.75,43.15) .. (52.75,40.47) .. controls (52.75,43.15) and (50.84,44.49) .. (47.01,44.49)(48.74,44.49) -- (47.01,44.49) .. controls (44.34,44.49) and (43,45.83) .. (43,48.5) ;
%Shape: Brace [id:dp2574743683702264] 
\draw   (74.5,98.5) .. controls (74.5,95.83) and (73.16,94.49) .. (70.49,94.49) -- (70.49,94.49) .. controls (66.66,94.49) and (64.75,93.15) .. (64.75,90.47) .. controls (64.75,93.15) and (62.84,94.49) .. (59.01,94.49)(60.74,94.49) -- (59.01,94.49) .. controls (56.34,94.49) and (55,95.83) .. (55,98.5) ;

% Text Node
\draw (28,10) node  [color={rgb, 255:red, 0; green, 0; blue, 0 }  ,opacity=1 ] [align=left] {{\scriptsize \textcolor[rgb]{0.91,0.44,0.32}{Step}}};
% Text Node
\draw (23,60) node  [color={rgb, 255:red, 0; green, 0; blue, 0 }  ,opacity=1 ] [align=left] {{\scriptsize \textcolor[rgb]{0.91,0.44,0.32}{Step}}};
% Text Node
\draw (18,109.9) node  [color={rgb, 255:red, 0; green, 0; blue, 0 }  ,opacity=1 ] [align=left] {{\scriptsize Data}};
% Text Node
\draw (63.5,32) node  [color={rgb, 255:red, 0; green, 0; blue, 0 }  ,opacity=1 ] [align=left] {{\scriptsize \textbf{16KB}}};
% Text Node
\draw (75.5,82) node  [color={rgb, 255:red, 0; green, 0; blue, 0 }  ,opacity=1 ] [align=left] {{\scriptsize \textbf{16KB}}};
% Text Node
\draw (58.5,10) node  [color={rgb, 255:red, 0; green, 0; blue, 0 }  ,opacity=1 ] [align=left] {{\scriptsize \textbf{16KB}}};

\end{tikzpicture}
    \caption{Tuned B-tree}
    \label{fig:structures:btree}
\end{subfigure}
% ~
% \begin{subfigure}[b]{\subfigwidth}
%     \centering
%     \input{figures/background/structures/2_band}
%     \caption{PWL Index}
%     \label{fig:structures:band}
% \end{subfigure}
~
\begin{subfigure}[b]{\subfigwidth}
    \centering
    \tikzset{every picture/.style={line width=0.75pt}} %set default line width to 0.75pt        

\begin{tikzpicture}[x=0.75pt,y=0.75pt,yscale=-0.75,xscale=0.75]
%uncomment if require: \path (0,122); %set diagram left start at 0, and has height of 122

%Shape: Rectangle [id:dp3718453341175926] 
\draw  [color={rgb, 255:red, 74; green, 144; blue, 226 }  ,draw opacity=1 ] (10,0) -- (80,0) -- (80,20.6) -- (10,20.6) -- cycle ;
%Shape: Rectangle [id:dp8898896595862034] 
\draw  [color={rgb, 255:red, 74; green, 144; blue, 226 }  ,draw opacity=1 ] (5,50) -- (85,50) -- (85,70.2) -- (5,70.2) -- cycle ;
%Shape: Rectangle [id:dp519434759465839] 
\draw  [color={rgb, 255:red, 0; green, 0; blue, 0 }  ,draw opacity=1 ] (0,99.8) -- (90,99.8) -- (90,120) -- (0,120) -- cycle ;
%Straight Lines [id:da40095108241834465] 
\draw [color={rgb, 255:red, 74; green, 144; blue, 226 }  ,draw opacity=1 ]   (40,20) -- (45.14,37.13) ;
\draw [shift={(46,40)}, rotate = 253.3] [fill={rgb, 255:red, 74; green, 144; blue, 226 }  ,fill opacity=1 ][line width=0.08]  [draw opacity=0] (5.36,-2.57) -- (0,0) -- (5.36,2.57) -- cycle    ;
%Straight Lines [id:da5864965621787549] 
\draw [color={rgb, 255:red, 74; green, 144; blue, 226 }  ,draw opacity=1 ]   (52,70) -- (56.27,87.09) ;
\draw [shift={(57,90)}, rotate = 255.96] [fill={rgb, 255:red, 74; green, 144; blue, 226 }  ,fill opacity=1 ][line width=0.08]  [draw opacity=0] (5.36,-2.57) -- (0,0) -- (5.36,2.57) -- cycle    ;
%Shape: Rectangle [id:dp2751526177254173] 
\draw  [color={rgb, 255:red, 74; green, 144; blue, 226 }  ,draw opacity=1 ] (43,52) -- (63,52) -- (63,68) -- (43,68) -- cycle ;
%Shape: Rectangle [id:dp1977996293808365] 
\draw  [color={rgb, 255:red, 0; green, 0; blue, 0 }  ,draw opacity=1 ] (55,102) -- (75,102) -- (75,118) -- (55,118) -- cycle ;
%Shape: Brace [id:dp9054953773273856] 
\draw   (62.5,48.5) .. controls (62.5,45.83) and (61.16,44.49) .. (58.49,44.49) -- (58.49,44.49) .. controls (54.66,44.49) and (52.75,43.15) .. (52.75,40.47) .. controls (52.75,43.15) and (50.84,44.49) .. (47.01,44.49)(48.74,44.49) -- (47.01,44.49) .. controls (44.34,44.49) and (43,45.83) .. (43,48.5) ;
%Shape: Brace [id:dp22794197032286712] 
\draw   (74.5,98.5) .. controls (74.5,95.83) and (73.16,94.49) .. (70.49,94.49) -- (70.49,94.49) .. controls (66.66,94.49) and (64.75,93.15) .. (64.75,90.47) .. controls (64.75,93.15) and (62.84,94.49) .. (59.01,94.49)(60.74,94.49) -- (59.01,94.49) .. controls (56.34,94.49) and (55,95.83) .. (55,98.5) ;

% Text Node
\draw (28,10) node  [color={rgb, 255:red, 0; green, 0; blue, 0 }  ,opacity=1 ] [align=left] {{\scriptsize \textcolor[rgb]{0.29,0.56,0.89}{PWL}}};
% Text Node
\draw (23,60) node  [color={rgb, 255:red, 0; green, 0; blue, 0 }  ,opacity=1 ] [align=left] {{\scriptsize \textcolor[rgb]{0.29,0.56,0.89}{PWL}}};
% Text Node
\draw (18,109.9) node  [color={rgb, 255:red, 0; green, 0; blue, 0 }  ,opacity=1 ] [align=left] {{\scriptsize Data}};
% Text Node
\draw (63.5,32) node  [color={rgb, 255:red, 0; green, 0; blue, 0 }  ,opacity=1 ] [align=left] {{\scriptsize \textbf{8KB}}};
% Text Node
\draw (75.5,82) node  [color={rgb, 255:red, 0; green, 0; blue, 0 }  ,opacity=1 ] [align=left] {{\scriptsize \textbf{10KB}}};
% Text Node
\draw (58.5,10) node  [color={rgb, 255:red, 0; green, 0; blue, 0 }  ,opacity=1 ] [align=left] {{\scriptsize \textbf{96B}}};

\end{tikzpicture}
    \caption{Tuned PWL Index}
    \label{fig:structures:tuned-band}
\end{subfigure}
~
\begin{subfigure}[b]{\subfigwidth}
    \centering
    \tikzset{every picture/.style={line width=0.75pt}} %set default line width to 0.75pt        

\begin{tikzpicture}[x=0.75pt,y=0.75pt,yscale=-0.75,xscale=0.75]
%uncomment if require: \path (0,122); %set diagram left start at 0, and has height of 122

%Shape: Rectangle [id:dp10465379654861418] 
\draw  [color={rgb, 255:red, 231; green, 111; blue, 81 }  ,draw opacity=1 ] (10,0) -- (80,0) -- (80,20.6) -- (10,20.6) -- cycle ;
%Shape: Rectangle [id:dp280512129169157] 
\draw  [color={rgb, 255:red, 74; green, 144; blue, 226 }  ,draw opacity=1 ] (5,50) -- (85,50) -- (85,70.2) -- (5,70.2) -- cycle ;
%Shape: Rectangle [id:dp4194751896834722] 
\draw  [color={rgb, 255:red, 0; green, 0; blue, 0 }  ,draw opacity=1 ] (0,99.8) -- (90,99.8) -- (90,120) -- (0,120) -- cycle ;
%Straight Lines [id:da12612557087706888] 
\draw [color={rgb, 255:red, 231; green, 111; blue, 81 }  ,draw opacity=1 ]   (40,20) -- (45.14,37.13) ;
\draw [shift={(46,40)}, rotate = 253.3] [fill={rgb, 255:red, 231; green, 111; blue, 81 }  ,fill opacity=1 ][line width=0.08]  [draw opacity=0] (5.36,-2.57) -- (0,0) -- (5.36,2.57) -- cycle    ;
%Straight Lines [id:da6093755468061124] 
\draw [color={rgb, 255:red, 74; green, 144; blue, 226 }  ,draw opacity=1 ]   (52,70) -- (56.27,87.09) ;
\draw [shift={(57,90)}, rotate = 255.96] [fill={rgb, 255:red, 74; green, 144; blue, 226 }  ,fill opacity=1 ][line width=0.08]  [draw opacity=0] (5.36,-2.57) -- (0,0) -- (5.36,2.57) -- cycle    ;
%Shape: Rectangle [id:dp4617992667101478] 
\draw  [color={rgb, 255:red, 74; green, 144; blue, 226 }  ,draw opacity=1 ] (43,52) -- (63,52) -- (63,68) -- (43,68) -- cycle ;
%Shape: Rectangle [id:dp5800153183778421] 
\draw  [color={rgb, 255:red, 0; green, 0; blue, 0 }  ,draw opacity=1 ] (55,102) -- (75,102) -- (75,118) -- (55,118) -- cycle ;
%Shape: Brace [id:dp15565274896437653] 
\draw   (62.5,48.5) .. controls (62.5,45.83) and (61.16,44.49) .. (58.49,44.49) -- (58.49,44.49) .. controls (54.66,44.49) and (52.75,43.15) .. (52.75,40.47) .. controls (52.75,43.15) and (50.84,44.49) .. (47.01,44.49)(48.74,44.49) -- (47.01,44.49) .. controls (44.34,44.49) and (43,45.83) .. (43,48.5) ;
%Shape: Brace [id:dp545447165545339] 
\draw   (74.5,98.5) .. controls (74.5,95.83) and (73.16,94.49) .. (70.49,94.49) -- (70.49,94.49) .. controls (66.66,94.49) and (64.75,93.15) .. (64.75,90.47) .. controls (64.75,93.15) and (62.84,94.49) .. (59.01,94.49)(60.74,94.49) -- (59.01,94.49) .. controls (56.34,94.49) and (55,95.83) .. (55,98.5) ;

% Text Node
\draw (28,10) node  [color={rgb, 255:red, 0; green, 0; blue, 0 }  ,opacity=1 ] [align=left] {{\scriptsize \textcolor[rgb]{0.91,0.44,0.32}{Step}}};
% Text Node
\draw (23,60) node  [color={rgb, 255:red, 0; green, 0; blue, 0 }  ,opacity=1 ] [align=left] {{\scriptsize \textcolor[rgb]{0.29,0.56,0.89}{PWL}}};
% Text Node
\draw (18,109.9) node  [color={rgb, 255:red, 0; green, 0; blue, 0 }  ,opacity=1 ] [align=left] {{\scriptsize Data}};
% Text Node
\draw (63.5,32) node  [color={rgb, 255:red, 0; green, 0; blue, 0 }  ,opacity=1 ] [align=left] {{\scriptsize \textbf{4KB}}};
% Text Node
\draw (75.5,82) node  [color={rgb, 255:red, 0; green, 0; blue, 0 }  ,opacity=1 ] [align=left] {{\scriptsize \textbf{2KB}}};
% Text Node
\draw (58.5,10) node  [color={rgb, 255:red, 0; green, 0; blue, 0 }  ,opacity=1 ] [align=left] {{\scriptsize \textbf{3KB}}};

\end{tikzpicture}
    \caption{\system}
    \label{fig:structures:system}
\end{subfigure}
~
\begin{subfigure}[b]{\subfigbarwidth}
    \begin{tikzpicture}
        \begin{axis}[
            height=35mm,
            width=\linewidth,
            axis lines=left,
            % axis y line*=none,
            % axis x line*=none,
            ymajorgrids,
            yminorgrids,
            ybar stacked,
            ymin=0,
            ymax=1500,
            enlarge x limits=0.3,
            xtick=data,
            xticklabels={(a),(b),(c),(d)},
            ylabel={Latency ($\mu$s)},
            % xticklabel style={rotate=18},
            % reverse legend,
            legend image code/.code={%
                \draw[#1, draw=none] (0cm,-0.1cm) rectangle (0.6cm,0.1cm);
            },
            legend style={
                % nodes={scale=0.65, transform shape}, 
                % at={(0.3,1.1)}, anchor=south,
                at={(1.0,1.0)}, anchor=north east,
                draw=black,
                font=\footnotesize},
            legend cell align={center},
            legend columns=3,
            every axis/.append style={font=\footnotesize},
        ]
        % \addplot [fill={rootcolor}] table [y=root, meta=label,x expr=\coordindex] {\testdata};
        % \addplot [fill={layeronecolor}] table [y=layerone, meta=label,x expr=\coordindex] {\testdata};
        % \addplot [fill={datalayercolor}] table [y=data, meta=label,x expr=\coordindex] {\testdata};
        % \legend{Root,Layer 1,Data}
        \addplot [fill={datalayercolor}] table [y=data, meta=label,x expr=\coordindex] {\testdata};
        \addplot [fill={layeronecolor}] table [y=layerone, meta=label,x expr=\coordindex] {\testdata};
        \addplot [fill={rootcolor}] table [y=root, meta=label,x expr=\coordindex] {\testdata};
        \legend{Data,Layer 1,Root}
        \end{axis}
    \end{tikzpicture}
    \vspace{-2mm}
    \caption{Lookup Speed Comparison}
    \label{fig:structures:latency}
\end{subfigure}
\vspace{-3mm}
\caption{Indexes built for \texttt{gmm} dataset stored on SSD (250$\mu$s latency, 175MB/s bandwidth) and their estimated latency. Bold numbers show the average read sizes of components in a query: root layer, partial first index layer, and partial data layer.}
\label{fig:structures}
% \vspace{-2mm}
\end{figure*}

\section{Motivation}
\label{sec:motivation}
\label{sec:bg}

We present why we need a new index builder that considers the end-to-end lookup latency
(\cref{sec:motivation:problem,sec:motivation:io,sec:motivation:layer}).
% Then, we give a high-level description
%     of how we can find an optimal index among exponentially many candidates (\cref{sec:motivation:opt}).

\subsection{Need for I/O-Aware Optimization}
\label{sec:motivation:io}

In this section, 
we motivate the need for environment-specific index optimization
    with concrete examples showing
there is no single dominant index structure (e.g., B-trees
        with fixed fanout) that can offer superior performance
    for all system environments.

\paragraph{Example}

We have two candidate B-tree structures: \sta and \stb
(Note: special cases of \ourmodel).
\sta consists of 4 KB nodes, each with 200 fanout. 
\stb consists of 100 KB nodes, each node with 5,000 fanout. 
Both \sta and \stb index the same dataset with one million distinct keys, stored in 4 KB pages.

To index the dataset,
\sta needs three layers (because the third layer can hold up to $200^3 = 8M$ pointers).
Likewise, \stb needs two layers (because the second layer can hold up to $5000^2 = 25M$ pointers). 
Note that while \stb is shallower than \sta, fetching each node of \stb takes longer because its page size is 25$\times$ larger.
\cref{fig:motivation:a,fig:motivation:b} depict these structures.

Interestingly, neither of these two indexes (i.e., \sta and \stb)
is superior to the other,
    if we compare their lookup latencies 
    based on a widely used formula: (data transfer time) $=$
(latency) + (data size) / (bandwidth);
that is, (1) \sta offers higher performance than \stb if we store data on \enva having 100 $\mu$s latency and 1 GB/s bandwidth,
and that (2) \stb offers higher performance than \sta if we store data on \envb having 100 ms latency and 100 MB/s bandwidth.
Specifically, \textbf{in \enva}, \stb is 21\% slower than \sta
because \sta needs to retrieve 3 nodes and 1 data page,
    taking 416 $\mu$s (=
3 $\times$ (100$\mu$s + 4KB / (1GB/s)) + (100$\mu$s + 4KB / (1GB/s)) )
while \stb needs to retrieve 2 nodes and 1 data page,
    taking 504 $\mu$s (=
2 $\times$ (100$\mu$s + 100KB / (1GB/s)) + (100$\mu$s + 4KB / (1GB/s)) ).
In contrast,
    \textbf{in \envb}, \sta is 32\% slower than \stb
        with \sta taking 400.16 ms
            and \stb taking 302.04 ms.
\cref{fig:motivation:c} summarizes this relative performance strength.
For each environment, the figure reports the \emph{relative} difference in 
end-to-end lookup time.
This shows that different index structures offer higher lookup
performance, depending on the storage device.

\subsection{Need for Layer-Wise Optimization}
\label{sec:motivation:layer}

In this section, we explain why we need to consider
    different types of branching functions for different layers.
That is,
    a homogeneous index---consisting of the same type of layers---may
offer poorer performance even with careful tuning
    in comparison to a tuned heterogeneous index---consisting
    of different types of layers.

\paragraph{General Lookup Process}

In general, we can consider a lookup process with a hierarchical index as
    a series of data fetch operations that proceed as follows:
the root (or the $i$-th) layer is fetched, based on which 
    we determine what data we need to fetch in the next (or the $(i-1)$-th) layer.
This iteration repeats until we reach the data layer.
This means that
    in each layer, we can use any type of monotonically increasing function 
        (with respect to search keys)
    that can tell us what data we need to fetch in the subsequent layer.
For example,
    a B-tree layer has a property such that
        for all the keys between two adjacent separators,
    we get the same child pointer (or the same range of data we need to fetch),
which can be expressed as a step function
    that \emph{jumps} at separators.
This gives rise to our \emph{unified index model}, \ourmodel,
    allowing different types of branching functions
        in different layers (\cref{sec:system-overall}).

Formally,
    we express such a design space with $\Theta_l$ for $l$-th layer,
and $\bfTheta = (\Theta_1, \ldots, \Theta_L)$ describes an entire index design.
The following examples demonstrate that
    by allowing $\Theta_i \ne \Theta_j$ for $i \ne j$,
we can construct an index with lower end-to-end latency.

\paragraph{Concrete Example}

Suppose two types of branching functions---step functions (Step) appearing in B-trees
    and piece-wise linear functions (PWL) employed in RMI~\cite{Kraska2018}.
Formally, a step function is a $p$-piece constant function that $\yh_{\text{step}}(x) = [b_i, b_{i+1})$, while a PWL is a $p$-piece band function (widen linear functions) $\yh_{\text{PWL}}(x) = [m_i x + c_i - \delta_i, m_i x + c_i + \delta_i)$ for $x \in [a_i, a_{i+1})$.
We present a case where
    by combining Step and PWL, we can build a higher-performing index than 
the ones solely comprising Step or PWL, respectively.

First, we construct an optimal B-tree index (\cref{fig:structures:btree}) by comparing
    the latencies of multiple B-trees with different node sizes (100 bytes--10 MB):
for the dataset we use, 16KB nodes offer the fastest lookup.
In \cref{fig:structures:latency},
    we show the costs of index traversal
    % by measuring the time it takes
        to fetch each layer
        % starting
        from the root to
            % we fetch a relevant key-value in
        the data layer to find the relevant key-value;
For the tuned B-tree, the end-to-end latency takes slightly longer than 1,000$\mu$s.
Likewise, we construct an optimal index solely consisting of PWL layers (\cref{fig:structures:tuned-band}).
The dataset we consider has its keys distributed favorably for PWL;
    thus, we need smaller layers than the optimal B-tree,
        needing to fetch 96B for the root, 8KB for Layer 1, 
    and 10KB for the data layer.
Accordingly,
    the overall latency shown in \cref{fig:structures:latency} is
        also lower than the B-tree.

In contrast, by combing Step and PWL,
    we can discover an index (denoted by \system) with significantly more compact layers overall
(\cref{fig:structures:system}).
Specifically, while \system's root layer is bigger in size
    compared to the tuned PWL index,
it allows fetching
    significantly smaller amounts of data for the other layers (i.e., Layer 1 and the data layer),
lowering the end-to-end latency.
In theory,
    an optimal index with heterogeneous layers
is guaranteed to deliver performance not worse
    than the optimal index with homogeneous layers.
Nevertheless,
    the amount of improvement we can offer with layer-wise optimization
well compensates for the increase in search effort,
    as we demonstrate with more empirical results in \cref{sec:exp}.

\begin{figure}[t]

\tikzset{
mynode/.style={
    draw=black,thick,font=\footnotesize\sf,
    minimum height=7mm,minimum width=18mm,
    align=center,
    % inner ysep=0.4mm,
}
}

\begin{tikzpicture}

\def\h{16mm}

\node[mynode] (R) {Lookup (\cref{sec:querying})};

\def\d{0.4}
\node[mynode,anchor=north west,minimum height=\h,minimum width=1mm] 
    (I) at ($(R.north east)+(\d,0)$) {Storage\\ Layer\\ Interface};
    
\node[mynode,anchor=south east] (B) at ($(I.south west)+(-\d,0)$) {\oursearch (\cref{sec:system-building})};

\node[mynode,anchor=north west,minimum height=\h,minimum width=10mm] 
    (S) at ($(I.north east)+(0.8,0)$) {Storage\\Layer\\(e.g., SSD,\\ NFS)};
    
\node[font=\bf,rotate=90] (N) at ($(R.west)+(-0.45,-0.5)$) {\system};
    
\node[mynode,fit={(N.north east) ($(I.south east)+(0.1,0)$)},
    thick,dashed,draw=black,
    inner xsep=0.5mm,inner ysep=1.5mm] (A) {};
    
\node[font=\normalsize\bf,anchor=east] (C) at ($(A.west)+(-0.6,0)$) {client};

% Arrows
\draw[to-to,ultra thick] (R.east) -- (R.east-|I.west);
\draw[to-to,ultra thick] (B.east) -- (B.east-|I.west);
\draw[to-to,ultra thick] (I.east) -- (S.west);
\draw[to-to,ultra thick] (C.east) -- (A.west);

\end{tikzpicture}

\vspace{-2mm}
\caption{\system Architecture. \system manages indexes on Storage Layer
    (e.g., SSD, NFS, cloud storage APIs)
    via its Storage Layer Interface that abstracts read/write operations.}
\label{fig:architecture}
% \vspace{-2mm}
\end{figure}
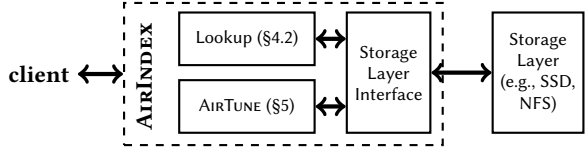

\subsection{Need for Novel Index Tuning}
\label{sec:motivation:problem}

\paragraph{New Search Space} 
% The search space that captures all possible index designs has recently grown exponentially.
    Because of the new connection between machine learning and indexing~\cite{Kraska2018}, 
    the search space has grown exponentially,
    including
    % a part of index designs is to determine the 
    models types (e.g., linear models to neural networks) 
    as well as their parameters (e.g., regression coefficients, weights), 
    hyperparameters (e.g., fitting methods, regularization weights), 
    and error correction algorithm (i.e., last mile search).
    The search space expands exponentially as each index sub-structure, like layer or node, can choose its model design individually.

    \paragraph{Lack of Predictability} Unlike traditional indexes, learned indexes have an unpredictable performance \emph{before} fitting them to the data, which consequently increases tuning costs. While we can rigorously analyze some traditional indexes like B-trees~\cite{Bayer1972,Comer1979} and skip-lists~\cite{pugh1990skip} to predict their worst-case or average-case performance given a set of hyperparameters (e.g., fanout, level fraction, key prefix size), analyzing learned indexes and learned models is much more challenging because:
    (1) learning performance depends on non-trivial statistics of dataset, in which the dependency may not yet be explainable,
    (2) learned models often comprise of many learning components whose existing theories may not combine together,
    and (3) learning may unreliably produce poorly fitted models due to randomization or numerical instability.
    As a result, learned index tuning needs to pay the price to fit models or restrict the search space to a limited set of well-understood learned models.

    \paragraph{Existing Tuning's Limitations} Existing tuning methods are either restricted, inconclusive, or inefficient.
    Traditional tuners like \cite{gray1997FiveMinuteRules} only apply to restricted sets of indexes prior to learned indexes, missing opportunities to fit better to data patterns and so save lookup latency.
    On the other hand, many existing learned index tuners inconclusively recommend many index designs, for example, \cdfshop~\cite{DBLP:conf/sigmod/MarcusZK20CDFShop} finds many Pareto efficient RMIs while \pgm~\cite{DBLP:journals/pvldb/FerraginaV20PGMIndex} and \plex~\cite{DBLP:journals/corr/Stoian2021PLEX} reduce their parameters into fewer hyperparameter(s), delegating selection or hyperparameter tuning to their users.
    % binary search: low cpu time, high real time (multiple passes sequentially on full dataset)
    Lastly, na\"ive methods such as brute-force search,
    % on index designs, 
    grid search,
    % on sub-structure designs, 
    or binary search
    % on an appropriate hyperparameter inefficiently 
    incur large overheads due to excessive fitting or multiple sequential passes on the full dataset.
    Later, our experiments (\cref{sec:exp-build-scale}) verify that these existing methods find suboptimal indexes and incur large tuning overheads.
    Therefore, a novel index tuning is needed to identify the optimal learned index from a much larger search space by efficiently balancing learning costs.

\section{System Overview}
\label{sec:overview}

We describe \system's core components (\cref{sec:overview:archi})
% the current scope of this work (\cref{sec:overview:scope}),
and a storage profile 
    representing I/O performance (\cref{sec:storage_model}).

% system operations

% \tofix{point lookup; why OK?}

% \tofix{why read-only OK? scope: read-only}

% \tofix{deployment under various cloud/storage interfaces}

% storage related

% \tofix{storage dominates latency}

% \tofix{storage model (abstraction)}

% \tofix{What is storage profile? function: sequential bytes -> time, Why generic function (not simple ones)}

% \tofix{What if storage profile significantly drifts over time?}

% \tofix{Variance in storage performance}

\subsection{Architecture}
\label{sec:overview:archi}

\system is an index library for sorted key-value data stored on a Storage Layer (e.g., SSD, NFS).
\system can work with various storage devices by extending its storage interface.
    % as described shortly.
\system stores indexes together with the data on Storage Layer
    while part of them may be cached.
Internally,
\system consists of three components---Lookup, Builder, and Storage Layer Interface---as
    depicted in \cref{fig:architecture}.
First, Storage Layer Interface provides the consistent abstraction over different storage interfaces/devices
    (e.g., virtual file systems (VFS), over SSD or object storage, cloud storage services)
    such as creating files
        and reading/writing serialized objects.
Based on this consistent interface, we can profile
    the time needed to read $\Delta$ bytes (\cref{sec:storage_model}).
Second, Lookup provides a querying interface.
    Given a key,
        the module traverses an index
    as it fetches necessary data (e.g., ranges within index layers) if not cached,
        caches the fetched data if space is available (\cref{sec:data_layout}),
    and makes inferences on what data it needs to fetch next (\cref{sec:querying}),
until it finds the value associated with the key (or if finds there are no such keys indexed).
Finally, Builder finds optimal index designs and
    builds/stores actual indexes on storage (\cref{sec:system-building-overall,sec:system-building-node-estimation,sec:system-building-selection}).
How Builder finds high-quality index designs efficiently
    is the key contribution of this work.

\begin{table}
    \centering
    \caption{Notations and their meaning.}
    \label{tab:notations}
    \vspace{-2mm}
    \footnotesize
    \begin{tabular}{ll}
        \toprule
        
        % \textbf{Symbol} & \textbf{Description} 
        
        \multicolumn{2}{l}{\textbf{Input to \system}} \\
        
        \midrule
        
        $x$ & Key \\
        % $\calX$ & Key distribution \\
        % $\E_{x \sim \calX}$ & Expectation over key distribution \\
        $y = [y^{-}, y^{+})$ & Position (range) on storage \\
        $D$ & Key-position dataset \\
        
        % \midrule
        
        $T(\Delta)$ & Storage profile, time to read $\Delta$ bytes from storage \\
        % $T_{\text{aff}}(\Delta)$ & Affine storage profile with latency $\ell$ and bandwidth $B$ \\
        % $\ell$ & \\
        % $B$ & \\

        \midrule

        \multicolumn{2}{l}{\textbf{Design Parameters (\cref{sec:data_layout})}} \\ 
        
        \midrule
        
        $L$ & Number of index layers \\
        $\bfTheta$ & Parameters across all index layers \\
        $\Theta_l$ & Parameters describing the $l$-th index layer \\
        % $\bfTheta^{*}$ & Optimal parameters \\
        $s(\Theta_l)$ & Size of the $l$-th index layer \\
        $\Delta(x; \Theta_l)$ & Read size of $x$ predicted by $l$-th index layer \\
        $\Delta(D; \Theta_l)$ & Average read size over keys in dataset $D$ \\
        $\yh = [\yh^{-}, \yh^{+})$ & A predicted position on storage \\
        % $(a_i, b_i)$ & \\
        % $(m, c, \delta)$ & \\
        
        % $\calL$ & \\
        $\calL_{SM}$ & Index lookup cost from storage model \\
        % $\calL_{\text{cache}}$ & \\
        % $\calL_{\text{compute}}$ & \\
        % $C$ & \\
        % $c_{\text{\text{compute}}}$ & \\
        
        \midrule

        \multicolumn{2}{l}{\textbf{Search Process (\cref{sec:system-building})}} \\ 

        \midrule
        
        % \midrule
        
        $F \in \calF$ & Node builder mapping a $D$ into a $\Theta$ \\
        % $\lambda_{GS}$ & \\
        % $\lambda_{GB}$ & \\
        % $\lambda_{EB}$ & \\
        $k$ & Number of top candidates to branch out \\
        $1 + \epsilon$ & exponential base for granularity exponentiation \\
        $\lambda_{low}, \lambda_{high}$ & Bounds for granularity exponentiation \\
        
        % \midrule
        
        $\tau(D)$ & Index complexity, the ideal index lookup cost \\
        $\hat{\tau}(D)$ & Step index complexity, an upper bound to $\tau(D)$ \\
        \bottomrule
    \end{tabular}
\end{table}

\subsection{Storage Model}
\label{sec:storage_model}

% \todo{Describe interface also. Is the storage supposed to be a file system-like interface?}

\system relies on \emph{storage performance profile} $T(\Delta)$, which represents the time taken for the storage layer to read consecutive data of size $\Delta$ bytes.
    Considering that the read time $\calT$ is probabilistic (e.g., due to variability, system loads, access paths, address alignment, lower-level optimizations), \system is interested in the conditional expectation $T(\Delta) = \E[ \calT | \Delta ]$.
    For example, $T(\Delta)$ can be an affine function $T_{\text{aff}}(\Delta) = \ell + \Delta / B$ with latency $\ell$ and bandwidth $B$.
    If the latency and bandwidth uniformly varies in $[\ell_0, \ell_1]$ and $[B_0, B_1]$, after calculating the expectation, the storage profile becomes $T_{\text{aff-uniform}}(\Delta) = \frac{\ell_1 + \ell_0}{2} + \frac{\Delta (\ln B_1 - \ln B_0)}{B_1 - B_0}$.
    While we implement the affine storage model $T_{\text{aff}}(\Delta)$ parameterized by $\ell$ and $B$, our optimization works with any monotonically increasing $T(\Delta)$.
% To obtain an accurate $T(\Delta)$, one could quantize $\Delta$ domain, measure latencies, and summarize the measurements with averages.

% To fit this affine storage profile, one could measure the latency $\ell$ using a small $\Delta$ and infer the bandwidth through another measurement of a bigger $\Delta$. 
% \cite{DBLP:journals/csur/Vitter2001ExternalMemoryModel} delves deeper into different models such as disk access model, parallel external memory model, and parallel disk model. 
% From this point on, we may refer to external memory and storage interchangeably.

Currently, we consider a deterministic and time-independent $T(\Delta)$,
        summarizing over other variables that are unnecessary for index tuning.
        % which are difficult to measure for index systems and 
    % Despite being out of scope,
    Future directions can extend $T(\Delta)$ from a deterministic function 
        to a distribution conditioned on the read size $\Delta$ to incorporate the randomness.
    % Apart from becoming more realistic, 
    Such storage models would enrich \system to tune on more complex goals such as fastest tail latency (e.g., lowest p99) or highest reliability (e.g., lowest latency variance).

\begin{figure}[t]
  \centering
%   \vspace{2mm}
  \resizebox{0.9\linewidth}{!}{%
  \input{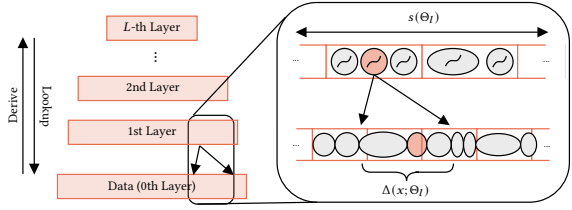}
  }
  \vspace{-2mm}
  \caption{A hierarchical index of $L$ layers. 
  % $L$ where index $l$-th layer derives $(l+1)$-th layer. 
  % Conversely, 
  The $l$-th layer looks up the data in the $(l-1)$-th layer. The close-up diagram layouts key-value pairs (ellipses) stored in pages. 
  It shows layer's size $s(\Theta_l)$ and precision $\Delta(x; \Theta_l)$ of a key whose node on $l$-th layer and relevant key-value on $(l-1)$-th are colored.} \label{fig:teaser}
  \vspace{-2mm}
\end{figure}

\subsection{Points of Applications}
\label{sec:apps}
\system applies to any index building in a system life cycle when lookup operations from storage need to be performant. \system primarily aims for two types of applications.
    (1) Immutable indexes' bulk loading: \system naturally builds high-speed indexes that do not change; nonetheless, the application can still support changing data with existing techniques such as gapped arrays and LSM-tree compaction.
    (2) Updatable indexes' initial design and maintenance: users can utilize \system to build the initial index structure, then follow any updatable index protocol compatible with the structure. After the index has mutated significantly, users can re-build the index using \system as a part of vacuum processes.
    % during downtime or a periodic maintenance.

    \system incurs computation overhead but is designed to minimize its real-time overhead. Although not required, a system with high parallelism is preferable. 
    Alternatively, users can trade off tuning accuracy for an overhead reduction through configurations.

% Despite being out of scope, non-deterministic storage models may apply to parts of our methodology with appropriate modifications to incorporate effects of randomness like optimizing for low-variance indexes. 
% All in all, we believe that $T(\Delta)$ is the right level of abstraction to reason about index structures.

% \input{figures/storage_profile}

% \begin{itemize}
%     \item{Hard Disk Drives}: Describe seek time, page read, behavior
%     \item{Solid-state Drives}: NAND, block-based, microsecond latency
%     \item{Network File Systems}: distributed file system, \href{https://www.semanticscholar.org/paper/The-Sun-Network-Filesystem\%3A-Design\%2C-Implementation-Sandberg/1b07afcb4e1b4d94dcb537589db1a2653f14e887}{SUN (1984)}
%     \item{Cloud Storage}: Blob storage, protocol and latencies. Examples in AWS, GCP, Azure.
%     \item{(Remote) Direct Memory Access?}
% \end{itemize}

% Latency and bandwidth both have physical and technological limitations. For one, the speed of light dictates the minimum latency accordingly to distance between compute and storage. Meanwhile, existing materials restricts the largest bandwidth the compute-storage connection can possess. \supawit{explain more on tech limits}. Therefore, external memory model and algorithms will be relevant for the foreseeable future, especially for large data systems. \supawit{generalize read requests} \supawit{$T$ as a summarization of random variable} \supawit{in-memory efforts: XIndex, but we are different}

%%%%%%%%%%%%%%%%%%%%%%%%%%%%%%%%%%%%%%%%%%%%%%%%%%%%%%%%%%%%%
%%%%%%%%%%%%%%%%%%%%%%%%%%%%%%%%%%%%%%%%%%%%%%%%%%%%%%%%%%%%%

\section{\ourmodel: Unified Index Model}
\label{sec:system-overall}
\label{sec:data_layout}

In this section,
    we mathematically model hierarchical indexes
        by representing them with design parameters $\bfTheta$,
    which are also used to express their lookup latencies.
These mathematical models
    are the foundations of our optimization process described in \cref{sec:system-building}.

\subsection{Hierarchical Indexes}
    \label{sec:system-overall-class-of-index}

A hierarchical index is a data structure that maintains data locations in a layered internal structure,
    consisting of explicit $L$ \emph{index layers}---$I_1, I_2, \ldots I_L$. 
    Here, $I_L$ refers to the root and $I_1$ refers to the bottom-most one, next to the data layer denoted $I_0$.
% Like other indexes, 
A data system can query a hierarchical index for the location of a desired item, 
    then retrieve the item from the location. 
Accordingly, a high-quality index
    must quickly find the location
    in a database, namely,    
% An index then must model the data's mapping from keys to their locations, namely, 
\emph{key-position collection} $D = \{(x_i, y_i)\}_{i=1}^n$, 
    where $x_i$ is the $i$-th key with data on a position $y_i = y(x_i) = [y^{-}(x_i), y^{+}(x_i))$ (i.e. range of data).
    % Unlike other generic indexes, however, a hierarchical index consists of an 
    % This layered structure imposes a clear query process (\cref{sec:querying}) which simplify reasonings for our tuning.

\paragraph{Index Layer}

Given a key, an index layer points to a range of data (in the next layer) 
that contains the information associated with the key. 
The system then can read the range of data from the next layer, 
    and continues the index traversal until it reaches the data layer.
% The next index layer then partially reads from this range of data, finds the associated information, and similarly points to another range of data. Therefore, one can follow this chain of indirections along index layers to search for an item associated with a key.
To support partial reads, an index layer consists of one or more \emph{nodes};
for a specific key, there is only one node associated with it.
    % where only one node in an index layer is associated with a key. 
Conceptually,
    a node is a function from a key to its position (as defined shortly),
% Considering a node as a smaller key-position function , 
and an index layer is a piecewise function comprising its node functions. 
% As a result, a hierarchical index is then able to 
This composite structure enables partial data reads and lower data access costs.%
% partially read relevant nodes rather than entire index layers and ultimately save storage access costs.
\begin{figure}[t]
  \centering
%   \vspace{2mm}
  % Adjust font size in `pretex=`
  % \includesvg[width=0.84\linewidth,pretex=\footnotesize]{figures/methodology/node_types}
%   \includegraphics[width=0.9\linewidth]{figures/methodology/node_types.png}
  \includegraphics[width=0.9\linewidth]{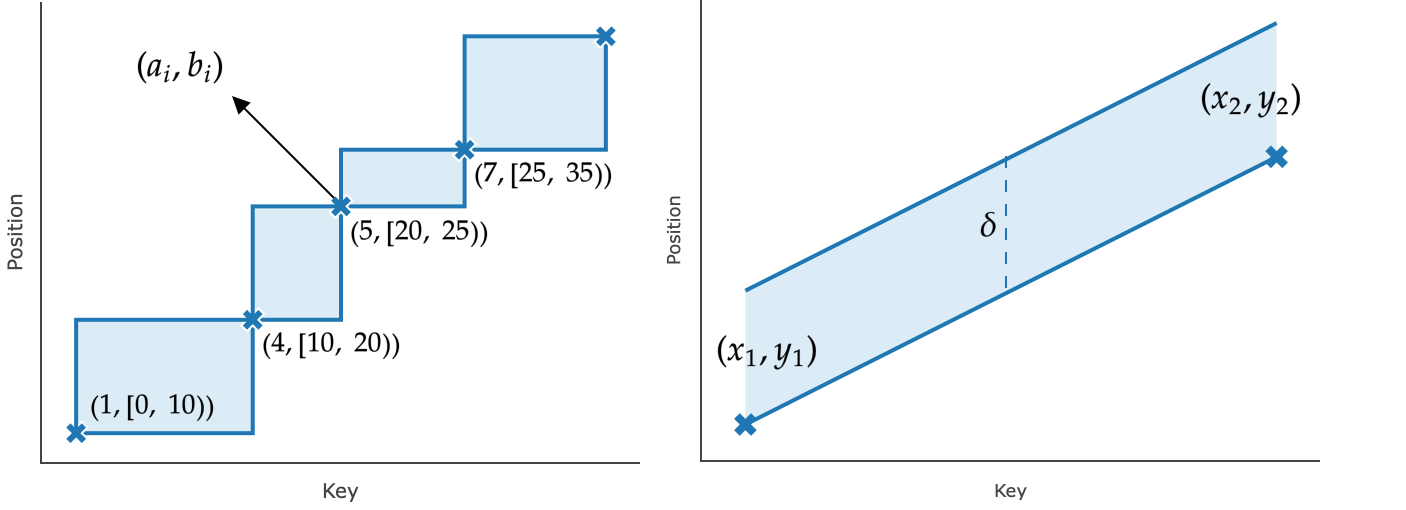}
  \vspace{-2mm}
  \caption{Two node types with key-position examples. Left: 5-piece \texttt{step}. Right: \texttt{band} with points and width.} \label{fig:node_types}
  \vspace{-2mm}
\end{figure}

\paragraph{Node} A node is the smallest data structure that maps a key to its position in the next layer.
% but only maps keys in a specific key range. 
Given a key $x$, a node predicts a position $\yh: \calX \rightarrow \calY$ in range that encompasses the actual position $y(x)$.
    \begin{equation} \label{eq:node-interface}
        \text{Node}\;\yh: \quad \yh(x) = [\yh^{-}(x), \yh^{+}(x)) \supseteq [y^{-}(x), y^{+}(x)) = y(x)
    \end{equation}

While our optimization framework can support any such function $\yh$ that satisfies \cref{eq:node-interface},
\system currently implements two types of nodes sketched in \cref{fig:node_types}. 
First, a step node (\texttt{step}) is a step function, or a $p$-piece constant function parameterized by partition keys $\bfa = (a_1, \dots, a_p)$ and partition positions $\bfb = (b_1, \dots, b_p)$. In other words, $\yh_{\texttt{step}}(x; \bfa, \bfb) = [b_i, b_{i+1})$ for $x \in [a_i, a_{i+1})$.
Second, a linear band node (\texttt{band}) is a thick linear function traveling through two key-position points $(x_1, y_1)$ and $(x_2, y_2)$ with a width $\delta$: $\yh_{\texttt{band}}(x; x_1, y_1, x_2, y_2, \delta) = [m x + c - \delta, m x + c + \delta)$ where $m = (y_2 - y_1) / (x_2 - x_1)$ and $c = y_1 - m x_1$. A step node is $16p$ bytes in size while a linear band node is $40$ bytes.
For example, consider a dataset with 4 key-position pairs: $D = \{(x, y)\} = \{ (1, [0, 10)), (4, [10, 20)), (5, [20, 25)), (7, [25, 35)) \}$ as in \cref{fig:node_types}. A $\texttt{step}$ node can represent this key-position collection with $p = 5$, $\bfa = (1, 4, 5, 7, \infty)$, and $\bfb = (0, 10, 20, 25, 35)$, guaranteeing the exact prediction $\yh_{\texttt{step}}(x; \bfa, \bfb) = y$ for all $(x, y) \in D$. Note that a $\texttt{step}$ node having fewer pieces can be valid but would have a higher error. Alternatively, a \texttt{band} node can approximately represent $D$ with $(x_1, y_1) = (1, 0), (x_2, y_2) = (4, 10), \delta = 15$. We encourage readers to verify that $\yh_{\texttt{band}}(x; 4, 0, 7, 25, 35) \supseteq y$ for all $(x, y) \in D$.
    % https://www.desmos.com/calculator/ogpq72mvds

Currently, these two node types are sufficient. Together, they fit sorted key-position collections accurately. \texttt{step} adapts to discontinuities while \texttt{band} regresses well with regularly sized key-value pairs~\cite{pmlr-v119-ferragina2020WhyLearnedIndexWork}. 
    They also possess many efficient fitting methods that pass over the entire key-position collection only once.
    Furthermore, with these types, \ourmodel can already incorporate many existing learned indexes such as PGM-Index~\cite{DBLP:journals/pvldb/FerraginaV20PGMIndex}, ALEX~\cite{DingMYWDLZCGKLK2020ALEX}, RadixSpline~\cite{DBLP:conf/sigmod/KipfMRSKK020RadixSpline}, SIndex~\cite{Wang2020SIndex}, XModel~\cite{Wei2020FastCacheXStoreXModel}.

\paragraph{Parameters (to be tuned)} The collection of all parameters $\bfTheta$ is a nested tuple of parameters that is sufficient to represent a hierarchical index instance. In particular, $\bfTheta$ specifies the number of layers $L$ followed by $L$ layer-wise parameters $\Theta_i$ which in turn specifies the node type, number of nodes $n_l$, and all node parameters.
    \begin{equation} \label{eq:bftheta_parameter}
        \bfTheta = (L, (\Theta_1, \dots, \Theta_L)), 
        \quad 
        \Theta_l = (\texttt{NodeType}, n_l, (\theta_1, \dots, \theta_{n_l}))
    \end{equation}

    Node-specific parameters $(\theta_1, \dots, \theta_{n_l})$ for $n_l$ nodes depend on the corresponding $\texttt{NodeType}$. 
    For example, if $\texttt{NodeType} = \texttt{step}(p)$, $\theta_i = (\bfa_i, \bfb_i)$ where $|\bfa_i| = |\bfb_i| = p$. If $\texttt{NodeType} = \texttt{band}$, $\theta_i = (x^{(i)}_1, y^{(i)}_1, x^{(i)}_2, y^{(i)}_2, \delta_i)$.
    We choose to specify a single node type per layer over multiple node types per layer because
    (1) it reduces the serialization overhead per node which reduces read volume and so overall lookup latency, and 
    (2) it simplifies our tuning (\cref{sec:system-building-node-estimation}).

\begin{algorithm}[t]
    \caption{\small \system Query Process, \texttt{Lookup}$(x; \; \yh_L, L)$}
    \label{algo:query-process}
    \small
    
    \DontPrintSemicolon
    
    \KwInput{Query key $x$, root position $\yh_L$, number of layers $L$}
    \KwOutput{Relevant key-value $(x, v)$}
    %   \KwData{Testing set $x$}
    \BlankLine
    % \Comment*[l]{Traverse layers by following relevant nodes}
    % $\yh_L(x) \leftarrow \yh_L$ \Comment*[r]{Always read whole root layer}
    \For{$l$ from $L$ to $0$} {
        $\{(x[i], v[i])\}_i \leftarrow \texttt{Read}(\yh_l(x))$ \Comment*[r]{Storage access}
        $v_l \leftarrow \texttt{Search}(x, \{(x[i], v[i])\}_i)$ \\
        \If{In index layer, $l \geq 1$} {
            $\yh_{l-1} \leftarrow \texttt{ReconstructNode}(v_l)$
        }
    }
    \Return $(x, v) = (x, v_0)$
\end{algorithm}

\paragraph{Examples}
    
    As a general class of indexes, hierarchical index examples include traditional indexes like B-tree and learned indexes like \rmi, \pgm, balanced \alex, and \plex. For example, 2-layer balanced B-tree with fanout 3 is a hierarchical index as in \cref{eq:btree-as-hindex} where $\theta_{i,j}$ and $\theta_{k}$ are appropriate step function parameters to encode keys and pointers in the leaf and root nodes respectively.
    \begin{equation} \label{eq:btree-as-hindex}
    \begin{gathered}
        \bfTheta_{\text{B-Tree}} = (2, (\Theta_1, \Theta_2)), \\
        \Theta_2 = (\texttt{step}, 3, (\theta_1,\theta_2,\theta_3)), \\
        \Theta_1 = (\texttt{step}, 3^2, (\theta_{i,j})_{i,j \in \{1, 2, 3\}})
    \end{gathered}
    \end{equation}
    
    \rmi with a cubic root node $ax^3 + bx^2 + cx + d$ and 8 linear intermediate nodes $(\alpha_i x + \beta_i) \pm \varepsilon_i$ is a hierarchical index as in \cref{eq:rmi-as-hindex}. \pgm, balanced \alex, and \plex can be similarly instantiated as hierarchical indexes by extending the node types.
    \begin{equation} \label{eq:rmi-as-hindex}
    \begin{gathered}
        \bfTheta_{\text{\rmi}} = (2, (\Theta_1, \Theta_2)), \\
        \Theta_2 = (\texttt{cubic}, 1, ((a, b, c, d))), \\
        \Theta_1 = (\texttt{linear}, 8, ((\alpha_i, \beta_i, \varepsilon_i))_{i \in \{1, \dots, 8\}})
    \end{gathered}
    \end{equation}

\subsection{Query Process}
    \label{sec:querying}

Functionally, a query process takes in a search key $x$ and outputs its relevant key-value $(x, v)$. It internally consults index layers inside a hierarchical index, starting from its root layer $L$, traversing down one index layer at a time to look up the relevant position of the query key, and finally retrieving the target value.
    
    \cref{algo:query-process} formalizes \system's overall query process, traversing through the index hierarchy of relevant node(s) $\yh_L, \dots, \yh_0$ to retrieve the relevant key-value $(x, v)$. There are mainly three steps in each iteration in an index layer:
    (1) \system reads potentially relevant raw bytes $\{(x[i], v[i])\}_i$,
    (2) it then searches for the relevant raw bytes $v_l$ based on the tagged key $\{x[i]\}_i$,
    and (3) \system deserializes $v_l$ to reconstruct the next relevant node $\yh_{l-1}$ and predicts the next position $\yh_{l-1}(x)$. 
    At the end when $l = 0$, \system returns the value with the query key $(x, v_0) = (x, v)$.

    Such a clearly defined query process allows us to estimate a hierarchical index performance for tuning. Given appropriate statistics on the hierarchical index and a storage model, we can translate \cref{algo:query-process} into the latency formula incorporating different lookup costs.

% \begin{table}
%     \centering
%     \caption{Optimization summary}
%     \vspace{-2mm}
%     \label{tab:opt_summary}
%     \small
%     \begin{tabular}{ll}
%         \toprule
        
%         Properties & \system \\
        
%         \midrule

%         Design variables & \makecell[tl]{
%             Index parameters $\bfTheta = (L, (\Theta_1, \dots, \Theta_L))$ \\
%             with each layer $\Theta_l = (\texttt{NodeType}, n_l, (\theta_1, \dots, \theta_{n_l}))$ \\
%             where $\theta_i$ depends on $\texttt{NodeType}$} \\

%         Fixed variables & Storage profile $T$, query key distribution $\calX$ \\

%         Constraints & Valid nodes $\yh(x) \supseteq y(x)$ for all existing keys $x$ \\

%         Cost function & Expected latency $\calL_{SM}(\calX; \bfTheta, T)$ (\cref{eq:emm-obj-airindex}) \\

%         Algorithm & \oursearch (\cref{sec:system-building}) \\
        
%         \bottomrule
%     \end{tabular}
% \end{table}

\begin{table}
    \centering
    \caption{Optimization summary}
    \vspace{-2mm}
    \label{tab:opt_summary}
    \small
    \begin{tabular}{ll}
        \toprule
        
        \multicolumn{2}{l}{\textbf{Design variables} $\bfTheta$:} \\
        
        % \midrule

        $L$ & Number of layers \\
        $\texttt{NodeType}_l$ & Node type in layer $l \in \{1, \dots, L\}$ \\
        $n_l$ & Number of nodes in layer $l \in \{1, \dots, L\}$ \\
        $\theta_{l, i}$ & Parameters of the $i$-th node in layer $l$, $i \in \{1, \dots, n_l\}$ \\
        
        \midrule
        
        \multicolumn{2}{l}{\textbf{Fixed variables}:} \\
        
        % \midrule

        $T$ & Storage profile \\
        $\calX$ & Query key distribution \\
        $D$ & Key-position collection $D = \{(x_i, y_i)\}_{i=1}^n$ \\

        \midrule
        
        \textbf{Constraints} & Valid index $\yh(x) \supseteq y$ for all $(x, y) \in D$ \\

        \midrule
        
        \textbf{Objective} & Minimize expected latency $\calL_{SM}(\calX; \bfTheta, T)$, \cref{eq:emm-obj-airindex} \\

        \midrule
        
        \textbf{Algorithm} & \oursearch (\cref{sec:system-building}) \\

        \bottomrule

    \end{tabular}
\end{table}

% =======================
% Design variable
% _______________________
% symbol & name
% symbol & name
% symbol & name
% symbol & name
% =======================
% Fixed variable...
% _______________________
% symbol & name
% symbol & name
% symbol & name
% symbol & name

\subsection{Latency Under Storage Model}
    \label{sec:objective}

    % Although a query process contains multiple steps, 
    For index traversals,
    the dominant costs are storage accesses (\cref{algo:query-process}, line 2), 
    compared to other internal computations including relevant value searching, data deserialization, and node prediction. 
    Following the iterations in \cref{algo:query-process}, there are exactly $L + 1$ sequential storage accesses corresponding to reading $L$ index layers and the data layer. Specifically, \system first reads the entire root layer of size $s(\Theta_L)$ bytes, then partially reads $\Delta(x; \Theta_{l+1}) = |\yh_l(x)|$ bytes from the next index layer, and so on until it reaches the data layer to read $\Delta(x; \Theta_{1}) = |\yh_0(x)|$ bytes. 
    Using a storage profile $T$ (\cref{sec:storage_model}), we express $\calL_{SM}(x; \bfTheta, T)$, the query latency to find a value for a key $x$ under the storage model, as follows:
    \begin{equation} \label{eq:emm-hierindex}
        \calL_{SM}(x; \bfTheta, T) = T(s(\Theta_L)) + \sum_{l=1}^L T(\Delta(x; \Theta_l))
    \end{equation}

To obtain expected latency over multiple keys, 
    we aggregate query latencies, each specific to key $x$, 
    % into the expected query latency 
    over a query key distribution $\calX$. 
    In this work, we set $\calX$ to be a uniform distribution over existing keys in the key-position collection $D$.
    \begin{equation}  \label{eq:emm-expectation}
        \calL_{SM}(\calX; \bfTheta, T) = \E_{x \sim \calX} \left[ T(s(\Theta_L)) + \sum_{l=1}^L T(\Delta(x; \Theta_l)) \right]
    \end{equation}

Given a key-position collection $D$, a storage profile $T$, and a query key distribution $\calX$, 
    our objective (\cref{eq:emm-obj-airindex}) is 
    to minimize this expected query latency 
    where $\bfTheta$ represents a hierarchical index valid over $D$. 
    \cref{tab:opt_summary} summarizes this optimization problem.
    \begin{equation}  \label{eq:emm-obj-airindex}
        \bfTheta^{*} = \argmin_{\bfTheta} \E_{x \sim \calX}\left[ T(s(\Theta_L)) + \sum_{l=1}^L T(\Delta(x; \Theta_l)) \right]
    \end{equation}

% Finding the optimal $\bfTheta$ is a complicated task.
% The possibilities of index structure expand a prohibitively large search space to be explored. Not only does \system need to decide the number of layers $L$, but it also needs to determine the node type as well as each node's parameters across all index layers.
% Furthermore, these parameters are dependent upon each other, as described in \cref{sec:intro}.
% % For example, a change in separator keys on a layer renders the parameters in layers above obsolete.
% % A particular pattern in separator keys can favor a particular node type, which is difficult to anticipate until after fitting.
% % Adding or removing the number of nodes can also burden or ease the upper layer respectively.

% % problem, cost, search space, hardness
% \subsection{Objective Function}
% \label{sec:system-building-objective}

% \system automatically tunes its index structure onto a data collection with respect to the underlying storage. An index structure is summarized by $\bfTheta = (\Theta_1, \dots, \Theta_L)$ where $\Theta_l$ represents the $l$-th index layer containing separator keys and parameters in all nodes. The structure possibility expand a prohibitively large search space to be explored. Not only does \system need to decide the number of layers $L$, but it also needs to determine per layer the node type as well as each node's parameters.

%%%%%%%%%%%%%%%%%%%%%%%%%%%%%%%%%%%%%%%%%%%%%%%%%%%%%%%%%%%%%
%%%%%%%%%%%%%%%%%%%%%%%%%%%%%%%%%%%%%%%%%%%%%%%%%%%%%%%%%%%%%

% \section{\oursearch: Optimization with Bounded Visits}
\section{\oursearch: Search with Bounded visits}
\label{sec:system-building}

This section describes \system's optimization algorithm, \oursearch, which solves the optimzation defined earlier (\cref{tab:opt_summary}).
    We first describe the overall process (\cref{sec:system-building-overall}).
    Then we explain important components in detail: index layer builders (\cref{sec:system-building-node-estimation}), pruning technique for computational efficiency (\cref{sec:system-building-selection}), and parallelization techniques (\cref{sec:system-building-parallelism}).
    Finally, we analyze its time complexity (\cref{sec:system-analysis}).

\begin{figure}[t]
  \centering
%   \vspace{2mm}
  \resizebox{0.9\linewidth}{!}{\input{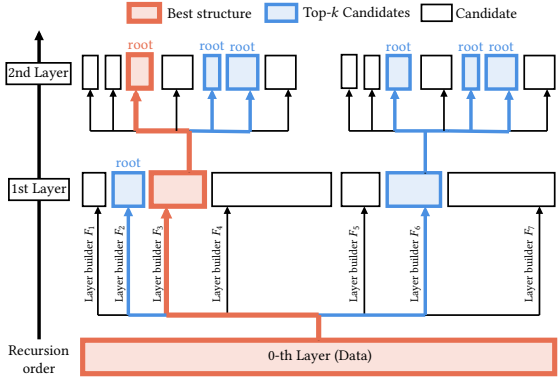}}
  \vspace{-2mm}
  \caption{An instance of \oursearch execution, starting from the bottom (data layer) to the top. The label ``root'' indicates that the candidate should be the root layer with no further index layer. In this illustration, the algorithm has $7$ layer builders and selects $k = 3$ top candidates to branch out.}
  \vspace{-1mm}
  \label{fig:branch_and_bound}
\end{figure}

\subsection{Guided Graph Search}
    \label{sec:system-building-overall}

    \oursearch is a guided graph search; it starts from an origin vertex, walks over edges to different vertices, and stops when a stopping criterion is satisfied. 
    Each vertex represents a layer in a hierarchical index.
    As the special case, the origin vertex $u_0$ represents the dataset being indexed. Each edge from a vertex $u$ to another vertex $u'$ represents a \emph{layer builder} building an index layer $u'$ based on $u$. A path from the origin represents a particular hierarchical index design. For example, a path $(u_0, u_1, u_2)$ is roughly equivalent to data and index layers $(I_0, I_1, I_2)$ where the index layers have parameters $\bfTheta = (2, (\Theta_1, \Theta_2))$. Two paths with common vertices represent two hierarchical index designs that share lower layers; thus, \oursearch can reuse layer-building results in those common lower layers.

    On a vertex, \oursearch first explores all outgoing edges, i.e. possible candidate index layers from all available layer builders (\cref{sec:system-building-node-estimation}, \cref{algo:main-tuning}, lines 3--6). Layer builder explorations are the most expensive step, but they are independent of one another and embarrassingly parallelizable. \oursearch leverages this observation to reduce the tuning time overhead (\cref{sec:system-building-parallelism}). 
    Upon receiving all candidates, it consults heuristic guidance to select only the top-$k$ candidates (\cref{sec:system-building-selection}, \cref{algo:main-tuning} line 7) to continue the search recursively (\cref{algo:main-tuning} lines 8--12). Selecting a few promising candidates is an important mechanism to limit the branching factor, avoiding exponential time complexity (see analysis in \cref{sec:system-analysis}). At the end of a recursive search, \oursearch compares all the options using the latency formula under a storage model (\cref{eq:emm-expectation}) and returns the best hierarchical index design.

    To decide when to stop searching, \oursearch determines whether an additional index layer will be beneficial. That is, if the \emph{ideal index layer}--the best (possibly impossible) layer we can build on top--does not reduce the query latency, \oursearch stops further exploration and declare the current vertex as the root index layer (\cref{algo:main-tuning}, lines 1--2). Specifically, an ideal index has the minimal size of $s(\Theta) = 1$ byte and the finest precision $\Delta(x; \Theta) = 1$ byte.

    \paragraph{Example} If we have $7$ layer builders and set $k = 3$, an execution of \oursearch could look like \cref{fig:branch_and_bound}. That is, \oursearch starts from the origin data-layer vertex (the long rectangle at the bottom) and explores all 7 layer builders, resulting in $7$ vertices (rectangles of varying sizes in the middle). The guidance then tells \oursearch to search deeper into $k = 3$ candidates of them (red and blue highlighted rectangles). \oursearch stops at one of the candidates because it is too small to gain any benefit from an ideal index layer. For the rest, \oursearch continues exploring, selecting top-$3$ candidates, and finally stops at 2nd layer candidates. After comparing all options, \oursearch reports the best path (all red highlighted rectangles), representing the fastest hierarchical index design.

\begin{algorithm}[t]
    \caption{\small \system Index Tuning, \oursearch$(D; T, \calF)$}
    \label{algo:main-tuning}
    \small
    
    \DontPrintSemicolon
    
    \KwInput{Key-position collection $D = \{(x_i, y_i)\}_{i=1}^n$, storage profile $T$, layer builders $\calF$}
    \KwOutput{Index structure $\bfTheta^{*}$}
    %   \KwData{Testing set $x$}
    \BlankLine
    \Comment*[l]{Check stopping criterion}
    \If{$\calL_{SM}(D; (), T) < \texttt{IdealLatencyWithIndex}(T)$} {
        \Return $()$  \Comment*[r]{Cannot improve with additional layer}
    }
    \BlankLine
    \Comment*[l]{Build multiple next layer candidates}
    \For{$F$ in $\calF$} {
        $\Theta_{\text{next}} \leftarrow F(D)$  \Comment*[r]{Build next layer (\cref{sec:system-building-node-estimation})}
        $D_{\text{next}} \leftarrow$ \texttt{Outline}$(\Theta_{\text{next}})$  \Comment*[r]{Turn into key-positions}
        $\calC \leftarrow \calC \oplus \{ (\Theta_{\text{next}}, D_{\text{next}}) \}$  \Comment*[r]{Append candidate}
    }
    \BlankLine
    \Comment*[l]{Select top-$k$ candidates (\cref{sec:system-building-selection})}
    $\calC \leftarrow $ \texttt{Select}$(\calC, k)$ \;
    \BlankLine
    \Comment*[l]{Build indexes on top-$k$ candidates}
    \For{$(\Theta_{\text{next}}, D_{\text{next}})$ in $\calC$} {
        $\bfTheta_{\text{next}}^{*} \leftarrow$ \oursearch$(D_{\text{next}}; T, \calF)$  \Comment*[r]{Call recursively}
        $\bfTheta_{\text{new}} \leftarrow (\Theta_{\text{next}}) \oplus \bfTheta_{\text{next}}^{*}$  \Comment*[r]{Prepend layer}
        \If{$\calL_{SM}(D; \bfTheta_{\text{new}}\;, T) < \calL_{SM}(D; \bfTheta^{*}, T)$} {
            $\bfTheta^{*} \leftarrow \bfTheta_{\text{new}}$  \Comment*[r]{Select the better structure}
        }
    }
    \Return $\bfTheta^{*}$
\end{algorithm}

\subsection{Layer Builders}
    \label{sec:system-building-node-estimation}

    A layer builder is a method to produce a valid index layer on top of existing index layer(s). 
    In other words, it is a mapping $F(D) = \Theta$ such that $\Theta$ satisfies a valid index layer $\yh(x) \supseteq y$ for all $(x, y) \in D$.

    % There are countless methods to build an index layer, even one having the same node type. 
    In theory, there is a large number of ways of building index layers.
    For example, a method $A_1$ can find the smallest collection of \texttt{band} that covers $D$ with error at most $\lambda$ bytes using $O(n^2)$ for $n$ key-position pairs. In $O(n)$, another method $A_2$ could quickly and mindlessly connect every other $m$ key-position points into a collection of \texttt{band}. To avoid exploring every possible method, we choose a set of good layer builders that (1) run quickly (say, time complexity $O(n)$), (2) build small and accurate index layers, and (3) synergetically cover different data patterns together. In the earlier examples, method $A_1$ builds the optimal \texttt{band} index layer but is too slow, while method $A_2$ is fast but builds a suboptimal \texttt{band} index layer. In addition, $\{A_1, A_2\}$ is also not a good set of methods, because they only cover the \texttt{band} node type.

    To cover the two types of nodes (\texttt{step} and \texttt{band}) on different data patterns, \system currently deploys three types of layer builders, each generating many layer builders by varying hyperparameters. (1) \textbf{Greedy Step} (\texttt{GStep}$(p, \lambda_{GS})$) builds $p$-piece \texttt{step} nodes with precision at most $\lambda_{GS}$ bytes by greedily packing key-position pairs. (2) \textbf{Greedy Band} (\texttt{GBand}$(\lambda_{GB})$) builds \texttt{band} nodes by greedily fitting as many key-position pairs as possible using the monotone chain convex hull~\cite{DBLP:journals/ipl/Andrew79ConvexHull}. (3) \textbf{Equal Band} (\texttt{EBand}$(\lambda_{EB})$) builds \texttt{band} nodes by grouping key-position pairs in equal-size position ranges. Please see our extended script~\cite{airindex_technical_report} for further details.

    \system generates the set of candidate layer builders $\calF$ by sampling the granularity exponentially: $\lambda_{low}, \lambda_{low} (1 + \epsilon), \lambda_{low} (1 + \epsilon)^2, \dots \lambda_{high}$ where $(\lambda_{low}, \lambda_{high})$ are the bounds and $\epsilon > 0$ controls the exponentiation base. For example, if $\lambda_{low} = 2^8$, $\lambda_{high} = 2^{20}$, and $1 + \epsilon = 2$ with $p = 16$, then 
    $\calF$ contains 39 builders in total:
    \begin{equation}
    \begin{split}
    \calF
        =& \{\texttt{GStep}(16, 2^8), \texttt{GStep}(16, 2^9), \dots, \texttt{GStep}(16, 2^{20})\} \\
        &\cup \{\texttt{GBand}(2^8), \texttt{GBand}(2^9), \dots, \texttt{GBand}(2^{20})\} \\
        &\cup \{\texttt{EBand}(2^8), \texttt{EBand}(2^9), \dots, \texttt{EBand}(2^{20})\}
    \end{split}
    \end{equation}

\subsection{Top-k Candidates by Index Complexity}
\label{sec:system-building-selection}

\input{figures/methodology/index_complexity}

Before branching out, \oursearch selects only top-$k$ candidates with the highest potential to be in the optimal design.
    For each candidate $(\Theta_i, D_i)$, \oursearch evaluates its quality as a summation of a ``remaining work'' heuristic function $\hat{\tau}(D; T)$ and its layer-specific lookup latency. Then, it selects top-$k$ candidates with $k$ lowest estimated costs ($\argmin^k$ denotes top-$k$ arguments of the minima).
    \begin{equation} \label{eq:selection-oracle}
        \{(\Theta_i, D_i)\}_{i=1}^k = \argmin_{(\Theta_i, D_i) \in \calC}{}^k \quad \hat{\tau}(D_i) + \E_{x \sim \calX} \left[ T(\Delta(x; \Theta_i)) \right]
    \end{equation}
    
    \paragraph{Choice of Heuristic Function} Ideally, if there exists an oracle that reveals the optimal search latency, we could simply select the best candidate and avoid branching out entirely. This optimal search latency of dataset $D$ under storage profile $T$ is called \textit{index complexity} $\tau(D; T)$. Unfortunately, $\tau(D; T)$ is unknown for a large class of indexes supported by \system.

    Instead, \oursearch estimates candidates' quality using an upper bound to the index complexity: \textit{step index complexity} $\hat{\tau}(D; T) \geq \tau(D; T)$. $\hat{\tau}(D; T)$ is the optimal search latency considering only \texttt{step} index layers (i.e. B-tree layers). Since the quality of \texttt{step}-based layers can be analytically computed, 
    we obtain an efficient algorithm that depends only on the collection size $s_D$ and storage profile $T$.
    % to solve for the optimal latency. 
    \cref{fig:index-complex} shows the general shape of $\hat{\tau}(D; T)$ solved with the algorithm.
    % Notice the sudden index complexity cliffs (such as those around $s_D = 1$MB in \cref{fig:index-complex-latency}), marking the boundaries between different chosen numbers of layers in \cref{eq:selection-step-complexity}.
    Please see our extended report~\cite{airindex_technical_report} for more details.

\subsection{Parallel Tuning}
\label{sec:system-building-parallelism}

\system is highly parallelizable from three sources of parallelisms as described below, 
    ordered from finest to coarsest layers. 
It is worth noting that, with a proper branching (\cref{sec:system-analysis-complexity}), 
    the node-building step is the primary target for parallelization.
    % in the building procedure 
% and is the main target for parallelization.

\paragraph{From Data Partitioning} \system partitions the key-position collections, uses a layer builder to build sub-range candidates, and merges them into a candidate. 
This is possible because \system's existing layer builders generate a piecewise function that can be merged together across different key ranges. 
By default, \system breaks a key-position collection into partitions, each containing 1 million key-position pairs. 
Thus, \system can scale with growing data sizes 
    by increasing the level of parallelisms accordingly.

\paragraph{From Across Layer Builders} 
Although \cref{algo:main-tuning} calls layer builders in loops, the invocations are independent of one another. 
\system conveniently turns the for-loop into a parallel mapping to produce $(\Theta_{\text{next}}, D_{\text{next}})$ and collecting into the candidate set $\calC$. With more parallelisms, \system can then scale along with the diversity of layer builders to capture wider key-position patterns.

% One may consider iterative building alternatives that choose the next layer builder setting based on past iterations. However, they will process through the key-position collection as many rounds as the number of iterations. In contrast, given enough parallelism, \system can generate multiple candidates using the same time as generating one candidate, equivalent to processing the key-position collection in one round. Since \system aims to handle a large data layer, processing in many rounds is slow and undesirable.

\paragraph{From Branching} 
\system can recursively call \cref{algo:main-tuning} in parallel and select the best index structure with the minimum storage model cost. This source of parallelism allows \system to explore more candidate branches and especially higher structures when favored by the storage profile (e.g., low bandwidth and low latency).

\subsection{Analysis}
\label{sec:system-analysis}

Branching recursive optimization requires a balancing between the branching factor and depth. If \oursearch branches out to too many candidates relative to the index depth, it would become uncontrollably slow. On the other extreme, it could miss the optimal candidate branch. We first analyze the tuning time complexity with respect to a choice of hyperparameter. Then, we attempt to provide an approximation factor of the automatic tuning. 
% \todo{difference from SMA* by upperbound --> faster, why it's ok?}

% \paragraph{Time Complexity}

\begin{theorem} \label{sec:system-analysis-complexity}
\textit{Time complexity}: Let there be $n$ key-position pairs, $L$ layers to be explored at most according to the storage profile $T$, and $|\calF|$ layer builders. If \oursearch selects at most $k \leq \sqrt[L+1\;]{n}$ candidates then the time complexity is at most $O((L + 1) |\calF| n)$.
\end{theorem}

\begin{proof}
Suppose the data $D$ consist of $n$ key-position pairs with total data size $s_D$. Let $L$ be the maximum number of layers expected to be explored, which has the upper bound $L_{\max} \geq L$: the number of layers chosen by the step index complexity.

Because our layer builders all process a $O(n)$ key-position collection in $O(n)$ time, the time to build all $|\calF|$ candidates is $O(|\calF| n)$. Next, considering the worst case compression ratio $O(n^{\frac{1}{L + 1}})$ and the number of branches $k$, we expect $O(n^{\frac{L}{L + 1}})$ key-position pairs and so the time complexity in the next layer is $O(|\calF| k n^{\frac{L}{L + 1}})$. The branching and compression of key-position pairs continue until the root layer. In summary, the total time complexity in $L$-layer branching recursion is as followed whose last step is the closed-form formula to the geometric series.
\begin{equation} \label{eq:system-time-complexity-full}
O\left(|\calF| \sum_{l=0}^{L} k^l n^{\frac{L + 1 - l}{L + 1}}\right)
    % = O\left(|\calF| \frac{n^{\frac{1}{L+1}}\left(n-k^{\left(L+1\right)}\right)}{\left(n^{\frac{1}{L+1}}-k\right)} \right)
    = O\left(|\calF| n \frac{1 - k^{L+1} / n}{1 - k / n^{\frac{1}{L+1}}} \right)
\end{equation}

Under $k^{L+1} / n \leq 1$ (i.e. $k \leq \sqrt[L+1\;]{n}$), this reduces to $O(|\calF| (L + 1) n)$ when $n \rightarrow \infty$.
\end{proof}

\subsection{Other Implementation Details}
\label{sec:system-implement}

We implement \system in Rust~\cite{airindex_repo}.
% ~\footnote{https://github.com/\githuborg/airindex-public}.
Like many data systems, it has two explicit levels in its memory hierarchy: 
    the underlying storage and its internal read-through cache. 
\system interacts with storage through an abstract interface, in which concrete implementations serve partial range reads at their best effort. 
Currently, \system's internal zero-copy read-through cache employs a first-in-first-out (FIFO) eviction policy due to its admission simplicity.
% and the abundance of cache space in our experiment settings.
Apart from customized node type format, \system serializes the metadata together with root layer as a byte array in the Postcard~\cite{postcard} format via Serde~\cite{serde}. 

\section{Extensions}
\label{sec:discussion}

\paragraph{Supporting Updates}
Although \system does not optimize for write workloads, it can tune and build an index that supports write operations.
For example, we can augment the data layer into a gapped array, allowing insertion into gaps and deletion without changing index layers. When gaps are filled or expected to be filled, we can enlarge data layer's gaps and build a new index with \system.
\system can also serves as the initial bulk loading of an updatable index (e.g. \alex). However, the updatable index may evolve suboptimally. To reduce the frequency of structural index updates, we can enlarge the position granularity from bytes to pages, or we can buffer writes similarly to LSM-trees~\cite{DBLP:journals/acta/ONeilCGO96LSMTree,DaiXGAKAA2020BourbonLSM}.

\paragraph{Cache-aware Optimization}
\system can find an optimal \emph{cache-aware} index design
% an index to a cache hierarchy by directly optimizing on the original cost (\cref{eq:original-cost}) with further specifications on 
by additionally considering
the distribution of cache hit $C_l$ and the cost of cache access $\calL_{\text{cache}}$. This modification then only affects the index complexity $\hat{\tau}$ and candidate selection in \oursearch.
While this work does not include explicit results for cache-aware optimization, \cref{fig:I_latency_curve} indicates that the indexes built with cache-pessimistic optimization already offer high performance than other existing methods across a wide range of cache warmness.

\paragraph{Pre-Search Assessment} Upon significant data or workload change, we can first assess the potential performance gain through the step index complexity (\cref{sec:system-building-selection}). Based on the assessment, users can better decide whether to tune the index. Step index complexity is a loose upper bound of the gain, however. Future works combining more accurate index complexity, what-if index design techniques, and data/workload trackers would help avoid unnecessary search costs for marginal performance gain.

% theoretical, hint: saturation

%%%%%%%%%%%%%%%%%%%%%%%%%%%%%%%%%%%%%%%%%%%%%%%%%%%%%%%%%%%%%
%%%%%%%%%%%%%%%%%%%%%%%%%%%%%%%%%%%%%%%%%%%%%%%%%%%%%%%%%%%%%

\section{Experiment}
\label{sec:exp}

We empirically study \system
    to demonstrate its faster search 
    (\cref{sec:exp-e2e}), benefits of automatic index designs (\cref{sec:exp-tune-accuracy}), adaptability under wide ranges of I/O profile (\cref{sec:exp-storage-impact}),
    and quick build time (\cref{sec:exp-build-scale}).

\input{figures/experiments/I_latency_cold}

\input{figures/experiments/I_latency_curve}

\subsection{Setup}
\label{sec:exp-setup}

% We benchmark lookup times across different systems. 
% From a bird's-eye view, 
The experiment locates on two physical components: compute and storage.
The former hosts benchmark scripts to execute queries against systems and measure their performance. 
These scripts together with required binaries are stored on local storage (i.e. Azure OS Disk).
Meanwhile, the latter stores both datasets and indexes. 
% In other words, dataset-specific information resides in the storage and 
%     non-dataset-specific information lives in the compute.
% The following paragraphs describe common setups in most of our experiments. Otherwise, the experiment sections will mention their distinction.
Our benchmark consists of 40 runs in each setting:
the $i$-th 
% In the $i$-th run, the benchmark 
prepares the environment (e.g., clear cache\footnote{For both NFS and SSD, we execute \texttt{sysctl vm.drop\_caches=3} on a VM, 
clearing Linux-related caches such as page cache, entries, and inodes. 
For NFS, we also unmount and re-mount the Azure Blob Storage NFS to reload its client.}), loads/executes 
the $i$-th list of 
    one million query keys sequentially,
        and measures the elapsed time.
We summarise those runtimes with average and standard deviation.
% sampled from the target dataset and shared across tested systems
% The benchmark then starts its stopwatch, reloads the system, then sequentially queries and measures elapsed times to complete the query list. 
% We summarize the measured times by their average and standard deviation.
% The benchmark resets the cache and storage adaptor at the beginning of each run. 

\paragraph{System Environment}
% Our experiments execute on 
We use Azure cloud platform~\cite{azure}, 
    specifically 
    % We reserve a 
\texttt{D8s\_v3} (8 vCPUs, 32 GiB RAM) with Ubuntu 20.04. 
The VM connects to two types of storage. (1) \textit{NFS}: Azure network file system~\cite{azure-nfs} hosted on Azure Blob Storage~\cite{azure-blob} (StorageV2, standard performance, zone-redundant storage, hot access tier). (2) \textit{SSD}: Azure Premium SSD~\cite{azure-ssd} with P20 performance tier (256 GiB, 2300 IOPS, 150 MBps, read/write host caching). 3) HDD: Azure Standard HDD~\cite{azure-ssd} (1024 GiB, 500 IOPS, 60 MBps, no host caching). All resources are allocated within the same East US region.

\paragraph{Baselines}
We compare \system to a traditional database index, 
    % two fashions of 
    learned indexes, and our manual configuration counterpart. Although many of them are in-memory indexes, we integrate them onto external storages whose implementations are in their respective forks~\cite{lmdb_repo,rmi_repo,pgm_repo,alex_repo,dcal_repo}.
We manually tune each of the baselines through microbenchmarks.
~
(1) \textbf{\lmdb}: 
\lmdb~\cite{lmdb} is a B-tree database that accesses its data on storage through \texttt{mmap}. 
~
(2) \textbf{\rmi}: 
\rmi~\cite{kipf2018learned,DBLP:conf/sigmod/MarcusZK20CDFShop} is a top-down learned index 
    with a compact two-layer structure 
where the top one contains only one perfectly accurate node 
    partitioning key space to the bottom nodes.
We utilize \cdfshop~\cite{DBLP:conf/sigmod/MarcusZK20CDFShop} to recommend function types and select the most accurate \rmi across all datasets.
~
(3) \textbf{\pgm}: \pgm~\cite{DBLP:journals/pvldb/FerraginaV20PGMIndex} is a learned index with bounded precision across all layers. 
% In contrast to \rmi's top-down building, 
\pgm partitions the key-position collection to build the next bottom-most layer towards the top.
~
(4) \textbf{\alex}: \alex~\cite{DingMYWDLZCGKLK2020ALEX} is an updatable learned index built top-down like \rmi but further arrange key-value pairs in its layout (notably, ``gapped array'' to buffer structural changes).
~
(5) \textbf{\plex}: \plex~\cite{DBLP:journals/corr/Stoian2021PLEX} is a learned index with compact Hist-Tree (CHT) layered on top of RadixSpline~\cite{DBLP:conf/sigmod/KipfMRSKK020RadixSpline} (RS). Although \plex optimizes most parameters, its user need to specify the maximum prediction error $\epsilon$. We select $\epsilon = 2048$ based on a benchmark on a setting (\cref{fig:VIIII_tuned_plex}).
~
(6) \textbf{\dcal}: \dcal~\cite{DBLP:conf/sigmod/IdreosZHKG2018DataCalc,DBLP:journals/corr/IdreosZHKG2018DataCalcInternal} is a data layout design engine that calculates the performance of a data structure. We follow its auto-completion and build its recommended data layout within \system's framework.
~
(7) \textbf{\btree}: A B-tree-like structure implemented using \system (4KB pages and 255 fanout), which serves
    as the most controlled baseline. 
    % where the only difference is 
    % \system's storage- and data-aware tuning.

\paragraph{Datasets} 
% Our datasets primarily come from the search on sorted data 
First, we use the SOSD benchmark~\cite{sosd-neurips,DBLP:journals/pvldb/MarcusKRSMK0K20BenchmarkLearned}, including \texttt{books} (800M), \texttt{fb} (200M), \texttt{osm} (800M), and \texttt{wiki} (200M). Each of these contains 200-800 million 64-bit integer keys stored consecutively in an array. Given a query integer key, the task is to find its offset position in the array. Equivalently, it asks the systems for the rank of the query integer. As an unusual dataset, \texttt{wiki} contains many duplicated keys in which the task is to find the smallest offset of the key.
Second, for more diverse data patterns,
    we also use a synthetic dataset, \texttt{gmm}, generated
        from a Gaussian mixture model (GMM) of 100 normal distribution clusters over 800 million keys.
% To diversify the overall data pattern, 
% we extend the list with \texttt{gmm} synthesized from a Gaussian mixture model (GMM) of \fixed{100} normal distribution clusters over 800 million keys.

\ignore{
\paragraph{\system Hyperparameters} \label{sec:exp-setup-hyperparam}
To demonstrate \system storage-aware tuning, we model the two storages after affine storage profile $T_{\text{aff}}(\Delta) = \ell + \Delta / B$, parameterized by latency and bandwidth. The numbers are collected and rounded from \texttt{ioping}~\cite{ioping} with different request sizes. In NFS, we set $\ell = 50$ms and $B = 12$MB/s, while in SSD, we set $\ell = 250$us and $B=175$MB/s. Beyond storage profiles, \system exponentiates the granularity hyperparameters in power of 2: $\lambda_{GS}, \lambda_{GB}, \lambda_{EB} \in \{256, 512, \dots, 1048576\}$ (39 node builders in total). It then selects top-5 ($k = 5$) candidates to branch out. In addition, each \texttt{step} node contains $16$ pieces with sufficiently fine granularity and compact byte representation.
}

%%%%%%%%%%%%%%%%%%%%%%%%%%%%%%%%%%%%%%%%%%%%%%%%%%%%%%%%%%%%%
%%%%%%%%%%%%%%%%%%%%%%%%%%%%%%%%%%%%%%%%%%%%%%%%%%%%%%%%%%%%%

\subsection{Faster End-to-end Lookup Speed}
\label{sec:exp-e2e}

We study cold-state and warm-state latencies separately. 
    Cold-state latency is useful for understanding the performance under short-lived executions (e.g., serverless, ad-hoc workloads) and very large data (e.g., many tables, large indexes). Afterward, we study warm-state latency curves over different warmnesses. \cref{sec:exp-storage-impact} later discusses index structures discovered by \system.

% Here the standard deviations of latency measurements vary from 20--50ms on NFS and 0.5--1.5ms
\paragraph{Cold-state Latency}
    
    \system is consistently one of the fastest methods at searching the first query, across datasets and storage (\cref{fig:I_latency_cold}).
    % A comparison between tuned index (\system) and traditional indexes (\lmdb and \btree) emphasizes the importance of storage adaptivity in index tuning. 
    Compared to 
    % traditional 
    \lmdb (B-tree),
    % solutions, 
        \system is $2.4\times$--$2.7\times$ faster 
    on NFS and $2.6\times$--$4.1\times$ faster on HDD.
            % ranging from . 
    \system is on par with \lmdb on SSD. 
    Similarly, \system is $2.0\times$--$2.4\times$ faster than 
    % the fixed structure 
    \btree on NFS
        but performs equally well on SSD on HDD. 
    This difference across storages suggests that \lmdb and \btree structures are tuned to disk-scale storage profiles like SSD's, 
        so they underperform on storage on a different scale like NFS.
% Normally, system administrator(s) would have to test many read sizes on expensive benchmarks 
%     for each data and storage setting. 
% On the other hand, \system is able to tune automatically, and so, avoid these costs in benchmark computation and human operation. 
% In addition, the distinction subtly suggests that the diversity in node types gives \system an edge to fit better with wider sets of data.

\begin{figure*}[t]

\pgfplotsset{
    accfig/.style={
        width=\linewidth,
        height=28mm,
        ybar,
        axis lines=left,
        bar width=1.6mm,
        bar shift=0pt,
        % xlabel=Corpus,
        % xmin=0,
        % xmax=7.5,
        enlarge x limits=0.1,
        xlabel=Number of Layers $L$, 
        ylabel near ticks,
        ylabel style={align=center},
        xlabel near ticks,
        xlabel shift=-2mm,
        xlabel style={yshift=0mm},
        xticklabel style={font=\scriptsize, text width=5mm, align=center},
        legend style={
            at={(1.02,1.00)},anchor=north west,column sep=2pt,
            draw=black,fill=white,line width=.5pt,
            /tikz/every even column/.append style={column sep=5pt}
        },
        legend cell align={left},
        legend columns=1,
        area legend,
        clip=false,
        every axis/.append style={
            font=\footnotesize
        },
        /pgfplots/log origin=infty,
        xtick style={draw=none},
        xtick pos=bottom,
        minor tick num=1,
        minor grid style=lightgray,
        ymajorgrids,
        yminorgrids,
        },
    accfig/.belongs to family=/pgfplots/scale,
}

\begin{subfigure}[t]{0.2\linewidth} \centering
    \begin{tikzpicture}
    
    \begin{axis}[accfig,
        ymin=0,
        ymax=290,
        xtick={1, ..., 5},
        xticklabels={
            % \textbf{1\quad(A)},  % 1
            1,
            2,
            3, 
            4, 
            5, 
        },
        ylabel=Latency (ms)]
    
    \addplot[fill={airindexcolor}]
    plot [error bars/.cd, y dir=both,y explicit,error bar style={color=black}]
    table[x=x,y=y, y error=std] { 
    x y std
    % 1 69.64 20.39
    1 69.64 20.39
    };
    
    \addplot[fill={btreecolor}]
    plot [error bars/.cd, y dir=both,y explicit,error bar style={color=black}]
    table[x=x,y=y, y error=std] { 
    x y std
    % 2 78.17 23.15
    % 3 111.30 27.47
    % 4 144.08 47.77
    % 5 194.71 149.19
    % 6 170.83 44.50
    % 2 102.63 22.18
    2 94.62 21.22
    3 120.78 34.81
    4 155.78 62.75
    5 190.77 65.34
    };
    
    \node[font=\scriptsize,anchor=south west,draw=airindexcolor,text=airindexcolor,
        thick,fill=white]  (O)
        at (axis cs: 0.8, 200) 
        {\textbf{\textsf{Tuned}}};
    \draw [to-,ultra thick,airindexcolor] (axis cs: 1,140) -- (1,200);
      
    \end{axis}
    
    \end{tikzpicture}
    \vspace{-2mm}
    \caption{Varying $L$ on NFS}
\end{subfigure}
~
\begin{subfigure}[t]{0.3\linewidth} \centering
    \begin{tikzpicture}
    
    \begin{axis}[accfig,
        ymin=0,
        ymax=290,
        ylabel=Latency (ms),
        xtick={1, 2, ..., 8},
        xticklabels={
            $2^{8}$, 
            $2^{10}$,
            $2^{12}$, 
            $2^{14}$, 
            $2^{16}$, 
            % \textbf{$\mathbf{2^{16.3}}$ (A)},  % 81,471.993 ~ 2^{16.3}
            $\mathbf{2^{16.3}}$,
            $2^{18}$, 
            $2^{20}$, 
        },
        xlabel=Granularity $\lambda$]
    
    \addplot[fill={airindexcolor}]
    plot [error bars/.cd, y dir=both,y explicit,error bar style={color=black}]
    table[x=x,y=y, y error=std] { 
    x y std
    6 69.64 20.39
    };
    
    \addplot[fill={btreecolor}]
    plot [error bars/.cd, y dir=both,y explicit,error bar style={color=black}]
    table[x=x,y=y, y error=std] { 
    x y std
    1 149.92 40.92
    2 112.46 35.64
    3 107.57 25.33
    4 89.64 25.90
    5 76.37 19.62
    7 76.50 26.46
    8 73.69 17.28
    };
    
    \node[font=\scriptsize,anchor=south west,draw=airindexcolor,text=airindexcolor,
        thick,fill=white]  (O)
        at (axis cs: 4.8, 200) 
        {\textbf{\textsf{Tuned}}};
    \draw [to-,ultra thick,airindexcolor] (axis cs: 6,140) -- (6,200);
    
    \end{axis}
    \end{tikzpicture}
    \vspace{-2mm}
    \caption{Varying $\lambda$ on NFS}
\end{subfigure}
~
\begin{subfigure}[t]{0.2\linewidth} \centering
    \begin{tikzpicture}
    
    \begin{axis}[accfig,
        xtick={1, ..., 5},
        xticklabels={
            1, 
            % \textbf{2\quad(A)}, % 2
            2,
            3, 
            4, 
            5, 
        },
        ymin=0,
        ymax=7.9,
        ylabel=Latency (ms)]
    
    \addplot[fill={airindexcolor}]
    plot [error bars/.cd, y dir=both,y explicit,error bar style={color=black}]
    table[x=x,y=y, y error=std] { 
    x y std
    2 2.90 0.52
    };
    
    \addplot[fill={btreecolor}]
    plot [error bars/.cd, y dir=both,y explicit,error bar style={color=black}]
    table[x=x,y=y, y error=std] { 
    x y std
    1 3.02 0.43
    % 3 3.23 0.85
    3 3.83 0.87
    4 4.83 0.99
    5 5.49 1.92
    };
    
    \node[font=\scriptsize,anchor=south west,draw=airindexcolor,text=airindexcolor,
        thick,fill=white]  (O)
        at (axis cs: 1.3, 6) 
        {\textbf{\textsf{Tuned}}};
    \draw [to-,ultra thick,airindexcolor] (axis cs: 2,4) -- (2,6);
    
    \end{axis}
    \end{tikzpicture}
    \vspace{-2mm}
    \caption{Varying $L$ on SSD}
\end{subfigure}
~
\begin{subfigure}[t]{0.3\linewidth} \centering
    \begin{tikzpicture}
    
    \begin{axis}[accfig,   
        ymin=0,
        ymax=7.9,
        xtick={1, 2, ..., 8},               
        xticklabels={
            $2^{8}$, 
            $2^{10}$,
            % \textbf{$\mathbf{2^{11.6}}$ (A)},  % 2,986.609 ~ 2^{11.6}
            $\mathbf{2^{11.6}}$,
            $2^{12}$, 
            $2^{14}$, 
            $2^{16}$, 
            $2^{18}$, 
            $2^{20}$, 
        },
        xlabel=Granularity $\lambda$,
        ylabel=Latency (ms)]
    
    \addplot[fill={airindexcolor}]
    plot [error bars/.cd, y dir=both,y explicit,error bar style={color=black}]
    table[x=x,y=y, y error=std] { 
    x y std
    3 2.90 0.52
    };
    
    \addplot[fill={btreecolor}]
    plot [error bars/.cd, y dir=both,y explicit,error bar style={color=black}]
    table[x=x,y=y, y error=std] { 
    x y std
    1 4.18 0.48
    2 3.85 1.59
    4 3.10 1.30
    5 2.67 0.85
    6 3.00 0.91
    7 2.68 0.82
    8 3.19 0.89
    };
    
    \node[font=\scriptsize,anchor=south west,draw=airindexcolor,text=airindexcolor,
        thick,fill=white]  (O)
        at (axis cs: 2.3, 6) 
        {\textbf{\textsf{Tuned}}};
    \draw [to-,ultra thick,airindexcolor] (axis cs: 3,4) -- (3,6);
      
    \end{axis}
    \end{tikzpicture}
    \vspace{-2mm}
    \caption{Varying $\lambda$ on SSD}
\end{subfigure}
    
    \vspace{-1mm}
    \caption{Comparison of average first-query latencies between \system-tuned designs (in red) and 
        manual alternatives across NFS and SSD
        (\texttt{fb} dataset)
        varying numbers of layers $L$ and granularities $\lambda$. 
    % Two index structure variables are the number of layers $L \in \{1, 2, 3, 4, 5\}$ and the granularity hyperparameter $\lambda \in \{ 2^{8}, 2^{10}, 2^{12}, 2^{14}, 2^{16}, 2^{18}, 2^{20} \}$. 
    The error bars display standard deviations.
    % \system's tuned structures on NFS and SSD have 1 and 2 index layers with average granularities of 81KB and 3KB respectively.
    }
    % \vspace{-1mm}
    \label{fig:III_variants}
\end{figure*}
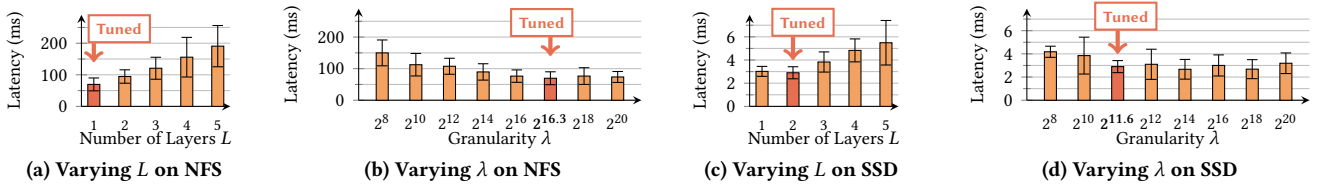

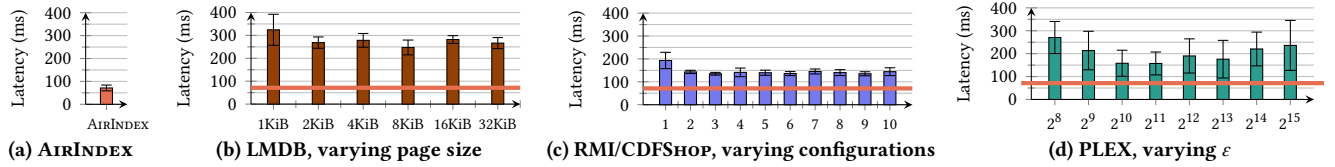
\begin{figure*}[t]

\pgfplotsset{
    accfig/.style={
        width=\linewidth,
        height=28mm,
        ybar,
        axis lines=left,
        bar width=1.6mm,
        bar shift=0pt,
        ymin=0,
        ymax=400,
        enlarge x limits=0.1,
        ylabel near ticks,
        ylabel style={align=center},
        xlabel style={yshift=0mm},
        xticklabel style={font=\scriptsize, text width=5mm, align=center},
        legend style={
            at={(1.02,1.00)},anchor=north west,column sep=2pt,
            draw=black,fill=white,line width=.5pt,
            /tikz/every even column/.append style={column sep=5pt}
        },
        legend cell align={left},
        legend columns=0,
        area legend,
        clip=false,
        every axis/.append style={
            font=\footnotesize
        },
        /pgfplots/log origin=infty,
        % xtick style={draw=none},
        xtick pos=bottom,
        minor tick num=1,
        minor grid style=lightgray,
        ymajorgrids,
        yminorgrids,
        },
    accfig/.belongs to family=/pgfplots/scale,
}

\begin{subfigure}[t]{0.12\linewidth} \centering
    \begin{tikzpicture}
    
    \begin{axis}[accfig,
        xmin=-1.5,
        xmax=1.5,
        xtick={0},
        xticklabels={\system},
        xlabel style={fill opacity=0.0},
        ylabel=Latency (ms)]
    
    \addplot[fill={airindexcolor}]
    plot [error bars/.cd, y dir=both,y explicit,error bar style={color=black}]
    table[x=x,y=y, y error=std] { 
    x y std
    0 71.31 12.63
    };

    \end{axis}
    \end{tikzpicture}
    \vspace{-2mm}
    \caption{\system}
\end{subfigure}
~
\begin{subfigure}[t]{0.29\linewidth} \centering
    \begin{tikzpicture}
    
    \begin{axis}[accfig,
        ylabel=Latency (ms),
        xtick={1, 2, ..., 8},
        xticklabels={
            % $2^{10}$,
            % $2^{11}$, 
            % $2^{12}$, 
            % $2^{13}$,
            % $2^{14}$, 
            % $2^{15}$,
            1KiB,
            2KiB,
            4KiB,
            8KiB,
            16KiB,
            32KiB,
        },
        % xticklabel style={rotate=90},
        % xlabel=Page Size,
        ]

    \addplot[fill={lmdbcolor}]
    plot [error bars/.cd, y dir=both,y explicit,error bar style={color=black}]
    table[x=x,y=y, y error=std] { 
    x y std
    1	324.4596723	67.91109521
    2	268.6082001	24.66235127
    3	278.491849	30.01338305
    4	247.3473173	32.39897421
    5	282.0126847	16.94066866
    6	266.3959743	24.1603186
    };
    
    % \node[font=\scriptsize,anchor=south west,draw=airindexcolor,text=airindexcolor,
    %     thick,fill=white]  (O)
    %     at (axis cs: 4.8, 200) 
    %     {\textbf{\textsf{Tuned}}};
    \draw [ultra thick,airindexcolor] (axis cs:\pgfkeysvalueof{/pgfplots/xmin},71.31) -- (axis cs:\pgfkeysvalueof{/pgfplots/xmax},71.31);
    
    \end{axis}
    \end{tikzpicture}
    \vspace{-2mm}
    \caption{\lmdb, varying page size}
    \label{fig:VIIII_tuned_lmdb}
\end{subfigure}
~
\begin{subfigure}[t]{0.29\linewidth} \centering
    \begin{tikzpicture}
    
    \begin{axis}[accfig,
        xtick={1, ..., 10},
        xticklabels={
            1, 
            2,
            3, 
            4, 
            5, 
            6, 
            7, 
            8, 
            9, 
            10, 
        },
        % ytick={0, 2, 4, ..., 10},
        % xlabel=Settings, 
        ylabel=Latency (ms)]
    
    \addplot[fill={rmicolor}]
    plot [error bars/.cd, y dir=both,y explicit,error bar style={color=black}]
    table[x=x,y=y, y error=std] { 
    x y std
    1	193.1716	35.67479144
    2	143.30365	7.052012713
    3	135.2525	6.012069557
    4	141.30455	18.87935006
    5	139.698825	11.13668577
    6	136.020375	8.702410281
    7	145.24525	10.84260051
    8	140.21565	12.96116394
    9	135.215625	8.789238363
    10	144.29925	17.12217923
    };
    \draw [ultra thick,airindexcolor] (axis cs:\pgfkeysvalueof{/pgfplots/xmin},71.31) -- (axis cs:\pgfkeysvalueof{/pgfplots/xmax},71.31);
    
    \end{axis}
    \end{tikzpicture}
    \vspace{-2mm}
    \caption{\rmi/\cdfshop, varying configurations}
    \label{fig:VIIII_tuned_rmi}
\end{subfigure}
~
\begin{subfigure}[t]{0.3\linewidth} \centering
    \begin{tikzpicture}
    
    \begin{axis}[accfig,
        xmode=log,
        xtick={
            256,
            512,
            1024,
            2048,
            4096,
            8192,
            16384,
            32768
        },
        xticklabels={
            $2^{8}$, 
            $2^{9}$,
            $2^{10}$, 
            $2^{11}$,
            $2^{12}$,
            $2^{13}$, 
            $2^{14}$, 
            $2^{15}$
            % 1KiB,
            % 4KiB,
            % 16KiB,
            % 64KiB,
            % 262KiB,
            % 1MiB,
        },
        minor y tick num=1,
        % xticklabel style={rotate=90},
        % % xlabel=Granularity $\lambda$,
        ylabel=Latency (ms)]
    
    \addplot[fill={plexcolor}]
    plot [error bars/.cd, y dir=both,y explicit,error bar style={color=black}]
    table[x=x,y=y, y error=std] { 
        x	y	std
        32768	235.99	109.07
        16384	220.42	73.65
        8192	175.91	82.16
        4096	190.02	74.66
        2048	157.15	50
        1024	158.01	57.04
        512	    213.66	84.05
        256	    270.57	70.07
    };
    \draw [ultra thick,airindexcolor] (axis cs:128,71.31) -- (axis cs:65536,71.31);

    \end{axis}
    \end{tikzpicture}
    \vspace{-2mm}
    \caption{\plex, varying $\varepsilon$}
    \label{fig:VIIII_tuned_plex}
\end{subfigure}
    
    \vspace{-1mm}
    % \caption{Comparison on the \texttt{books} dataset under NFS between \system-tuned design (left) and other methods with varying knobs: \lmdb, \rmi, and \btree. \rmi's settings are recommended by its optimizer with the least accurate (fewer models) on the left and the most accurate (more models) on the right. The error bars display standard deviations of the latency.}
    \caption{Comparison of average first-query latencies on the \texttt{books} dataset in NFS between \system-tuned design (left) and other methods with varying knobs. \rmi's settings are recommended by \cdfshop with the least accurate (fewer models) on the left and the most accurate (more models) on the right. The error bars display standard deviations of the latency.}
    \vspace{-2mm}
    \label{fig:VIIII_tuned}
\end{figure*}

% With various node types, \rmi is a formidable competitor. Considering its perfectly accurate and minimal root layer, \rmi, unfortunately, 
Compared to learned index baselines, \system is reliably faster without unexpectedly long latency arising from tuning difficulties.
    While \rmi performs reasonably well,
        it has two limitations.
    First, its fixed two-layer structure makes it rigid to the underlying storage.
    Second, because \rmi's top-down building assigns a disproportionate amount of data to intermediate nodes, the second-layer precision varies substantially. This later effect is more pronounced in \texttt{gmm}. 
    Overall, \system delivers lower latencies more reliably than \rmi, being $1.2\times$--$2.6\times$ faster on NFS, $1.0\times$--$2.2\times$ faster on SSD, and $1.4\times$--$5.9\times$ faster on HDD. 
    ~
    \pgm suffers on \texttt{books}, \texttt{fb}, and \texttt{osm}, but is competitive on \texttt{wiki} and \texttt{gmm}. Upon closer inspection, \pgm fits poorly with the former three, creating larger indexes than those in the latter two.
    % Because \texttt{MappedPGMIndex} fetches all index layers and partially reads only the data layer, model fitness or index size is a crucial performance indicator in \pgm. In contrast, \system agnostically addresses this unpredictable node type's preference by exploring different choices independently.
    \system is on par with \pgm on \texttt{wiki} and \texttt{gmm}, but outperforms on other datasets with $3.0\times$--$7.2\times$, $5.8\times$--$11.7\times$, and $9.0\times$--$15.6\times$ speedup on NFS, SSD, and HDD respectively.
    ~
    Similarly, \system is faster than \alex in all settings except in \texttt{gmm} SSD, with $1.7\times$--$10.1\times$, $1.3\times$--$46.3\times$, and $3.0\times$--$22.2\times$ speedup on NFS, SSD, and HDD. \alex performs poorly in \texttt{osm} because its root node holds 2M child pointers with more than 15MB of data.
    % , forcing it to read over 15MB of data initially.
    ~
    Lastly, \system is faster than \plex on NFS and HDD with $1.5\times$--$2.2\times$, and $1.7\times$--$3.3\times$ speedup. Both methods perform equally well on SSD: \system is $1.7\times$ faster at searching \texttt{osm}, but $1.4\times$ slower at \texttt{gmm}. Upon closer inspection, \plex's compact histogram tree fit \texttt{osm} poorly (762KB in size) but fit \texttt{gmm} exceptionally better (0.6KB in size).

% Because our implementation of \dcal can only consider indexes with \texttt{step} nodes, 
Compared to \dcal,
    \system searches in a richer set of indexes and so it is consistently faster: $1.4\times$--$2.0\times$, $1.2\times$--$1.5\times$, and $1.0\times$--$1.4\times$ speedup on NFS, SSD, and HDD. On the bright side, \dcal generally has a faster lookup than \btree because of its storage-aware tuning and \btree's fixed structure.

\paragraph{Warm-state Latency Curve} 

% If we let these systems fill up their caches, 
As we continue querying (and caching more),
    % they will serve subsequent queries 
    the queries become progressively faster shown as per-query latency curves in \cref{fig:I_latency_curve}. 
Here, a point $(x, y)$ in the latency curve implies that a system completes $x$ queries in $x \times y$ seconds. 
The differences in accelerations across methods can be explained by their index structures. In hierarchical indexes, shorter and narrower indexes accelerate more aggressively than taller and wider indexes because fewer (random) queries are needed to touch all nodes in a narrower index layer. For example, in \texttt{osm} dataset, \alex accelerates faster than \lmdb because of its shorter index with a one-node root as opposed to a full B-tree.
% All latency curves have a similar downward step pattern where they alternate between fast and slow accelerations. 
% Fast regions occurs when the cache is filling up almost all of an index layer that cache hits become increasingly more beneficial. 
% In \system, this index layer corresponds to the topmost one that is not totally cached. 
% If the experiment proceeded further, say 10--100 million queries, we would observe the last plateau where all index and data layers are in memory. 
% In general, a $L$-layer hierarchical index would have $L + 1$ plateaus, for example, noting that the last plateau is not visible, \system's indexes in the first three plots of \cref{fig:I_latency_curve} have $1$ layers while \system's index in \cref{fig:I_latency_curve_ssd_osm} has $2$ layers.
Even though \system optimizes for cold-state latency, the tuned structure is still faster for warm-state latency,
% . Not only does \system search the first query faster, but it also continues to do so up until certain numbers of queries
ranging from 100 to 100K queries. Such a range of warmness is useful for a large collection of datasets, short-lived search sessions, or limited memory environments.

% \tofix{Write possibly no free lunch across query}

%%%%%%%%%%%%%%%%%%%%%%%%%%%%%%%%%%%%%%%%%%%%%%%%%%%%%%%%%%%%%
%%%%%%%%%%%%%%%%%%%%%%%%%%%%%%%%%%%%%%%%%%%%%%%%%%%%%%%%%%%%%

\ignore{
\subsection{Latency Breakdown}
\label{sec:exp-breakdown}

\def\numberyshift{10.0}
\def\subfigurewidth{0.23\linewidth}
\def\subfigurelegendwidth{0.31\linewidth}
\def\legendwidth{0.08\linewidth}

\pgfplotstableread{
    queries	setting	method	root	layer_2	layer_1	data	CHECKSUM	total	speed	yshift
    1	nfs	AirIndex	34.5271	0.0000	0.0000	48.4262	82.9533	83.0275	83.0275344	{\numberyshift}
    10	nfs	AirIndex	3.3104	0.0000	0.0000	25.1918	28.5022	285.5671	28.5567098	{\numberyshift}
    100	nfs	AirIndex	0.3465	0.0000	0.0000	10.7126	11.0591	1109.1088	11.0910879	{\numberyshift}
    1000	nfs	AirIndex	0.0325	0.0000	0.0000	10.3366	10.3691	10395.0381	10.39503806	{\numberyshift}
    10000	nfs	AirIndex	0.0040	0.0000	0.0000	10.3967	10.4007	104270.7217	10.42707217	{\numberyshift}
    100000	nfs	AirIndex	0.0005	0.0000	0.0000	8.7604	8.7608	878595.9146	8.785959146	{\numberyshift}
    1000000	nfs	AirIndex	0.0001	0.0000	0.0000	2.3466	2.3466	2366450.4612	2.366450461	{\numberyshift}
}\AirindexNFS

\pgfplotstableread{
    queries	setting	method	root	layer_2	layer_1	data	CHECKSUM	total	speed	yshift
    1	nfs	B-tree	42.5475	21.0059	51.9127	61.6137	177.0798	177.1482	177.1481976	{\numberyshift}
    10	nfs	B-tree	3.7547	5.6846	26.0170	46.3382	81.7944	818.4797	81.84797339	{\numberyshift}
    100	nfs	B-tree	0.3361	0.6180	15.9270	47.2940	64.1751	6420.2773	64.20277329	{\numberyshift}
    1000	nfs	B-tree	0.0398	0.0703	14.7843	46.8981	61.7925	61812.0694	61.81206939	{\numberyshift}
    10000	nfs	B-tree	0.0051	0.0069	3.0544	42.0192	45.0856	451035.5964	45.10355964	{\numberyshift}
    100000	nfs	B-tree	0.0005	0.0032	0.2108	22.1394	22.3540	2237249.4847	22.37249485	{\numberyshift}
    1000000	nfs	B-tree	0.0001	0.0001	0.0155	3.9664	3.9822	4000381.8391	4.000381839	{\numberyshift}
}\BtreeNFS

\pgfplotstableread{
    queries	setting	method	root	layer_2	layer_1	data	CHECKSUM	total	speed	yshift
    1	local	AirIndex	0.7846	0.0000	0.3722	0.6387	1.7955	1.8446	1.8446016	{\numberyshift}
    10	local	AirIndex	0.0806	0.0000	0.1713	0.3649	0.6168	6.3412	0.63411884	{\numberyshift}
    100	local	AirIndex	0.0086	0.0000	0.1116	0.2865	0.4067	41.4568	0.4145683815	{\numberyshift}
    1000	local	AirIndex	0.0008	0.0000	0.0473	0.2738	0.3219	326.1493	0.3261493397	{\numberyshift}
    10000	local	AirIndex	0.0001	0.0000	0.0049	0.2708	0.2757	2797.7870	0.279778696	{\numberyshift}
    100000	local	AirIndex	0.0000	0.0000	0.0008	0.2740	0.2748	27923.8097	0.2792380973	{\numberyshift}
    1000000	local	AirIndex	0.0000	0.0000	0.0001	0.1368	0.1369	143080.9193	0.1430809193	{\numberyshift}
}\AirindexSSD

\pgfplotstableread{
    queries	setting	method	root	layer_2	layer_1	data	CHECKSUM	total	speed	yshift
    1	local	B-tree	0.8119	0.1492	0.4109	0.6748	2.0468	2.0973	2.0972648	{\numberyshift}
    10	local	B-tree	0.0972	0.0684	0.3100	0.3559	0.8315	8.4901	0.849007395	{\numberyshift}
    100	local	B-tree	0.0084	0.0080	0.1592	0.3058	0.4814	48.8935	0.4889347795	{\numberyshift}
    1000	local	B-tree	0.0009	0.0008	0.1166	0.2826	0.4009	405.8741	0.4058741094	{\numberyshift}
    10000	local	B-tree	0.0001	0.0001	0.0245	0.2818	0.3065	3106.4899	0.3106489902	{\numberyshift}
    100000	local	B-tree	0.0000	0.0000	0.0043	0.2842	0.2886	29412.6522	0.2941265221	{\numberyshift}
    1000000	local	B-tree	0.0000	0.0000	0.0004	0.1328	0.1333	143544.9902	0.1435449902	{\numberyshift}
}\BtreeSSD

\pgfplotsset{
    breakdown/.style={
        width=\linewidth,
        height=35mm,
        stack plots=y,
        area style,
        % xmode=log,  % does not work in style
        % ymode=log,  % does not work in style
        bar width=1.8mm,
        axis lines=left,
        % axis line style={-},
        % y axis line style={draw opacity=1}
        % x axis line style={draw opacity=1},
        minor tick num=1,
        xmajorgrids,
        ymajorgrids,
        yminorgrids,
        ytick style={draw=none},
        xtick={1, 10, 100, 1000, 10000, 100000, 1000000},
        xticklabels={1, \;, $10^2$, \;, $10^4$, \;, $10^6$},
        xlabel near ticks,
        ylabel near ticks,
        ylabel=Latency (ms/op),
        xlabel=Number of Queries,
        ylabel shift=-1mm,
        xlabel shift=-1mm,
        every axis/.append style={font=\footnotesize},
        reverse legend,
        legend columns=1,
        legend style={
            font=\scriptsize,
            at={(1.05,0.5)}, anchor=west,
            draw=black,
            fill=white
        },
        legend cell align={left},
        every node near coord/.append style={yshift=0pt}
    },
}

\pgfplotsset{
    breakdownnfs/.style={
        breakdown,
        ymin=0,
        ymax=199,
        xmax=2000000,
    },
}

\pgfplotsset{
    breakdownssd/.style={
        breakdown,
        ymin=0,
        ymax=2.4,
        xmax=2000000,
    },
}

\begin{figure*}[t]
    \centering
    
    \begin{subfigure}[b]{\subfigurewidth}
        \centering
        \begin{tikzpicture}
            \begin{axis}[
                breakdownnfs,
                % ymode=log,
                xmode=log
            ]
                \addplot[
                    fill={rootcolor},
                    text opacity=1] table [x=queries, y=root] {\BtreeNFS} \closedcycle;
                \addplot[
                    fill={layeronecolor},
                    text opacity=1] table [x=queries, y=layer_2] {\BtreeNFS} \closedcycle;
                \addplot[
                    fill={layertwocolor},
                    text opacity=1] table [x=queries, y=layer_1] {\BtreeNFS} \closedcycle;
                \addplot[
                    fill={datalayercolor},
                    text opacity=1] table [x=queries, y=data] {\BtreeNFS} \closedcycle;
                % \addplot [
                %     opacity=0.0,
                %     text opacity=1.0,
                %     nodes near coords,
                %     point meta=explicit
                % ] table [x=queries, y=yshift, meta=speed] {\BtreeNFS} \closedcycle;
            \end{axis}
        \end{tikzpicture}
        \vspace{-2mm}
        \caption{NFS (\btree)}
        \label{fig:II_breakdown_by_layer_nfs_btree}
    \end{subfigure}
    ~
    \begin{subfigure}[b]{\subfigurewidth}
        \centering
        \begin{tikzpicture}
            \begin{axis}[
                breakdownnfs,
                % ymode=log,
                xmode=log
            ]
                \addplot[
                    fill={rootcolor},
                    text opacity=1] table [x=queries, y=root] {\AirindexNFS} \closedcycle;
                \addplot[
                    fill={layeronecolor},
                    text opacity=1] table [x=queries, y=layer_2] {\AirindexNFS} \closedcycle;
                \addplot[
                    fill={layertwocolor},
                    text opacity=1] table [x=queries, y=layer_1] {\AirindexNFS} \closedcycle;
                \addplot[
                    fill={datalayercolor},
                    text opacity=1] table [x=queries, y=data] {\AirindexNFS} \closedcycle;
                % \addplot [
                %     opacity=0.0,
                %     text opacity=1.0,
                %     nodes near coords,
                %     point meta=explicit
                % ] table [x=queries, y=yshift, meta=speed] {\AirindexNFS} \closedcycle;
            \end{axis}
        \end{tikzpicture}
        \vspace{-2mm}
        \caption{NFS (\system)}
        \label{fig:II_breakdown_by_layer_nfs_airindex}
    \end{subfigure}
    ~
    \begin{subfigure}[b]{\subfigurewidth}
        \centering
        \begin{tikzpicture}
            \begin{axis}[
                breakdownssd,
                % ymode=log,
                xmode=log
            ]
                \addplot[
                    fill={rootcolor},
                    text opacity=1] table [x=queries, y=root] {\BtreeSSD} \closedcycle;
                \addplot[
                    fill={layeronecolor},
                    text opacity=1] table [x=queries, y=layer_2] {\BtreeSSD} \closedcycle;
                \addplot[
                    fill={layertwocolor},
                    text opacity=1] table [x=queries, y=layer_1] {\BtreeSSD} \closedcycle;
                \addplot[
                    fill={datalayercolor},
                    text opacity=1] table [x=queries, y=data] {\BtreeSSD} \closedcycle;
                % \addplot [
                %     opacity=0.0,
                %     text opacity=1.0,
                %     nodes near coords,
                %     point meta=explicit
                % ] table [x=queries, y=yshift, meta=speed] {\BtreeSSD} \closedcycle;
            \end{axis}
        \end{tikzpicture}
        \vspace{-2mm}
        \caption{SSD (\btree)}
        \label{fig:II_breakdown_by_layer_ssd_btree}
    \end{subfigure}
    ~
    \begin{subfigure}[b]{\subfigurelegendwidth}
        \centering
        \begin{tikzpicture}
            \begin{axis}[
                breakdownssd,
                width=0.74\linewidth,
                % ymode=log,
                xmode=log
            ]
                \addplot[
                    fill={rootcolor},
                    text opacity=1] table [x=queries, y=root] {\AirindexSSD} \closedcycle;
                \addplot[
                    fill={layeronecolor},
                    text opacity=1] table [x=queries, y=layer_2] {\AirindexSSD} \closedcycle;
                \addplot[
                    fill={layertwocolor},
                    text opacity=1] table [x=queries, y=layer_1] {\AirindexSSD} \closedcycle;
                \addplot[
                    fill={datalayercolor},
                    text opacity=1] table [x=queries, y=data] {\AirindexSSD} \closedcycle;
                % \addplot [
                %     opacity=0.0,
                %     text opacity=1.0,
                %     nodes near coords,
                %     point meta=explicit
                % ] table [x=queries, y=yshift, meta=speed] {\AirindexSSD} \closedcycle;
                \legend{Root Layer, Layer 2, Layer 1, Data Layer}
            \end{axis}
        \end{tikzpicture}
        \vspace{-2mm}
        \caption{SSD (\system)}
        \label{fig:II_breakdown_by_layer_ssd_airindex}
    \end{subfigure}
    
    \vspace{-3mm}
    \caption{Latency breakdown of \btree and \system by the time spent reading different layers. The \texttt{books} dataset is used. We vary the warmness by the number of queries 
    from one to 1 million
    % in $\{1, 10, 10^2, 10^3, \dots, 10^6\}$ 
    from left to right in logarithmic scale. Layer-wise latencies are stacked from bottom up in their retrieval order (root index layer to data layer). For visual purposes, the plots then interpolate linearly and fill areas in between with different colors indicating different layers.}
    \label{fig:II_breakdown_by_layer}
    \vspace{-3mm}
\end{figure*}

We decompose the end-to-end latency into 
    the times spent 
    % in retrieving 
    for each layer, 
which includes I/O and non-I/O operations (e.g., in-memory caching, data structure deserialization, node prediction, relevant key-value finding). These non-I/O operations only account for up to $1.0\%$ on NFS and $9.0\%$ on SSD, even after 1 million queries.
Apart from some small exceptions, 
    latency measurements across layers roughly follow the storage profiles (\cref{sec:exp-setup-hyperparam} for specific hyperparameters): 
    larger root size $s(\Theta_L)$ and coarser precision $\Delta_l(x; \Theta_l)$ reflects in a longer latency spent in the corresponding layer. 
Over a number of queries, 
    we also observe the alternating acceleration phenomenon in more detail. 
That is, the fast acceleration region indicates that the topmost partially cached index layer is becoming fully cached. 
For example, in \cref{fig:II_breakdown_by_layer_ssd_btree} between $10^3$ and $10^4$ queries, \btree searches faster because of the speedup in its layer-1 index.

This breakdown also reveals factors unaccounted for in \system. 
    For a prominent example, in the first query under SSD (\cref{fig:II_breakdown_by_layer_ssd_btree,fig:II_breakdown_by_layer_ssd_airindex}),
    both \system and \btree spend more time reading their root and data layers as opposed to reading other index layers.
This is because index layers and data layers are stored in separate directories, 
    forcing the file system (\texttt{ext4}) to slowly walk through different paths of directory entries (dentry) to fetch index-layer and data-layer inodes. 
Consequently, 
    reading subsequent index layers stored in the same directory is then significantly faster than expected. 
Although these missing characteristics are crucial for future works towards maximally fast indexes, 
    we believe that \system's storage profile is 
    at an appropriate level of abstraction to adapt to diverse types of storage.
}

%%%%%%%%%%%%%%%%%%%%%%%%%%%%%%%%%%%%%%%%%%%%%%%%%%%%%%%%%%%%%
%%%%%%%%%%%%%%%%%%%%%%%%%%%%%%%%%%%%%%%%%%%%%%%%%%%%%%%%%%%%%

% \subsection{Skewed Workload}
% \label{sec:exp-skew}

% Although \system's objective (\cref{eq:emm-obj-airindex}) considers the query distribution $\calX$, future query distributions may change unexpectedly. This experiment (\cref{fig:VI_skew}) builds \system on the uniform distribution $\calX$ but requests keys sampled from a Zipf distribution with parameters 0.5 (least skewed), 1.0, and 2.0 (most skewed). The more skewed the query is, the faster all methods can respond at warm state (\cref{fig:VI_skew_later}). However, the skew does not affect first-query latency as much across all methods (\cref{fig:VI_skew_first}). Because \system is tuned for cold-state latency, higher skewness results in a quicker takeover. For example, it takes 12k uniform queries for any methods (\pgm) to take over \system, but only 70, 41, and 725 queries in 0.5, 1.0, and 2.0 Zipf query.

%%%%%%%%%%%%%%%%%%%%%%%%%%%%%%%%%%%%%%%%%%%%%%%%%%%%%%%%%%%%%
%%%%%%%%%%%%%%%%%%%%%%%%%%%%%%%%%%%%%%%%%%%%%%%%%%%%%%%%%%%%%

\def\subfigurewidth{0.3\linewidth}
\def\imagewidth{1.0\linewidth}

% \vspace{-2mm}
\begin{figure*}[t]
    \centering
    % \input{figures/heatmap.pgf}
    % \input{old_heatmap_files/heatmap_cost.pgf}
    
    % Try this for resize
    %   \resizebox{\linewidth}{!}{%
    %   \input{figures/methodology/structure_tikz}
    %   }

    %% 3 plots in one image
    % \resizebox{0.8\linewidth}{!}{%
    %   \input{heatmap_cost.pgf}
    % }
    % \begin{subfigure}{\subfigurewidth}
    %     \vspace{-10mm}
    %     \caption{Number of Layers}
    % \end{subfigure}
    % ~
    % \begin{subfigure}{\subfigurewidth}
    %     \vspace{-10mm}
    %     \caption{Total Read Volume}
    % \end{subfigure}
    % ~
    % \begin{subfigure}{\subfigurewidth} 
    %     \vspace{-10mm}
    %     \caption{Optimal Cost}
    % \end{subfigure}

    %% 3 plots in 3 images
    \begin{subfigure}{\subfigurewidth}
        \resizebox{\imagewidth}{!}{\input{heatmap_cost_1.pgf}}
        \vspace{-9mm}
        \caption{Number of layers}
        \label{fig:IV_storage_impact_layers}
    \end{subfigure}
    ~
    \begin{subfigure}{\subfigurewidth}
        \resizebox{\imagewidth}{!}{\input{heatmap_cost_2.pgf}}
        \vspace{-9mm}
        \caption{Total Read volume}
        \label{fig:IV_storage_impact_readvolume}
    \end{subfigure}
    ~
    \begin{subfigure}{\subfigurewidth} 
        \resizebox{\imagewidth}{!}{\input{heatmap_cost_3.pgf}}
        \vspace{-9mm}
        \caption{Optimal cost}
        \label{fig:IV_storage_impact_cost}
    \end{subfigure}
    ~
    \begin{subfigure}[t]{0.07\linewidth} 
        \resizebox{\imagewidth}{!}{\input{heatmap_cost_legend.pgf}}
        % \vspace{-8mm}
    \end{subfigure}
    
    \vspace{-3mm}
    \caption{
    Impact of storage latency/bandwidth on \system's index design. The \texttt{fb} dataset is used.
    % \system's index structures tuned to varying storage profile's latency and bandwidth, on the \texttt{fb} dataset. 
    % Bandwidth and latency are sampled from $\{2^{10}, 2^{11}, \dots, 2^{40}\}$ B/s and ns respectively. 
    Note the logarithm scales covering 1KB/s -- 1TB/s bandwidth and 1$\mu$s -- 1000s latency. 
    The number of layers $L$, total read volume $s(\Theta_L) + \sum_{l=1}^L \E_{x \sim \calX} \Delta(x; \Theta_l)$, and the optimal costs are displayed in color annotated in the sidebar. NFS, SSD, and HDD performances are marked accordingly.
    % Due to limited sampling resolutions, we interpolate the pixels according to the nearest neighbor.
    }
    \label{fig:IV_storage_impact}
    \vspace{-2mm}
\end{figure*}

\begin{figure}[t]
    \centering

    \def\figwidth{0.49\linewidth}
    \def\figheight{32mm}

    \input{figures/data/storage_variability}
    
    \pgfplotsset{
        suboptcurves/.style={
            height=\figheight,
            width=\linewidth,
            % ymode=log,
            axis lines=left,
            enlarge x limits=0.1,
            xtick={0.001,0.01,0.1,1,10,100,1000},
            xticklabels={$0.001\times$,,,$1\times$,,,$1000\times$},
            xlabel=Relative Latency,
            xlabel near ticks,
            % xlabel shift=-2mm,
            ylabel=Rel. Slowdown,
            ylabel near ticks,
            % ylabel shift=-2mm,
            ylabel style={align=center},
            legend style={
                at={(0.5,1.05)},anchor=south,column sep=2pt,
                draw=black,fill=white,line width=.5pt,
                font=\scriptsize,
                /tikz/every even column/.append style={column sep=5pt}
            },
            legend columns=2,
            colormap name=bright,
            every axis/.append style={font=\footnotesize},
            % xminorgrids,
            minor x tick num=0,
            xmajorgrids,
            ymajorgrids,
            yminorgrids,
        },
    }
    
    \begin{subfigure}[b]{\linewidth} \centering
    %% DUMMY PICTURE TO ADD LEGEND.
    \begin{tikzpicture}
        \begin{axis}[
                ticks=none,
                width=\linewidth,
                hide axis,
                ymin=0,
                ymax=1,
                xmin=0,
                xmax=1,
                legend style={
                    at={(0.5,0)},anchor=center,
                    column sep=2pt,
                    draw=black,fill=white,line width=.5pt,
                    font=\scriptsize,
                    /tikz/every even column/.append style={column sep=5pt}
                },
                legend image code/.code={%
                    \draw[mark repeat=2,mark phase=2]
                    plot coordinates {
                        (0cm,0cm)
                        (0.2cm,0cm)        %% default is (0.3cm,0cm)
                        (0.4cm,0cm)         %% default is (0.6cm,0cm)
                    };%
                },
                % legend image code/.code={%
                %     \draw[#1, draw=black] (0cm,-0.1cm) rectangle (0.6cm,0.1cm);
                % },
                legend columns=6,
        ]
        \node[align=center, opacity=0] {
            % \addlegendimage{lmdbcolor,mark=x,thick}
            % \addlegendentry{\lmdb};
            % \addlegendimage{rmicolor,mark=square,thick}
            % \addlegendentry{\rmi};
            % \addlegendimage{pgmcolor,mark=o,thick}
            % \addlegendentry{\pgm};
            % \addlegendimage{alexcolor,mark=pentagon,thick}
            % \addlegendentry{\alex};
            % \addlegendimage{dcalcolor,mark=star,thick}
            % \addlegendentry{\dcal};
            % \addlegendimage{btreecolor,mark=triangle,thick}
            % \addlegendentry{\btree};

            \addlegendimage{empty legend}
            \addlegendentry{\textbf{Rel. Bandwidth:}};
            \addlegendimage{vlowcolor,mark=asterisk,thin}
            \addlegendentry{$0.01\times$};
            
            \addlegendimage{lowcolor,mark=x,thick}
            \addlegendentry{$0.1\times$};
            
            \addlegendimage{normalcolor, mark=triangle, thick, mark options=solid}
            \addlegendentry{$1\times$};
            
            \addlegendimage{highcolor, mark=square, thick, mark options=solid}
            \addlegendentry{$10\times$};
            
            \addlegendimage{vhighcolor, mark=o, thick, mark options=solid}
            \addlegendentry{$100\times$};
        };
        \end{axis}
    \end{tikzpicture}
    \end{subfigure}
    \begin{subfigure}[b]{\figwidth}
        \centering
        \begin{tikzpicture}
            \begin{axis}[
                suboptcurves,
                xmode=log,
                ymode=log,
                ymin=1,
                ymax=250,
                ytick={1,10,100},
                yticklabels={$1\times$,$10\times$,$100\times$},
                % minor y tick num=1,
            ]
            
            % \addplot+[mark=x, color={vvlowcolor}, thin, mark repeat=10, mark options=solid]
            % table[x=rellat,y=la] {\nfsvariability};
            
            \addplot+[mark=asterisk, color={vlowcolor}, thin, mark repeat=10, mark options=solid]
            table[x=rellat,y=lb] {\nfsvariability};
            
            \addplot+[mark=x, color={lowcolor}, thick, mark repeat=10, mark options=solid]
            table[x=rellat,y=lc] {\nfsvariability};
            
            \addplot+[mark=triangle, color={normalcolor}, thick, mark repeat=10, mark options=solid]
            table[x=rellat,y=ld] {\nfsvariability};
            
            \addplot+[mark=square, color={highcolor}, thick, mark repeat=10, mark options=solid]
            table[x=rellat,y=le] {\nfsvariability};
            
            \addplot+[mark=o, color={vhighcolor}, thin, mark repeat=10, mark options=solid]
            table[x=rellat,y=lf] {\nfsvariability};
            
            % \addplot+[mark=square, color={vvhighcolor}, thin, mark repeat=10, mark options=solid]
            % table[x=rellat,y=lg] {\nfsvariability};

            \end{axis}
        \end{tikzpicture}
        \vspace{-2mm}
        \caption{Variability on NFS}
        \label{fig:iv_storage_variability_nfs}
    \end{subfigure}
    ~
    \begin{subfigure}[b]{\figwidth}
        \begin{tikzpicture}
            \begin{axis}[
                suboptcurves,
                xmode=log,
                ymode=log,
                ymin=1,
                ymax=15,
                ytick={1,10},
                yticklabels={$1\times$,$10\times$},
                % ymin=1,
                % ymax=8,
                % ytick={1,3,5,7},
                % yticklabels={$1\times$,$3\times$,$5\times$,$7\times$},
                % minor y tick num=1,
            ]
            
            % \addplot+[mark=x, color={vvlowcolor}, thin, mark repeat=10, mark options=solid]
            % table[x=rellat,y=la] {\ssdvariability};
            
            \addplot+[mark=asterisk, color={vlowcolor}, thin, mark repeat=10, mark options=solid]
            table[x=rellat,y=lb] {\ssdvariability};
            
            \addplot+[mark=x, color={lowcolor}, thick, mark repeat=10, mark options=solid]
            table[x=rellat,y=lc] {\ssdvariability};
            
            \addplot+[mark=triangle, color={normalcolor}, thick, mark repeat=10, mark options=solid]
            table[x=rellat,y=ld] {\ssdvariability};
            
            \addplot+[mark=square, color={highcolor}, thick, mark repeat=10, mark options=solid]
            table[x=rellat,y=le] {\ssdvariability};
            
            \addplot+[mark=o, color={vhighcolor}, thin, mark repeat=10, mark options=solid]
            table[x=rellat,y=lf] {\ssdvariability};
            
            % \addplot+[mark=square, color={vvhighcolor}, thin, mark repeat=10, mark options=solid]
            % table[x=rellat,y=lg] {\ssdvariability};

            \end{axis}
        \end{tikzpicture}
        \vspace{-2mm}
        \caption{Variability on SSD}
        \label{fig:iv_storage_variability_ssd}
    \end{subfigure}
    \vspace{-3mm}
    \caption{Extreme errors ($\pm 3$ and $\pm 2$ magnitude differences in latencies/bandwidths) expectedly make tuned indexes suboptimal compared to correctly tuned ones. Left: the index is tuned for NFS (50 ms, 12 MB/s). Right: the index is tuned for SSD (250 $\mu$s, 175 MB/s). \texttt{fb} is the underlying dataset.}
    \label{fig:iv_storage_variability}
    \vspace{-2mm}
\end{figure}

\subsection{Layer-wise Optimization Helps}
\label{sec:exp-tune-accuracy}

\paragraph{Speedup over Hierarchical Indexes} 
To empirically verify that \system tunes accurately and finds a fast index, 
    we compare \system's tuned index designs against manually configured ones
        in terms of their first-query latencies.
    % we vary index configurations to build fixed index structures. We then benchmark and compare their first-query latencies with those from \system's tuned index structures.
\cref{fig:III_variants} presents the comparison 
    % results 
on numbers of layers $L$ and granularity hyperparameter $\lambda$ across the two storage within the same dataset \texttt{fb}. 
In all settings and variable dimensions, 
    \system consistently finds the fastest index designs.
Inspecting the trends, we observe that $\lambda$ forgivingly admits a larger optimal region, 
    even in the logarithmic scale, than the number of layers $L$. 
This allows \system to select a coarse granularity exponentiation base $1 + \eps$ without risking suboptimality.

We have also experimented with other dimensions such as the granularity exponentiation base $1 + \eps$ and the set of node estimators to discover any trade-off. Lower bases $1 + \eps$ result in faster indexes but with only insignificant gain for a higher cost in a longer tuning time. 
A wider set of node estimators and types provides some fitness advantages.
% , leading to smaller index layers and smaller latency in general. 
The impact is clear when the data pattern is exclusive to a node type, for example, \texttt{band} nodes fit perfectly on a uniformly random key set (\texttt{uden64} from \cite{sosd-neurips}) while \texttt{step} nodes do not.

%%%%%%%%%%%%%%%%%%%%%%%%%%%%%%%%%%%%%%%%%%%%%%%%%%%%%%%%%%%%%
%%%%%%%%%%%%%%%%%%%%%%%%%%%%%%%%%%%%%%%%%%%%%%%%%%%%%%%%%%%%%

\paragraph{Speedup over Well-tuned Baselines}
\label{sec:exp-well-tuned}

\system is faster than all baseline configurations, and so is faster than optimal baselines. \cref{fig:VIIII_tuned} varies all permissible page sizes of \lmdb,
% (64KiB is possible but requires enabling development mode which adds non-trivial latency), 
all 10 \rmi settings recommended by its optimizer, and 8 chosen $\varepsilon$ in \plex to cover the optimal valley. 
% As a side note, this \btree's valley resembles a B-tree latency estimate: $\lceil \log_{\lambda / s}(n) \rceil (\ell + \lambda / B)$ for datasize $n = 8 \times 10^{8}$, key-pointer size $s = 16$ bytes, latency $\ell = 50$ms, and bandwidth $B = 12$MB/s. Anyhow, 
We observe that \system is $2.7\times$, $1.5\times$, and $1.7\times$ faster that the optimal \lmdb, \rmi, and \plex, respectively. Our similar experiment with \btree by varying $\lambda$ granularity shows $1.3\times$ speed up. These gaps from optimal baselines suggest the benefit to consider a larger class of indexes (i.e. \ourmodel).
% Besides, while we now know the optimal settings for these baselines on \texttt{books} dataset under NFS, \system is able to find a fast index without the expensive benchmarking.

%%%%%%%%%%%%%%%%%%%%%%%%%%%%%%%%%%%%%%%%%%%%%%%%%%%%%%%%%%%%%
%%%%%%%%%%%%%%%%%%%%%%%%%%%%%%%%%%%%%%%%%%%%%%%%%%%%%%%%%%%%%

\subsection{Adaptive to I/O Performance}
\label{sec:exp-storage-impact}

% e2e latency of affine -- read/layer, if deviate from accurate --> doesn't regress too much
% graceful way to handle error

\paragraph{On Testing Storages} Indeed, \system discovers different optimal index structures for the NFS and SSD/HDD storages in previous experiments. NFS indexes have $L=1$ index layer with only \texttt{band} node types, while SSD/HDD indexes have $L=2$ layers with a mix of \texttt{band}-\texttt{band}, \texttt{band}-\texttt{step}, and \texttt{step}-\texttt{band} node types. The sizes of root layers $s(\Theta_L)$ and precisions $\Delta$ range from 36KB to 328KB in NFS and 864B to 16KB in SSD/HDD, depending on dataset size and complexity. Among 5 datasets, \texttt{osm} is the most challenging one, reflecting the same observation from \cite{DBLP:journals/pvldb/MarcusKRSMK0K20BenchmarkLearned}. Please see our extended manuscript~\cite{airindex_technical_report} for specific index structures.

\paragraph{On Latency-Bandwidth Spectrum} If we have a storage with latency $\ell$ and bandwidth $B$, what would the fastest index look like? We answer this question as a whole, on a wide spectrum of latency $\ell \in [1 \mu\text{s}, 1000 \text{s}]$ and bandwidth $B \in [1 \text{KB/s}, 1 \text{TB/s}]$. \cref{fig:IV_storage_impact} shows \system adapting its index to the diverse range of storage profiles.  
    Higher bandwidth or latency promotes shallower indexes with coarser precision (larger total read volume). 
    In the extreme, \system proposes no index at all, i.e. fetching the entire data layer to search locally. 
    On the other hand, lower bandwidth or latency promotes taller indexes with finer precision.
    Although this trend is similar to a well-known tuning rule of thumb for B-tree, \system offers a more complete tuning on a much larger class of indexes, for any data pattern, and storage profile.

% We would like to note that \system have reached an implementation limit towards lower latency and bandwidth (at the bottom-left corner) where \system could have but cannot select taller indexes with finer precision due to the smallest serialized node size (40 bytes). To travel beyond, \system would need to either decrease the effective node size (e.g., by means of compression) or consider implicit hierarchical index that allows multiple storage accesses per layer.

% \ignore{  % in favor of space...
%     The right plot from \cref{fig:IV_storage_impact} confirms a simple fact that storage with lower latency and higher bandwidth supports a faster search (top left corner) and vice versa. Beyond the top left corner whose storage model cost is lower, future index tuners should consider the internal computation in their problem formulation. More surprisingly, the cost landscape is smooth. Despite the sharp index structure frontiers mentioned earlier, the optimal index selected by \system transitions between structures as the storage profile moves across a frontier. 
%     In a fixed index structure, if we were to trace contour lines between two areas (e.g., green 10ms and yellow 100ms), 
%         we would expect to see L-shape lines similar to those in $z = \max \{ x, -y \}$. 
%     However, the optimized costs bend the contour lines down, indicating the benefits of our principled index tuning.
% }

\paragraph{With Storage Variability} Storage performance may vary. If it varies within a magnitude, \system's tuned index mostly stays optimal. In fact, it does so if the performance remains within the same band (i.e. same color in \cref{fig:IV_storage_impact_layers} and \cref{fig:IV_storage_impact_readvolume}) as the profiled performance. However, if the actual performance $T'$ varies across many magnitudes, the index tuned with inaccurate storage profile $T$ can be suboptimal as shown in \cref{fig:iv_storage_variability} through the relative slowdown of the index $\bfTheta$ tuned with the inaccurate profile $T$ based on the index $\bfTheta'$ tuned with the actual profile $T'$ in retrospect: $\calL_{SM}(\calX; \bfTheta, T) / \calL_{SM}(\calX; \bfTheta', T')$. For example, if the actual NFS latency is 0.001 times smaller than the profiled NFS latency (i.e., 50$\mu$s instead of 5ms), the originally tuned index would be 35 times slower than the accurately tuned index.

% As a reminder, these observations are sound under affine storage profiles where we can distill the performance of an index into two dimensions: number of layers and total read volume.

%%%%%%%%%%%%%%%%%%%%%%%%%%%%%%%%%%%%%%%%%%%%%%%%%%%%%%%%%%%%%
%%%%%%%%%%%%%%%%%%%%%%%%%%%%%%%%%%%%%%%%%%%%%%%%%%%%%%%%%%%%%

\subsection{Competitive Build Time}
\label{sec:exp-build-scale}

\begin{figure*}[t]
\pgfplotstableread{
    data_size	Airindex RMI	PGM	LMDB BTree ALEX dcal PLEX
    % 200M 15 15 13 63 5 51 34 21
    % 400M 29 28 27 105 11 100 66 41
    % 600M 41 42 40 165 15 153 109 63
    % 800M 56 55 57 212 21 242 130 81
    200M 15 86 13 63 5 51 34 21
    400M 29 169 27 105 11 100 66 41
    600M 41 255 40 165 15 153 109 63
    800M 56 328 57 212 21 242 130 81
}\scaleData
% 1.0 RMI 5.9 LMDB 3.8 ALEX 4.3 Dcal 2.3

\pgfplotstableread{
    data_size	Airindex RMI	dcal
    % 200M 15 71 30
    % 400M 29 141 59
    % 600M 41 213 99
    % 800M 56 273 117
    200M 10 71 30
    400M 18 141 59
    600M 26 213 99
    800M 35 273 117
}\scaleDataSearch
% 7.8 3.3

% 10 17 27 33
% 17 --> 22 --> 31 --> 44 --> 62
% 13 --> 18 --> 29 --> 44 --> 65
% 5, 11, 15, 21
% 33%, 38%, 36%, 37%

\pgfplotsset{
    timebars/.style={
        width=\linewidth,
        height=33mm,
        axis lines=left,
        xlabel={Data Size (Keys)},
        xlabel shift=0mm,
        enlarge x limits=0.2,
        symbolic x coords={200M, 400M, 600M, 800M},
        xtick = data,
        xticklabels={200M, 400M, 600M, 800M},
        xticklabel style={inner sep=0pt, below},
        ybar=1pt,
        bar width=5pt,
        area legend,
        legend image code/.code={%
            \draw[#1, draw=none] (0cm,-0.1cm) rectangle (0.6cm,0.1cm);
        },
        every axis/.append style={font=\footnotesize},
        minor grid style=lightgray,
        minor x tick num=0,
        minor y tick num=1,
        ymajorgrids,
        yminorgrids,
        legend cell align={left},
        legend columns=4,
        transpose legend,
        legend style={
            nodes={scale=0.63, transform shape},
            at={(0.02,1.0)},anchor=north west,
            column sep=2pt,draw=black,fill=white,line width=.5pt,
            font=\footnotesize,
            /tikz/every even column/.append style={column sep=5pt},
        }
    },
}

\begin{subfigure}[t]{0.55\linewidth} \centering
    \begin{tikzpicture}
        \begin{axis}[
            timebars,
            ymin=0,
            ymax=450,
            ylabel={Build Time (s)},
            ytick={0,100,...,400}
        ]

        \addplot[fill={lmdbcolor}]
        plot [error bars/.cd, y dir=plus,y explicit,error bar style={color={black}}]
        table[x=data_size,y=LMDB] {\scaleData};
        
        \addplot[fill={rmicolor}]
        plot [error bars/.cd, y dir=plus,y explicit,error bar style={color=black}]
        table[x=data_size,y=RMI] {\scaleData};
        
        \addplot[fill={pgmcolor}]
        plot [error bars/.cd, y dir=plus,y explicit,error bar style={color=black}]
        table[x=data_size,y=PGM] {\scaleData};
        
        % \addplot[draw=none] coordinates {(1,1)};
        
        \addplot[fill={alexcolor}]
        plot [error bars/.cd, y dir=plus,y explicit,error bar style={color={black}}]
        table[x=data_size,y=ALEX] {\scaleData};
        
        \addplot[fill={plexcolor}]
        plot [error bars/.cd, y dir=plus,y explicit,error bar style={color={black}}]
        table[x=data_size,y=PLEX] {\scaleData};
        
        \addplot[fill={dcalcolor}]
        plot [error bars/.cd, y dir=plus,y explicit,error bar style={color={black}}]
        table[x=data_size,y=dcal] {\scaleData};
        
        \addplot[fill={btreecolor}]
        plot [error bars/.cd, y dir=plus,y explicit,error bar style={color={black}}]
        table[x=data_size,y=BTree] {\scaleData};
        
        \addplot[fill={airindexcolor}]
        plot [error bars/.cd, y dir=plus,y explicit,error bar style={color={black}}]
        table[x=data_size,y=Airindex] {\scaleData};
        
        % \addplot[fill={airindexcolor}]
        % plot [error bars/.cd, y dir=plus,y explicit,error bar style={color=black}]
        % table[x=data_size,y=airindex] {\localData};
        
        \legend{\lmdb, \rmi/\cdfshop, \pgm, \alex, \plex, \dcal, \btree, \system}
            
        \end{axis}
    \end{tikzpicture}
    \vspace{-2mm}
    \caption{Total Time to Build Index}
    \label{fig:V_build_comparison_build}
\end{subfigure}
~
\begin{subfigure}[t]{0.34\linewidth} \centering
    \begin{tikzpicture}
        \begin{axis}[
            timebars,
            ymin=0,
            ymax=325,
            ylabel={Search Overhead (s)},
            ytick={0,50,100,...,300}
        ]
        
        \addplot[fill={rmicolor}]
        plot [error bars/.cd, y dir=plus,y explicit,error bar style={color=black}]
        table[x=data_size,y=RMI] {\scaleDataSearch};
        
        \addplot[fill={dcalcolor}]
        plot [error bars/.cd, y dir=plus,y explicit,error bar style={color={black}}]
        table[x=data_size,y=dcal] {\scaleDataSearch};
        
        \addplot[fill={airindexcolor}]
        plot [error bars/.cd, y dir=plus,y explicit,error bar style={color={black}}]
        table[x=data_size,y=Airindex] {\scaleDataSearch};
        
        \legend{\rmi/\cdfshop, \dcal, \system}
        \end{axis}
    \end{tikzpicture}
    \vspace{-2mm}
    \caption{Search Overhead Time}
    \label{fig:V_build_comparison_search}
\end{subfigure}

\vspace{-4mm}
\caption{Index build and search overhead for different data sizes (200, 400, 600, and 800 million keys) from the \texttt{gmm} dataset. 
    Build times for \lmdb, \rmi, \pgm, \alex, and \plex do \textit{not} include their manual tuning time.}
\label{fig:V_build_comparison}
\vspace{-4mm}
\end{figure*}

\paragraph{Total Build Time} \cref{fig:V_build_comparison_build} measures index build times on a machine with 2 AMD EPYC 7552 48-Core Processors (192 CPUs in total). Build times in \lmdb, \rmi, \pgm, and \alex only account for data loading, inserting into the system, and writing to files,
    excluding their manual hyperparameter tuning. 
    \dcal's build time includes parallelized autocompletion and index building. 
    All methods use all available cores.
    % Because our \alex integration to external storage may increase its build time, we decide to 
    % For \alex, we measure its build time based on its implementation to SOSD~\footnote{\url{https://github.com/learnedsystems/SOSD/blob/master/competitors/alex.h}} on \texttt{books} dataset of equal sizes.

    Thanks to its parallel tuning (\cref{sec:system-building-parallelism}), \system's total build time is competitive with other baselines. \system is $3.8\times$, $5.9\times$, and $4.3\times$ faster than \lmdb, \rmi/\cdfshop, and \alex. while \system both tunes and builds as fast as \rmi (excluding \cdfshop time), \pgm, and \plex build their indexes. Compared to \dcal's autocompletion and building, \system is $2.3\times$ faster despite exploring a larger class of indexes. Nonetheless, \system tuning and building are $2.7\times$ slower compared to its fixed structure counterpart \btree. If needed to be faster, \system can relax some hyperparameters (e.g. number of candidates, granularity base and bounds) to trade its speed with its tuning accuracy.
% In a bigger picture, \system focuses on the best trade-off between tuning accuracy and time complexity; future works should holistically consider other costs such as CPU time and memory requirement.

\paragraph{Search Overhead} \cref{fig:V_build_comparison_search} measures search overheads---the differences between total build time and build time given a known configuration (i.e., \cdfshop's search procedure on \rmi structures, \dcal's autocompletion, \system's \oursearch excluding the time for building the optimal index). \system incurs non-negligible search overhead (around 50 ns/key on the 192-core machine, or 9.6 $\mu$s/key on single-core machines); however, this search overhead is lower compared to other methods. Because of its parallelization and top-$k$ candidate selection limiting branching, \system finds its index structure $7.8\times$ and $3.3\times$ faster than learned index tuning method \cdfshop and traditional index tuning method \dcal, respectively. In contrast, \dcal's parallelized autocompletion is slow because it tries all design combinations similarly to a grid search. We note that \cdfshop outputs multiple index structures on a Pareto front.

% \fixed{
%     \paragraph{Search Overhead Analysis} \system incurs CPU and real time overheads, but these overheads can be acceptable depending on the allowance, even when we ignore the benefit of index tuning. To estimate the search overhead, we subtract \system's total build time by \system's build time given a fixed index configuration. The former suggests \system's pace at 75 ns/key while measurements for the later are similar to \btree's total build time in \cref{fig:V_build_comparison_build}, having a pace at 25 ns/key. Therefore, \system's search overhead is 50 ns/key on our machine (192 CPUs) or 9.6 $\mu$s/key on single-core machines; in other words, its search throughput is 20M key/s on our machine or 10.4K key/s on single-core machines. This can be tolerable for some applications. For example, if a database on a 192-core machine tolerates 5\% time overhead, it can use \system to support write throughputs up to 1M key/s. On a single-core machine, this database with the same allowance can support up to 5.2K key/s workloads.
%     For petabyte-scale and larger key-position collections that do not fit in a single machine, \system's index building throughput may decrease due to the need for taller indexes and the costs to distribute across machines. Evaluations on large-scale index building are future directions.
% }

%%%%%%%%%%%%%%%%%%%%%%%%%%%%%%%%%%%%%%%%%%%%%%%%%%%%%%%%%%%%%
%%%%%%%%%%%%%%%%%%%%%%%%%%%%%%%%%%%%%%%%%%%%%%%%%%%%%%%%%%%%%

\subsection{Applicable to Read-Write Workloads}
\label{sec:exp-readwrite}

\begin{figure}[t]
    \pgfplotstableread{
        name	workload	lmdb	lmdbstd	alex	alexstd	airindex	airindexstd
        ronly	1	2341.876336	64.5996254	2345.586414	102.1663167	2708.289906	122.5866845
        rw	2	2302.712841	82.77363235	2370.495834	88.26915968	2682.374414	110.0693488
        wheavy	3	2280.016155	107.7179671	2277.480629	94.55901468	2817.449962	132.0361394
        wonly	4	2276.849883	87.91999881	2201.471693	85.11207397	2950.154365	137.9558966
    }\rwData
    
    \begin{subfigure}[t]{1.0\linewidth} \centering
        \begin{tikzpicture}
            \begin{axis}[
                width=0.85\linewidth,
                height=31mm,
                axis lines=left,
                ymin=0,
                ymax=3500,
                ylabel={Tput. (op/s)},
                ytick={0,1000,...,3000},
                enlarge x limits=0.15,
                xtick={1, 2, 3, 4},
                xticklabels={\texttt{Read-Only}, \texttt{Read-Write}, \texttt{Write-Heavy}, \texttt{Write-Only}},
                ybar=1pt,
                bar width=4pt,
                area legend,
                legend image code/.code={%
                    \draw[#1, draw=none] (0cm,-0.1cm) rectangle (0.6cm,0.1cm);
                },
                every axis/.append style={font=\footnotesize},
                minor tick num=1,
                minor grid style=lightgray,
                minor x tick num=0,
                ymajorgrids,
                yminorgrids,
                legend cell align={left},
                legend columns=1,
                % transpose legend,
                legend style={
                    nodes={scale=0.63, transform shape},
                    at={(0.95, 1.0)},anchor=north west,
                    column sep=2pt,draw=black,fill=white,line width=.5pt,
                    font=\footnotesize,
                    /tikz/every even column/.append style={column sep=5pt},
                }
            ]
            
            \addplot[fill={lmdbcolor}]
            plot [error bars/.cd, y dir=both,y explicit,error bar style={color={black}}]
            table[x=workload,y=lmdb,y error=lmdbstd] {\rwData};

            \addplot[fill={alexcolor}]
            plot [error bars/.cd, y dir=both,y explicit,error bar style={color={black}}]
            table[x=workload,y=alex,y error=alexstd] {\rwData};
            
            \addplot[fill={airindexcolor}]
            plot [error bars/.cd, y dir=both,y explicit,error bar style={color={black}}]
            table[x=workload,y=airindex,y error=airindexstd] {\rwData};
            
            \legend{\lmdb, \alex, \system}
                
            \end{axis}
        \end{tikzpicture}
    \end{subfigure}
    
    \vspace{-4mm}
    \caption{Average throughputs of \lmdb, \alex, and \system across read-write workloads on \texttt{osm} dataset in SSD.}
    \label{fig:X_readwrite_workload}
    \vspace{-4mm}
\end{figure}

We implement a proof-of-concept updatable \system based on the gapped array~\cite{DingMYWDLZCGKLK2020ALEX} that allocates empty gaps on data layer for \system to insert a key-value at any available gap within the predicted position $\yh(x)$.
    Our read-write benchmark follows that of \cite{DingMYWDLZCGKLK2020ALEX}. It initially inserts 100M keys sampled from \texttt{osm} and measures the time to complete 10K queries consisting of cycles of $r$ read and $w$ write operations. Four workloads vary the read/write proportion: (1) \texttt{Read-Only}: $(r, w) = (1, 0)$, (2) \texttt{Read-Write}: $r = 19, w = 1$, (3) \texttt{Write-Heavy}: $r = 1, w = 1$, (4) \texttt{Write-Only}: $r = 0, w = 1$. The benchmark samples read keys uniformly from the inserted key set and samples write keys from the non-inserted key set.

    In \cref{fig:X_readwrite_workload}, our prototype remains the fastest compared to updatable baselines (\lmdb and \alex) across all workloads, confirming that tuning for lookup speed is relevant to both read and write performances. Apart from the direct relation to read operations, lookup speed is relevant to write operations because all methods need to first look up the insertion position from their indexes before writing the target key-value pair.

%%%%%%%%%%%%%%%%%%%%%%%%%%%%%%%%%%%%%%%%%%%%%%%%%%%%%%%%%%%%%
%%%%%%%%%%%%%%%%%%%%%%%%%%%%%%%%%%%%%%%%%%%%%%%%%%%%%%%%%%%%%

% \subsection{Top-$k$ Candidate Parameter Sweep}
% \label{sec:exp-topk}

% Across all experiments, we set $k = 5$ as an arbitrary constant greater than one and less than the number of node builders ($|\calF| = 45$). In this experiment, we vary this hyperparameter $k$ to verify our understanding: as $k$ increases, build time should increase in $L$-degree polynomial ($L = 2$ in this setting) while the optimized cost should monotonically decrease. \cref{fig:VII_topk} reaffirms this hypothesis, but also shows that the available parallelism (192 CPUs) is able to hide the polynomial build time more than we had expected (taking around 50 seconds up until $k = 20$), implying that we could have selected a higher $k$ to get a faster index at no additional build time cost.

% \input{figures/experiments/V_build_comparison}

% \input{figures/experiments/V_build_scalability}

%%%%%%%%%%%%%%%%%%%%%%%%%%%%%%%%%%%%%%%%%%%%%%%%%%%%%%%%%%%%%
%%%%%%%%%%%%%%%%%%%%%%%%%%%%%%%%%%%%%%%%%%%%%%%%%%%%%%%%%%%%%

% \clearpage  % REMOVE
\section{Related Work}
\label{sec:related}

Our work is built on top of the vast amount of existing research on index design and optimization as summarized below;
however, our unified model (\ourmodel) and efficient search (\oursearch)
    enables
    a high-quality data and I/O-aware hierarchical index.

\paragraph{Storage-aware Indexes} 
Besides the original B-trees~\cite{Bayer1972,Comer1979},
    many works have studied unique storage properties to design indexes specifically optimized for certain storage,
such as CPU cache~\cite{Graefe2001}, SRAM cache~\cite{Chen2001}, disk~\cite{Chen2002}, NVMe~\cite{Wang2020}, and distributed cloud~\cite{Wu2010, Zhou2014, Bin2014}.
% For example, B-Tree~\cite{Bayer1972,Comer1979} has always been motivated and optimized across memory hierarchies:  
As the most generic of all, \cite{bender2000CacheObliviousBtree} designs B-Trees that perform well for any I/O page size $P$; however, its storage profile (cache-oblivious model) only limits to $T(\Delta) = O(\lfloor \Delta / P \rfloor)$.
Also,
% On the other hand, 
skip list
% ---an index structure based on a linked list with skip connections---
has been adapted to various settings, such as multi-core~\cite{Dick2017}, cache-sensitive for range queries~\cite{Sprenger2017}, non-uniform access~\cite{Daly2018}, and distributed nodes~\cite{He2018}.
% Perhaps the most common feature is the realization of a natural data transfer size (e.g., disk page size, network packet size) and specialization of their index parameters to the size. 
In contrast, \system takes a general approach by 
    composing an optimization problem in consideration of storage profiles,
    % encoding data transfer speed into 
    % the storage profile which composes a general optimization problem. 
% In this way, \system can automatically 
    which makes it possible to adapt its structure 
without re-evaluating the parameters when the transfer size changes.
% Many of them leverage storage-specific accelerations such as hardware prefetching and asynchronous I/O instructions. While some would affect \system's cost formulation, some can be applied behind \system's storage interface to improve the storage profile further.

\paragraph{Index Tuning} 

\system automates index designs
    by improving on long-standing heuristics
such as ``use larger pages for larger bandwidth''~\cite{gray1997FiveMinuteRules, AppuswamyGBA2019FiveMin30Years, lomet98BtreePageWithCaching}.
% Common B-Tree tuning practices resonate with analogous findings from \system, for example, larger pages (coarser precision in \system) are recommended when disk bandwidth is larger, seek time is longer, or when the cache is cold.
% \cite{gray1997FiveMinuteRules} makes a recommended range of B-Tree page sizes based on the storage trend at the time (recently re-evaluated by \cite{AppuswamyGBA2019FiveMin30Years}). Shortly after, \cite{lomet98BtreePageWithCaching} extends the study to include partially caching B-Trees. However, \system considers a wider class of indexes, motivated by recent learned index works, with more parameters beyond that of B-Trees.
%
Rather than deciding \textit{what} index to build, other index tuning techniques determine \textit{when and where} to build index, as a well-known index selection problem (ISP)~\cite{lum1971attributeindicator,stonebraker1974partialcombined,schkolnick1975optimalindex,frankON1992adaptiveISP,chaudhuri1998AutoAdmin,JimenezSTP2012Kaizen,DashPA2011CoPhy},
    which are orthogonal to our work.

% Starting from the 1970s, \cite{lum1971attributeindicator,stonebraker1974partialcombined,schkolnick1975optimalindex} formulates and solves ISPs based on many factors such as storage performance, workload, and data characteristics given sufficient statistics under some circumstances.
% \cite{frankON1992adaptiveISP,chaudhuri1998AutoAdmin,JimenezSTP2012Kaizen,DashPA2011CoPhy} and many more of later literature emphasize the pragmatic aspect of index tuning to aid system administrators. They propose and build automatic, semi-automatic, or interactive index tuning tools that minimize administrators' effort while providing high-quality recommendations.
% \system shares the same goal with these later works but tackles an orthogonal index tuning question: what is the best index structure if we desire to build one?
% As learned indexes continue to expand the possibilities of index structures, this question is becoming more challenging and influential to overall performance.

% Bearing a resemblance to one another,
\dcal~\cite{DBLP:conf/sigmod/IdreosZHKG2018DataCalc,DBLP:journals/corr/IdreosZHKG2018DataCalcInternal} 
helps designing efficient data structures
by evaluate the cost of a structure in a what-if fashion
in relation to workload and hardware.
% and \system share the same interest: to design efficient data structures. 
% Both systems recognize common challenges in massive design space and dependency on workload and hardware. 
% While \dcal aims to list \emph{all data structures} and evaluate the cost of a structure for interactive what-if design, \system aims to efficiently tune and build \emph{learned indexes}. In other words, \dcal is more comparable to our hierarchical indexes and cost model in \cref{sec:system-overall}. 
However, its auto-completion search---recursively trying all possible designs
($|\calE| = 10^{16}$ discretized designs)---
% Although \dcal proposes an auto-completion procedure, the algorithm recursively tries all possible data layout designs ($|\calE| = 10^{16}$ discretized designs),
scales poorly, which worsens 
    % and even more so 
if we extend \dcal's periodic table 
    and cost synthesis flowchart with learned indexes.
\system solves this challenge
    for index design.

\vspace{1mm}
\paragraph{Learned Indexes}

% A few works propose techniques to automatically tune learned indexes but only under the main memory context.
Previous works largely focus on tuning in-memory indexes.
\cite{DBLP:conf/sigmod/MarcusZK20CDFShop} demonstrates an interactive model tool that allows users to modify RMI configuration (per-layer node type and branching factor) and observe the resulting model fitness. Although the tool provides automatic tuning, it measures the lookup latency by benchmarking each configuration, which can be expensive on a larger scale.
\cite{DBLP:journals/corr/Stoian2021PLEX} formulates a lookup cost function and tunes a single maximum error hyperparameter to build a combination of RadixSpline~\cite{DBLP:conf/sigmod/KipfMRSKK020RadixSpline} and a compact histogram tree. In contrast, \system encompasses a larger index design space, and more importantly, targets a different cost setting where I/O cost is dominant.

Many manually tuned learned indexes have shown success stories in the context of external storage.
\cite{DaiXGAKAA2020BourbonLSM} studies favorable conditions to learn data patterns and integrates learned indexes as an optimization into an LSM-based storage system on disk.
For NVM storages, \cite{ChenC2021ALEXonNVMe} adapts 
% a mutable learned index 
ALEX~\cite{DingMYWDLZCGKLK2020ALEX} to cooperate with the preferred access pattern.
Instead of predicting locations of written records, \cite{abulibdeh2020LearnedBigTable} uses learned indexes to distribute data into blocks in BigTable~\cite{DBLP:conf/osdi/ChangDGHWBCFG06BigTable}.
Similarly, \cite{DBLP:conf/sigmod/Li0ZY020LISA} deploys learned indexes to organize on-disk spatial data into shards and pages, reducing I/O costs over other spatial trees.
With the learning paradigm at the core, \cite{DBLP:conf/cidr/KraskaABCKLMMN19SageDB} re-designs a database system that supports data persistence on disk but is only evaluated in the in-memory mode without disk accesses.
Many of these works have already involved parameter tuning but as a manual tuning step for each setting.
While we share common tuning principles, \system can be seen as a fine-grained automatic optimization of learned index structures.

\vspace{-0.5mm}
\section{Conclusion}
\label{sec:con}

This work presents a novel index-building technique, \system,
    that can build high-speed hierarchical indexes
    by learning from
    both data and I/O characteristics---the first of its kind.
% Our current technique focuses on
    % read-only indexes stored in secondary storage.
% \system tunes hierarchical indexes to both data patterns and storage performance profile.
To achieve its goal, \system formulates an optimization problem 
    consisting of a large hierarchical index search space  and a lookup latency objective function (\ourmodel).
To overcome the computational challenges rising from the inter-dependency between index layers and exponentially many candidate designs, 
    \system explores the search space using
        a purpose-built graph-search method (\oursearch). 
Our experiments verify that \system accurately finds the optimal configuration and 
    provides performance gains over
        conventional indexes as well as 
    state-of-the-art learned indexes.
In many applications,
    the decisions on data placements---local disks, cloud storage, network file system---are relatively fixed,
which makes \system's data-and-I/O-aware optimization
    appealing to achieve significantly faster lookup
compared to the ones not specifically optimized.
% In addition, our studies assert the relevance of cold-state index optimization, confirm the scalability of parallelizable tuning, and reveal tuned indexes over the storage profile landscape for a broader understanding of the search problem.

% \begin{acks}
% This work is supported in part by Microsoft Azure.
% %  This work was supported by the [...] Research Fund of [...] (Number [...]). Additional funding was provided by [...] and [...]. We also thank [...] for contributing [...].
% \end{acks}

\begin{acks}
This work is supported in part by Microsoft Azure.
\end{acks}

\clearpage

\bibliographystyle{ACM-Reference-Format}
\bibliography{bib/airindex}

\appendix
\section{\system in Detail}

\subsection{Layer Builders}

A layer builder turns the target key-position collection $D$ into an index layer $\Theta$. In other words, it is a function $F(D) = \Theta$ where $D = \left\{ (x_i, y_i) \right\}_{i=1}^n$ is a collection of keys $x_i$ together with their positions $y_i$, and $\Theta = \{ (z_j, \theta_j) \}_{j=1}^{n^{+}}$ is the index layer containing the per-node parameters $\theta_j$ with per-node key range $[z_j, z_{j+1})$. 
To build two types of nodes (\texttt{step} and \texttt{band}, sketched in \cref{fig:node_types}), 
    \system currently deploys three types of layer builders below.
\begin{enumerate}[leftmargin=5mm,]
    \item \textbf{Greedy Step} (\texttt{GStep}$(p, \lambda_{GS})$) builds a $p$-piece step function with precision at most $\lambda_{GS}$ bytes. It iterates over each key-position pair $(x_i, y_i)$ and greedily determines whether to create the next constant function if $y^{+}_i - b_k > \lambda_{GS}$ where $(a_k, b_k)$ represents the current constant function. If so, the next constant function has a partition key $a_{k+1} = x_i$ and partition position $b_{k+1} = y_i$. Once it reaches $p$ pieces of constant functions, \texttt{GStep}$(p, \lambda_{GS})$ generates a $p$-piece \texttt{step} node.
    
    \item \textbf{Greedy Band} (\texttt{GBand}$(\lambda_{GB})$) builds linear band nodes by greedily fitting as many key-position pairs as possible, using a method called monotone chain convex hull~\cite{DBLP:journals/ipl/Andrew79ConvexHull}, until we have $\Delta(x_i) > \lambda_{GB}$. Then, it generates a \texttt{band} node with $(x_1, y_1, x_2, y_2, \delta)$. 
    % Rebuilding the linear function for each new key-position would be inefficient; instead, \texttt{GBand}$(\lambda_{GB})$ maintains a convex hull of key-position pairs included so far and queries the convex hull whether it can feasibly fit in a linear band function with a width $\delta \leq \lambda_{GB}$. 
    % This procedure is also known as the monotone chain convex hull~\cite{DBLP:journals/ipl/Andrew79ConvexHull}. 
    % In the total processing of 
    For $n$ key-position pairs, the convex hull inserts a key-position pair in $O(n)$ time ($O(1)$ amortized insertion) and answers a feasibility query in $O(n \log m)$ time ($O(\log m)$ amortized query) where $m$ is the average number of key-position pairs included in one linear band. Typically, $m$ is small when $\lambda_{GB}$ is small.
    
    \item \textbf{Equal Band} (\texttt{EBand}$(\lambda_{EB})$) builds linear band nodes by grouping key-position pairs in equal-size position ranges. That is, each group $\{(x_l, y_l)$, $(x_{l+1}, y_{l+1})$, $\dots,$ $(x_r, y_r)\}$ has a bounded position range $|y^{-}_l - y^{+}_r| \leq \lambda_{EB}$. It then fits a linear band function to each group and creates a \texttt{band} node. Note that the precision $\Delta(x_i)$ can vary depending on how ``linear'' the group is. \texttt{EBand}$(\lambda_{EB})$ groups by position ranges rather than key ranges so that the worst-case precision is controlled by $\lambda_{EB}$.
\end{enumerate}

\noindent
First, \texttt{GStep}$(p, \lambda_{GS})$ is equivalent to bulk indexing in a sparse B-tree with a fanout $p$ and page size $\lambda_{GS}$ bytes.
Second, \texttt{GBand}$(\lambda_{GB})$ is a generalization from step functions to linear functions with precision $\Delta(x_i) \leq \lambda_{GB}$.
Third, \texttt{EBand}$(\lambda_{EB})$ is another generalization focusing on the key-position group size $|y^{-}_l - y^{+}_r| \leq \lambda_{EB}$.

\paragraph{Granularity Exponentiation} $\lambda_{GS}$, $\lambda_{GB}$, and $\lambda_{EB}$ are called \textit{granularity},  which roughly control layer builders' tendency to split key-position pairs apart. Because they correlate with the resulting node's precision $\Delta(x; \Theta)$, \system needs to determine the appropriate granularity for each node type.

\system creates many candidate layer builders $\calF$ by sampling the granularity on an exponential grid: $\lambda_{low}, \lambda_{low} (1 + \epsilon), \lambda_{low} (1 + \epsilon)^2, \dots \lambda_{high}$ where $(\lambda_{low}, \lambda_{high})$ are the bounds and $\epsilon > 0$ controls the exponentiation base. Smaller $\epsilon$ implies a finer search which improves optimization accuracy but increases tuning time.

%%%%%%%%%%%%%%%%%%%%%%%%%%%%%%%%%%%%%%%%%%%%%%%%%%%%%%
%%%%%%%%%%%%%%%%%%%%%%%%%%%%%%%%%%%%%%%%%%%%%%%%%%%%%%

\subsection{Cache}

\system uses a read-through cache in local memory. That is, if a read range is present in the cache, \system gets the layer view directly from it. Otherwise, \system reads from storage and fills in the corresponding cache page(s). 
Filling a cache page and getting a layer view are both zero-copy operations: 
    they do not copy the raw bytes but only their references. 
As the cache fills up, \system avoids more expensive storage reads, offering faster lookup speed;
    nevertheless, to offer consistent performance,
our optimization minimizes the worst-case latency when there is no cached data.
    % so its query speed tends to increase. 
% We defer the formulation and optimization around long-term caching, prefetching, memory hierarchy, and workload effects on cache to future works.

% \paragraph{Caching}

% \tofix{Read through cache, eviction policy}

% \tofix{Behavior: state as things may get cached more, query speed tends to increase. However, in a highly elastic environment, caches are frequently evicted (all together) because compute nodes may quickly churn.}

% \paragraph{Layer View}
% \tofix{Layer view details (avoid copy)}

%%%%%%%%%%%%%%%%%%%%%%%%%%%%%%%%%%%%%%%%%%%%%%%%%%%%%%
%%%%%%%%%%%%%%%%%%%%%%%%%%%%%%%%%%%%%%%%%%%%%%%%%%%%%%

\subsection{Index Complexity}

\def\numberyshift{10.0}
\def\subfigurewidth{0.23\linewidth}
\def\widersubfigurewidth{0.4\linewidth}
\def\subfigurelegendwidth{0.31\linewidth}
\def\legendwidth{0.08\linewidth}

\pgfplotstableread{
    queries	setting	method	io	cache	deserialize	predict	find	other	CHECKSUM	total	speed	yshift
    1	nfs	AirIndex	0.996458	0.000758	0.002592	0.000026	0.000124	0.000041	87.2400	87.3019	87.3019	{\numberyshift}
    10	nfs	AirIndex	0.998602	0.000665	0.000482	0.000028	0.000171	0.000052	49.5390	495.8175	49.5817	{\numberyshift}
    100	nfs	AirIndex	0.998864	0.000801	0.000076	0.000030	0.000177	0.000053	46.3499	4636.5204	46.3652	{\numberyshift}
    1000	nfs	AirIndex	0.998791	0.000922	0.000032	0.000030	0.000174	0.000051	45.2119	45216.2084	45.2162	{\numberyshift}
    10000	nfs	AirIndex	0.998653	0.001053	0.000029	0.000032	0.000180	0.000054	43.6075	436097.7764	43.6098	{\numberyshift}
    100000	nfs	AirIndex	0.997599	0.001829	0.000051	0.000061	0.000367	0.000093	18.8312	1883262.1873	18.8326	{\numberyshift}
    1000000	nfs	AirIndex	0.990401	0.007175	0.000178	0.000205	0.001677	0.000364	2.7769	2777625.9476	2.7776	{\numberyshift}			
}\AirindexNFS

\pgfplotstableread{
    queries	setting	method	io	cache	deserialize	predict	find	other	CHECKSUM	total	speed	yshift
    1	local	AirIndex	0.910317	0.019239	0.065823	0.000753	0.002929	0.000939	2.4707	2.5192	2.5192	{\numberyshift}							
    10	local	AirIndex	0.947863	0.018540	0.025208	0.000960	0.005847	0.001582	0.7401	7.6220	0.7622	{\numberyshift}							
    100	local	AirIndex	0.966923	0.020864	0.003540	0.000964	0.006028	0.001680	0.5974	60.3497	0.6035	{\numberyshift}							
    1000	local	AirIndex	0.966421	0.023786	0.001191	0.000962	0.005934	0.001706	0.5574	559.1749	0.5592	{\numberyshift}							
    10000	local	AirIndex	0.963833	0.026172	0.000966	0.000997	0.006228	0.001804	0.5522	5530.7677	0.5531	{\numberyshift}							
    100000	local	AirIndex	0.953325	0.024643	0.002812	0.003028	0.013008	0.003184	0.2802	28075.9689	0.2808	{\numberyshift}							
    1000000	local	AirIndex	0.910319	0.049003	0.005158	0.005506	0.024303	0.005712	0.1435	143995.4859	0.1440	{\numberyshift}							
}\AirindexSSD

\pgfplotstableread{
    queries	setting	method	io	cache	deserialize	predict	find	other	CHECKSUM	total	speed	yshift
    1	nfs	B-tree	0.999195	0.000307	0.000293	0.000019	0.000138	0.000048	178.3774	178.4442	178.4442	{\numberyshift}							
    10	nfs	B-tree	0.998707	0.000381	0.000393	0.000035	0.000439	0.000046	81.5901	816.3398	81.6340	{\numberyshift}							
    100	nfs	B-tree	0.998492	0.000475	0.000369	0.000042	0.000570	0.000051	61.5403	6155.5376	61.5554	{\numberyshift}							
    1000	nfs	B-tree	0.998282	0.000608	0.000390	0.000047	0.000617	0.000056	53.8104	53814.3902	53.8144	{\numberyshift}							
    10000	nfs	B-tree	0.998338	0.000682	0.000341	0.000051	0.000538	0.000050	45.4777	454796.4564	45.4796	{\numberyshift}							
    100000	nfs	B-tree	0.994733	0.002341	0.000992	0.000172	0.001607	0.000155	11.1081	1110940.3694	11.1094	{\numberyshift}							
    1000000	nfs	B-tree	0.990379	0.004379	0.001883	0.000418	0.002629	0.000311	4.0055	4006294.2371	4.0063	{\numberyshift}							
}\BtreeNFS

\pgfplotstableread{
    queries	setting	method	io	cache	deserialize	predict	find	other	CHECKSUM	total	speed	yshift
    1	local	B-tree	0.970672	0.011555	0.010421	0.000787	0.004496	0.002069	2.8170	2.8643	2.8643	{\numberyshift}							
    10	local	B-tree	0.954730	0.011272	0.016600	0.001444	0.014527	0.001426	1.0496	10.6906	1.0691	{\numberyshift}							
    100	local	B-tree	0.944880	0.012724	0.018985	0.002376	0.019353	0.001681	0.6332	63.8681	0.6387	{\numberyshift}							
    1000	local	B-tree	0.934196	0.017277	0.021295	0.002591	0.022619	0.002022	0.5231	524.7225	0.5247	{\numberyshift}							
    10000	local	B-tree	0.925451	0.023838	0.021759	0.003329	0.023263	0.002361	0.3789	3796.5128	0.3797	{\numberyshift}							
    100000	local	B-tree	0.906415	0.034415	0.023946	0.004426	0.027664	0.003134	0.2965	29702.0064	0.2970	{\numberyshift}							
    1000000	local	B-tree	0.841016	0.067449	0.035984	0.008388	0.041592	0.005571	0.1488	149307.5307	0.1493	{\numberyshift}							
}\BtreeSSD

%%%%%%%%%%%%%%%%%%%%%%%%%%%%%%%%%%%%%%%%%%%%%%%%%
%%% Normalized IO vs Non-IO

\pgfplotstableread{
    queries	setting	method	io	nonio	CHECKSUM	total	speed	yshift
    1	nfs	AirIndex	0.996458	0.003542	1.0000	87.3019	87.3019368	{\numberyshift}
    10	nfs	AirIndex	0.998602	0.001398	1.0000	495.8175	49.58174551	{\numberyshift}
    100	nfs	AirIndex	0.998864	0.001136	1.0000	4636.5204	46.3652041	{\numberyshift}
    1000	nfs	AirIndex	0.998791	0.001209	1.0000	45216.2084	45.21620841	{\numberyshift}
    10000	nfs	AirIndex	0.998653	0.001347	1.0000	436097.7764	43.60977764	{\numberyshift}
    100000	nfs	AirIndex	0.997599	0.002401	1.0000	1883262.1873	18.83262187	{\numberyshift}
    1000000	nfs	AirIndex	0.990401	0.009599	1.0000	2777625.9476	2.777625948	{\numberyshift}
}\AirIndexNFSIOs

\pgfplotstableread{
    queries	setting	method	io	nonio	CHECKSUM	total	speed	yshift
    1	local	AirIndex	0.910317	0.089683	1.0000	2.5192	2.519232775	{\numberyshift}
    10	local	AirIndex	0.947863	0.052137	1.0000	7.6220	0.7622043825	{\numberyshift}
    100	local	AirIndex	0.966923	0.033077	1.0000	60.3497	0.6034966228	{\numberyshift}
    1000	local	AirIndex	0.966421	0.033579	1.0000	559.1749	0.5591748547	{\numberyshift}
    10000	local	AirIndex	0.963833	0.036167	1.0000	5530.7677	0.553076775	{\numberyshift}
    100000	local	AirIndex	0.953325	0.046675	1.0000	28075.9689	0.2807596894	{\numberyshift}
    1000000	local	AirIndex	0.910319	0.089681	1.0000	143995.4859	0.1439954859	{\numberyshift}
}\AirIndexSSDIOs

\pgfplotsset{
    breakdown/.style={
        width=\linewidth,
        height=35mm,
        stack plots=y,
        area style,
        % xmode=log,  % does not work in style
        % ymode=log,  % does not work in style
        bar width=1.8mm,
        axis lines=left,
        % axis line style={-},
        % y axis line style={draw opacity=1}
        % x axis line style={draw opacity=1},
        minor tick num=1,
        xmajorgrids,
        ymajorgrids,
        yminorgrids,
        ytick style={draw=none},
        scaled y ticks=false,
        yticklabel={\pgfmathparse{\tick*100}\pgfmathprintnumber{\pgfmathresult}\%},
        xtick={1, 10, 100, 1000, 10000, 100000, 1000000},
        xticklabels={1, \;, $10^2$, \;, $10^4$, \;, $10^6$},
        xlabel near ticks,
        ylabel near ticks,
        ylabel=Latency Fraction,
        xlabel=Number of Queries,
        ylabel shift=-1mm,
        xlabel shift=-1mm,
        every axis/.append style={font=\footnotesize},
        reverse legend,
        legend columns=1,
        legend style={
            font=\scriptsize,
            at={(1.05,0.5)}, anchor=west,
            draw=black,
            fill=white
        },
        legend cell align={left},
        every node near coord/.append style={yshift=0pt}
    },
}

\pgfplotsset{
    breakdownnfs/.style={
        breakdown,
        ymin=0,
        xmax=2000000,
    },
}

\pgfplotsset{
    breakdownssd/.style={
        breakdown,
        ymin=0,
        xmax=2000000,
    },
}

\pgfplotsset{
    iolines/.style={
        width=\linewidth,
        height=35mm,
        ymin=0,
        % xmode=log,  % does not work in style
        % ymode=log,  % does not work in style
        bar width=1.8mm,
        axis lines=left,
        minor tick num=4,
        xmajorgrids,
        ymajorgrids,
        yminorgrids,
        ytick style={draw=none},
        scaled y ticks=false,
        xtick={1, 10, 100, 1000, 10000, 100000, 1000000},
        xticklabels={1, \;, $10^2$, \;, $10^4$, \;, $10^6$},
        ytick={0, 0.05, 0.1},
        yticklabels={0\%, 5\%, 10\%},
        enlarge x limits=0.1,
        xlabel near ticks,
        ylabel near ticks,
        ylabel=Latency Fraction,
        xlabel=Number of Queries,
        ylabel shift=-1mm,
        xlabel shift=-1mm,
        every axis/.append style={font=\footnotesize},
        reverse legend,
        legend style={
            at={(0.5,0.55)},anchor=south,
            column sep=2pt,draw=black,fill=white,line width=.5pt,
            font=\scriptsize,
            /tikz/every even column/.append style={column sep=5pt}
        },
        legend columns=1,
        legend cell align={left},
        every node near coord/.append style={yshift=0pt}
    },
}

\begin{figure*}[t]
    \centering
    
    \begin{subfigure}[b]{\widersubfigurewidth}
        \centering
        \begin{tikzpicture}
            \begin{axis}[
                iolines,
                % width=0.80\linewidth,
                % ymode=log,
                ymax=0.11,
                xmode=log
            ]
                \addplot+[
                    color={nfscolor},
                    mark=x,
                    ultra thick,
                    text opacity=1] table [x=queries, y=nonio] {\AirIndexNFSIOs};
                \addplot+[
                    color={ssdcolor},
                    mark=o,
                    ultra thick,
                    text opacity=1] table [x=queries, y=nonio] {\AirIndexSSDIOs};
                % \addplot [
                %     opacity=0.0,
                %     text opacity=1.0,
                %     nodes near coords,
                %     point meta=explicit
                % ] table [x=queries, y=yshift, meta=speed] {\BtreeNFS} \closedcycle;
                \legend{NFS, SSD}
            \end{axis}
        \end{tikzpicture}
        \vspace{-2mm}
        \caption{Non-I/O Fraction}
        \label{fig:II_breakdown_by_step_fraction}
    \end{subfigure}
    ~
    % \begin{subfigure}[b]{\subfigurewidth}
    %     \centering
    %     \begin{tikzpicture}
    %         \begin{axis}[
    %             breakdownnfs,
    %             % ymode=log,
    %             xmode=log
    %         ]
    %             \addplot[
    %                 fill={iocolor},
    %                 text opacity=1] table [x=queries, y=io] {\BtreeNFS} \closedcycle;
    %             \addplot[
    %                 fill={cachecolor},
    %                 text opacity=1] table [x=queries, y=cache] {\BtreeNFS} \closedcycle;
    %             \addplot[
    %                 fill={deserializecolor},
    %                 text opacity=1] table [x=queries, y=deserialize] {\BtreeNFS} \closedcycle;
    %             \addplot[
    %                 fill={predictcolor},
    %                 text opacity=1] table [x=queries, y=predict] {\BtreeNFS} \closedcycle;
    %             \addplot[
    %                 fill={findcolor},
    %                 text opacity=1] table [x=queries, y=find] {\BtreeNFS} \closedcycle;
    %             \addplot[
    %                 fill={othercolor},
    %                 text opacity=1] table [x=queries, y=other] {\BtreeNFS} \closedcycle;
    %             % \addplot [
    %             %     opacity=0.0,
    %             %     text opacity=1.0,
    %             %     nodes near coords,
    %             %     point meta=explicit
    %             % ] table [x=queries, y=yshift, meta=speed] {\BtreeNFS} \closedcycle;
    %         \end{axis}
    %     \end{tikzpicture}
    %     \caption{NFS (\btree)}
    %     \label{fig:II_breakdown_by_layer_nfs_btree}
    % \end{subfigure}
    % ~
    \begin{subfigure}[b]{\subfigurewidth}
        \centering
        \begin{tikzpicture}
            \begin{axis}[
                breakdownnfs,
                % ymode=log,
                ymax=0.011,
                xmode=log
            ]
                % \addplot[
                %     fill={iocolor},
                %     text opacity=1] table [x=queries, y=io] {\AirindexNFS} \closedcycle;
                \addplot[
                    fill={cachecolor},
                    text opacity=1] table [x=queries, y=cache] {\AirindexNFS} \closedcycle;
                \addplot[
                    fill={deserializecolor},
                    text opacity=1] table [x=queries, y=deserialize] {\AirindexNFS} \closedcycle;
                \addplot[
                    fill={predictcolor},
                    text opacity=1] table [x=queries, y=predict] {\AirindexNFS} \closedcycle;
                \addplot[
                    fill={findcolor},
                    text opacity=1] table [x=queries, y=find] {\AirindexNFS} \closedcycle;
                \addplot[
                    fill={othercolor},
                    text opacity=1] table [x=queries, y=other] {\AirindexNFS} \closedcycle;
                % \addplot [
                %     opacity=0.0,
                %     text opacity=1.0,
                %     nodes near coords,
                %     point meta=explicit
                % ] table [x=queries, y=yshift, meta=speed] {\AirindexNFS} \closedcycle;
            \end{axis}
        \end{tikzpicture}
        \vspace{-2mm}
        \caption{Non-I/O Breakdown on NFS}
        \label{fig:II_breakdown_by_step_nonio_nfs}
    \end{subfigure}
    ~
    % \begin{subfigure}[b]{\subfigurewidth}
    %     \centering
    %     \begin{tikzpicture}
    %         \begin{axis}[
    %             breakdownssd,
    %             % ymode=log,
    %             xmode=log
    %         ]
    %             \addplot[
    %                 fill={iocolor},
    %                 text opacity=1] table [x=queries, y=io] {\BtreeSSD} \closedcycle;
    %             \addplot[
    %                 fill={cachecolor},
    %                 text opacity=1] table [x=queries, y=cache] {\BtreeSSD} \closedcycle;
    %             \addplot[
    %                 fill={deserializecolor},
    %                 text opacity=1] table [x=queries, y=deserialize] {\BtreeSSD} \closedcycle;
    %             \addplot[
    %                 fill={predictcolor},
    %                 text opacity=1] table [x=queries, y=predict] {\BtreeSSD} \closedcycle;
    %             \addplot[
    %                 fill={findcolor},
    %                 text opacity=1] table [x=queries, y=find] {\BtreeSSD} \closedcycle;
    %             \addplot[
    %                 fill={othercolor},
    %                 text opacity=1] table [x=queries, y=other] {\BtreeSSD} \closedcycle;
    %             % \addplot [
    %             %     opacity=0.0,
    %             %     text opacity=1.0,
    %             %     nodes near coords,
    %             %     point meta=explicit
    %             % ] table [x=queries, y=yshift, meta=speed] {\BtreeSSD} \closedcycle;
    %         \end{axis}
    %     \end{tikzpicture}
    %     \caption{SSD (\btree)}
    %     \label{fig:II_breakdown_by_layer_ssd_btree}
    % \end{subfigure}
    % ~
    \begin{subfigure}[b]{\subfigurelegendwidth}
        \centering
        \begin{tikzpicture}
            \begin{axis}[
                breakdownssd,
                width=0.74\linewidth,
                % ymode=log,
                ymax=0.11,
                xmode=log
            ]
                % \addplot[
                %     fill={iocolor},
                %     text opacity=1] table [x=queries, y=io] {\AirindexSSD} \closedcycle;
                \addplot[
                    fill={cachecolor},
                    text opacity=1] table [x=queries, y=cache] {\AirindexSSD} \closedcycle;
                \addplot[
                    fill={deserializecolor},
                    text opacity=1] table [x=queries, y=deserialize] {\AirindexSSD} \closedcycle;
                \addplot[
                    fill={predictcolor},
                    text opacity=1] table [x=queries, y=predict] {\AirindexSSD} \closedcycle;
                \addplot[
                    fill={findcolor},
                    text opacity=1] table [x=queries, y=find] {\AirindexSSD} \closedcycle;
                \addplot[
                    fill={othercolor},
                    text opacity=1] table [x=queries, y=other] {\AirindexSSD} \closedcycle;
                % \addplot [
                %     opacity=0.0,
                %     text opacity=1.0,
                %     nodes near coords,
                %     point meta=explicit
                % ] table [x=queries, y=yshift, meta=speed] {\AirindexSSD} \closedcycle;
                % \legend{I/O, Cache, Deser., Predict, Find, Other}
                \legend{Cache, Deser., Predict, Find, Other}
                % \legend{Cache, Predict, Find, Other}
            \end{axis}
        \end{tikzpicture}
        \vspace{-2mm}
        \caption{Non-I/O Breakdown on SSD}
        \label{fig:II_breakdown_by_step_nonio_ssd}
    \end{subfigure}
    \vspace{-3mm}
    \caption{Latency breakdown of \system's non-I/O operations over different levels of warmness 
    (indicated by the numbers of queries from one to 1 million).
    % in $\{1, 10, 10^2, 10^3, \dots, 10^6\}$). 
    Left: the fraction of non-I/O operations---the rest is for I/O operations.
    Middle and right: a deeper breakdown of latency spent on different types of I/O operations on NFS and SSD, respectively.}
    \label{fig:II_breakdown_by_step}
\end{figure*}
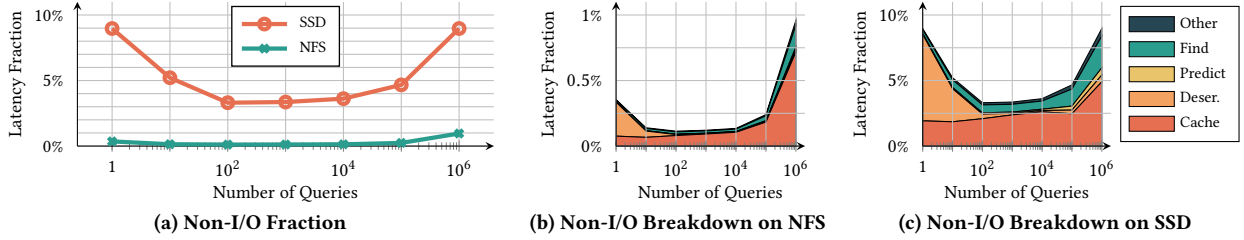

Before branching out, \system selects only top-$k$ candidates with the highest potential to be in the optimal design. Suppose there exists an oracle that minimizes the objective function on a key-position collection $D$ under the storage profile. We call the optimal search cost under storage profile $T$: \textit{index complexity} $\tau(D; T)$. If $\tau(D; T)$ is known, selection of candidates $\calC = \{(\Theta_i, D_i)\}_{i=1}^{|\calF|}$ would be as straightforward as \cref{eq:selection-oracle_a}. That is, we could simply select the top-1 candidate and avoid branching out.
\begin{equation} \label{eq:selection-oracle_a}
    (\Theta^{*}, D^{*}) = \argmin_{(\Theta_i, D_i) \in \calC} \tau(D_i) + \E_{x \sim \calX} \left[ T(\Delta(x; \Theta_i)) \right]
\end{equation}

Unfortunately, $\tau(D; T)$ is unknown for a large class of indexes supported by \system. Instead, \system uses a surrogate upper bound to the index complexity: \textit{step index complexity} $\hat{\tau}(D; T) \geq \tau(D; T)$. Specifically, $\hat{\tau}(D; T)$ imitates a step-function hierarchical index that partitions the key-position collection $D$ into groups with arbitrary position ranges. Let $s_D = y^{+}_{|D|} - y^{-}_1$ be the total size of the candidate layer, $\hat{\tau}(D; T)$ tries to build ideal step indexes with different numbers of layers $L \in \{0, 1, \dots, O(\log s_D)\}$. For each $L$, it perfectly balances both root layer size and subsequent precisions 
so that $s(\Theta_L) = \Delta(x; \Theta_l) = \sqrt[L+1]{s_D s_{step}^{L}}$.
% (\cref{eq:selection-step-balance}).
This 
% takes into account of 
considers the total size $s_D$ and the ideal size for each 1-piece step node $s_{step}$ (e.g., 16 bytes for 8-byte key and position types). 
% \begin{equation} \label{eq:selection-step-balance}
%     % s(\Theta_L) \approx \Delta(x; \Theta_l) = s_D^{1 / (L + 1)} s_{step}^{L / (L + 1)} \qquad, l \in \{1, \dots, L\}
%     s(\Theta_L) \approx \Delta(x; \Theta_l) = \sqrt[L+1]{s_D s_{step}^{L}} \qquad \text{for } l \in \{1, \dots, L\}
% \end{equation}
% (data_size as f64).powf(1.0 / (num_layers + 1) as f64)
%                   * (STEP_SIZE as f64).powf(num_layers as f64 / (num_layers + 1) as f64);
Lastly, $\hat{\tau}(D; T)$ then outputs the lowest storage model cost as the step index complexity (\cref{eq:selection-step-complexity_a}).
\begin{equation} \label{eq:selection-step-complexity_a}
    % s(\Theta_L) \approx \Delta(x; \Theta_l) = s_D^{1 / (L + 1)} s_{step}^{L / (L + 1)} \qquad, l \in \{1, \dots, L\}
    \hat{\tau}(D; T) = \min_{L \in \{0, 1, \dots, O(\log s_D)\}} (L + 1) \times T \left( \sqrt[L+1]{s_D s_{step}^{L}} \right)
\end{equation}

Note that $\hat{\tau}(D; T)$ is only interested in the integer $s_D$ (not the distribution of $D$);
    thus, $\hat{\tau}(D; T)$ can be arithmetically computed (hence cheap). 
    % and all calculations are arithmetic; 
\cref{fig:index-complex} shows the general shape of $\hat{\tau}(D; T)$ with respect to the collection size $s_D$, under an affine storage profile $T$. Notice the sudden index complexity cliffs (such as those around $s_D = 1$MB in \cref{fig:index-complex-latency}), marking the boundaries between different chosen numbers of layers. The technique to minimize an objective based on its cheaper upper bound is related to majorize-minimization algorithms~\cite{DBLP:books/daglib/Lange2016MMalgorithms}.

%%%%%%%%%%%%%%%%%%%%%%%%%%%%%%%%%%%%%%%%%%%%%%%%%%%%%%
%%%%%%%%%%%%%%%%%%%%%%%%%%%%%%%%%%%%%%%%%%%%%%%%%%%%%%

\section{Baselines}

We compare \system to a traditional database index, 
    % two fashions of 
    learned indexes, and our manual configuration counterpart, hosted at their respective forks~\footnote{\url{https://github.com/\githuborg/lmdb}, \url{https://github.com/\githuborg/RMI}, \url{https://github.com/\githuborg/PGM-index}, \url{https://github.com/\githuborg/ALEX_ext}, \url{https://github.com/\githuborg/airindex-public/tree/main/src/bin/data_calculator.rs}}.

\paragraph{\lmdb} 
\lmdb~\cite{lmdb} is a B-tree database that accesses its data on storage through \texttt{mmap}. 
% Among other choices of databases, 
We have also tested PostgreSQL~\cite{postgresql} and RocksDB~\cite{rocksdb} 
but decided to present \lmdb due to its competitive performance in our setting.

\paragraph{\rmi} 
\rmi~\cite{kipf2018learned} is a top-down learned index 
    with a compact two-layer structure 
where the top one contains only one perfectly accurate node 
    partitioning key space to the bottom nodes.
To build one, we execute its provided optimizer for each dataset and select the most accurate configuration describing index size and model types.
We then integrate \rmi onto external memory setting by \texttt{mmap}-ing its parameter arrays so \rmi can access its parameters through the OS buffer cache.

\paragraph{\pgm} \pgm~\cite{DBLP:journals/pvldb/FerraginaV20PGMIndex} is a learned index with bounded precision across all layers. 
% In contrast to \rmi's top-down building, 
\pgm partitions the key-position collection to build the next bottom-most layer towards the top.
To benchmark in our settings, 
we use the \texttt{MappedPGMIndex} variant that operates on file systems.
Although its tuner does not satisfy our target (fastest index, regardless of size), we adjust its error level $\varepsilon \in \{16, 32, \dots, 1024\}$ to microbenchmark on \texttt{wiki} (chosen one arbitrarily) and finally pick $\varepsilon = 128$.

\paragraph{\alex} \alex~\cite{DingMYWDLZCGKLK2020ALEX} is an updatable learned index built top-down like \rmi but further arrange key-value pairs in its layout (notably, ``gapped array'' to buffer structural changes). We integrate \alex onto external memory setting by \texttt{mmap}-ing its serialized node objects, key arrays, and value arrays.

\paragraph{\plex} \plex~\cite{DBLP:journals/corr/Stoian2021PLEX} is a learned index with compact Hist-Tree (CHT) layered on top of RadixSpline~\cite{DBLP:conf/sigmod/KipfMRSKK020RadixSpline} (RS). We integrate \plex onto external memory setting through \texttt{mmap} similarly to \alex. Although \plex optimizes most parameters, its user need to specify the maximum prediction error $\epsilon$. We select $\epsilon = 2048$ based on a benchmark on a setting (\cref{fig:VIIII_tuned_plex}).

\paragraph{\dcal} \dcal~\cite{DBLP:conf/sigmod/IdreosZHKG2018DataCalc,DBLP:journals/corr/IdreosZHKG2018DataCalcInternal} is a data layout design engine that calculates the performance of a data structure. For index learning, we extend \dcal to auto-complete \texttt{recursion allowed} (number of layers), \texttt{fanout}, \texttt{key partitioning}. By following its cost synthesis flowcharts (Fig 5 in \cite{DBLP:conf/sigmod/IdreosZHKG2018DataCalc} and Fig 29 in \cite{DBLP:journals/corr/IdreosZHKG2018DataCalcInternal}), we identify data access primitives, profile them on SSD and NFS for cost models, and execute a parallelized auto-completion flow. Later, the selected data layout is built within \system's framework.

\paragraph{\btree} A B-tree-like structure implemented using \system's framework. It has 255-piece \texttt{step} nodes built with $\texttt{GStep}(p = 255, \lambda_{GS} = 4096)$, which is equivalent to a B-tree with 4KB node pages and 255 fanout factors. This is the most controlled baseline where the only difference is \system's storage- and data-aware tuning.

%%%%%%%%%%%%%%%%%%%%%%%%%%%%%%%%%%%%%%%%%%%%%%%%%%%%%%
%%%%%%%%%%%%%%%%%%%%%%%%%%%%%%%%%%%%%%%%%%%%%%%%%%%%%%

\section{Extended Experiments}

\input{figures/experiments/VI_skew}

%%%%%%%%%%%%%%%%%%%%%%%%%%%%%%%%%%%%%%%%%%%%%%%%%%%%%%
%%%%%%%%%%%%%%%%%%%%%%%%%%%%%%%%%%%%%%%%%%%%%%%%%%%%%%

\subsection{Latency Breakdown}
\label{sec:exp-breakdown}

% I/O cost is the dominating factor, but how much is spent for each object during a search? Does \system's storage profile accurately represent the true cost? As we have observed earlier, cache warmness shifts from I/O usage into compute and memory. How hot of a cache should we reconsider I/O dominance assumption? 

We investigate the effect of our optimization
    % with finer measurements~\cite{tracing} 
    by studying latency breakdowns, by layers and by operations, respectively.
    % by zooming in on \cref{fig:I_latency_curve_nfs_books,fig:I_latency_curve_ssd_books}.
% The breakdown is only limited to indexes under \system for a fair instrumentation and comparison via tracing ecosystem. 
% Our experiments take two different slices of latency breakdown, .

\paragraph{By Layer} 
We first decompose the end-to-end latency into 
    the times spent in retrieving each layer which includes I/O, cache read, and all internal computation. 
Apart from some small exceptions, 
    latency measurements across layers roughly follow the storage profiles: 
    larger root size $s(\Theta_L)$ and coarser precision $\Delta_l(x; \Theta_l)$ reflects in a longer latency spent in the corresponding layer. 
Over numbers of queries, 
    we also observe the alternating acceleration phenomenon in more details. 
That is, the fast acceleration region indicate that the topmost partially cached index layer is becoming fully cached. 
For example, in \cref{fig:II_breakdown_by_layer_ssd_btree} between $10^3$ and $10^4$ queries, \btree searches faster because the speedup in its layer-1 index.

This breakdown also reveals factors unaccounted for in \system. 
    For a prominent example, in the first query under SSD (\cref{fig:II_breakdown_by_layer_ssd_btree,fig:II_breakdown_by_layer_ssd_airindex}),
    both \system and \btree spend more time reading their root and data layers as opposed to reading other index layers.
This is because index layers and data layer are stored in separate directories, 
    forcing the file system (\texttt{ext4}) to slowly walk through different paths of directory entries (dentry) to fetch index-layer and data-layer inodes. 
Consequently, 
    reading subsequent index layers stored in the same directory is then significantly faster than expected. 
Although these missing characteristics are crucial for future works towards maximally fast indexes, 
    we believe that \system's storage profile is 
    at an appropriate level of abstraction to adapt to diverse types of storage.

\paragraph{By Operation} 

We categorize \system's operations into I/O and non-I/O groups. 
The latter group contains in-memory caching, data structure deserialization, node prediction, relevant key-value finding, and other negligible steps. 
\cref{fig:II_breakdown_by_step_fraction} shows that these non-I/O operations only account for up to $1.0\%$ on NFS and $9.0\%$ on SSD, even after 1 million queries. 
These small non-I/O fractions matches with our expectation. 
Because of its 4KB--5KB average precisions of tuned structures on NFS and SSD, 
    \system needs more than a million query to completely cache the 6.4GB data layer. 
As an improvement on its warm-state performance, 
    \system can prefetch to warm up its cache more aggressively.

\cref{fig:II_breakdown_by_step_nonio_nfs,fig:II_breakdown_by_step_nonio_ssd} delve deeper into non-I/O operations to put their significance into perspective. 
We found that caching occupies non-I/O latencies increasingly over warmness. 
In the first query, deserialization dominates because it needs to deserialize the root-metadata file, 
    send the root layer to cache, and reconstruct a nested data structure for query processing. 
These two observations explain the U-shape fraction of non-I/O operations on SSD. 
Aside from those, finding operations (i.e. the last mile binary search) take a considerable portion within non-I/O latency, but incomparably small compared to I/O. If they were to grow larger, \system would have to consider smoothly transitioning to in-memory index rather than a binary search on cache. 
Lastly, as expected, node prediction is insignificant given that our current node types are simple.

%%%%%%%%%%%%%%%%%%%%%%%%%%%%%%%%%%%%%%%%%%%%%%%%%%%%%%
%%%%%%%%%%%%%%%%%%%%%%%%%%%%%%%%%%%%%%%%%%%%%%%%%%%%%%

\subsection{Skewed Workload}
\label{sec:exp-skew}

Although \system's objective (\cref{eq:emm-obj-airindex}) considers the query distribution $\calX$, future query distributions may change unexpectedly. This experiment (\cref{fig:VI_skew}) builds \system on the uniform distribution $\calX$ but requests keys sampled from a Zipf distribution with parameters 0.5 (least skewed), 1.0, and 2.0 (most skewed). The more skewed the query is, the faster all methods can respond at warm state (\cref{fig:VI_skew_later}). However, the skew does not affect first-query latency as much across all methods (\cref{fig:VI_skew_first}). Because \system is tuned for cold-state latency, higher skewness results in a quicker takeover. For example, it takes 12k uniform queries for any methods (\pgm) to take over \system, but only 70, 41, and 725 queries in 0.5, 1.0, and 2.0 Zipf query.

%%%%%%%%%%%%%%%%%%%%%%%%%%%%%%%%%%%%%%%%%%%%%%%%%%%%%%
%%%%%%%%%%%%%%%%%%%%%%%%%%%%%%%%%%%%%%%%%%%%%%%%%%%%%%

% \input{figures/experiments/V_build_scalability}

%%%%%%%%%%%%%%%%%%%%%%%%%%%%%%%%%%%%%%%%%%%%%%%%%%%%%%
%%%%%%%%%%%%%%%%%%%%%%%%%%%%%%%%%%%%%%%%%%%%%%%%%%%%%%

\subsection{Top-k Candidate Parameter Sweep}
\label{sec:exp-topk}

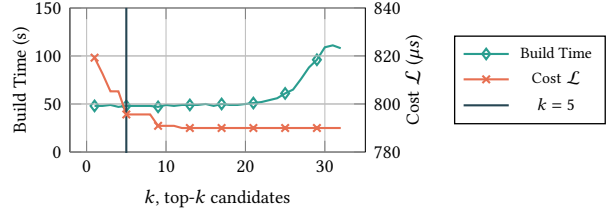
\begin{figure}[t]
    \centering
    \pgfplotstableread{
        dataset	storage	k	cost_us	build_s
        books_800M_uint64	ssd	1	819.256	48
        books_800M_uint64	ssd	2	812.621	48
        books_800M_uint64	ssd	3	805.255	49
        books_800M_uint64	ssd	4	805.255	47
        books_800M_uint64	ssd	5	795.638	48
        books_800M_uint64	ssd	6	795.638	48
        books_800M_uint64	ssd	7	795.638	48
        books_800M_uint64	ssd	8	795.638	48
        books_800M_uint64	ssd	9	790.896	47
        books_800M_uint64	ssd	10	790.896	49
        books_800M_uint64	ssd	11	790.896	48
        books_800M_uint64	ssd	12	789.992	49
        books_800M_uint64	ssd	13	789.992	49
        books_800M_uint64	ssd	14	789.992	49
        books_800M_uint64	ssd	15	789.992	50
        books_800M_uint64	ssd	16	789.992	48
        books_800M_uint64	ssd	17	789.992	50
        books_800M_uint64	ssd	18	789.992	49
        books_800M_uint64	ssd	19	789.992	49
        books_800M_uint64	ssd	20	789.992	50
        books_800M_uint64	ssd	21	789.992	51
        books_800M_uint64	ssd	22	789.992	52
        books_800M_uint64	ssd	23	789.992	54
        books_800M_uint64	ssd	24	789.992	56
        books_800M_uint64	ssd	25	789.992	61
        books_800M_uint64	ssd	26	789.992	65
        books_800M_uint64	ssd	27	789.992	76
        books_800M_uint64	ssd	28	789.992	88
        books_800M_uint64	ssd	29	789.992	96
        books_800M_uint64	ssd	30	789.992	109
        books_800M_uint64	ssd	31	789.992	111
        books_800M_uint64	ssd	32	789.992	108
    } \paramKData
    \def\subplotwidthThree{0.65\linewidth}
    \def\height{35mm}
    \def\hspaceGap{0.04\linewidth}
    \pgfplotsset{
        accfigLeft/.style={
            height=\height,
            width=\linewidth,
            % xmode=log,
            ymode=linear,
            axis y line*=left,
            xlabel={$k$, top-$k$ candidates},
            ylabel=Build Time (s),
            ylabel style={align=center},
            legend columns=2,
            colormap name=bright,
            every axis/.append style={font=\footnotesize},
            xmajorgrids,
            ymajorgrids,
            xlabel near ticks,
            ylabel near ticks,
            ylabel shift=-1.5mm,
        },
        accfigLeft/.belongs to family=/pgfplots/scale,
    }
    \pgfplotsset{
        accfigRight/.style={
            height=\height,
            width=\linewidth,
            % xmode=log,
            % ymode=log,
            axis x line=none,
            axis y line*=right,
            xlabel={$k$, top-$k$ candidates},
            xlabel near ticks,
            ylabel={Cost $\calL$ ($\mu s$)},
            ylabel near ticks,
            ylabel style={align=center},
            legend style={
                % at={(0.5,1.05)},anchor=south,column sep=2pt,
                at={(1.3,0.5)},anchor=west,column sep=2pt,
                draw=black,fill=white,line width=.5pt,
                font=\scriptsize,
                /tikz/every even column/.append style={column sep=5pt}
            },
            legend columns=1,
            colormap name=bright,
            every axis/.append style={font=\footnotesize},
            every non boxed x axis/.append style={x axis line style=-},
            xmajorgrids,
            ymajorgrids,
            ylabel near ticks,
            ylabel shift=-1.5mm,
        },
        accfigRight/.belongs to family=/pgfplots/scale,
    }
    % \begin{subfigure}[b]{\linewidth}
    %     %% DUMMY PICTURE TO ADD LEGEND.
    %     \begin{tikzpicture}
    %         \begin{axis}[
    %                 ticks=none,
    %                 width=\linewidth,
    %                 hide axis,
    %                 xmin=10,
    %                 xmax=50,
    %                 ymin=0,
    %                 ymax=0.4,
    %                 legend style={
    %                     at={(0.55,0)},
    %                     anchor=center,
    %                     column sep=2pt,
    %                     draw=black,fill=white,line width=.5pt,
    %                     font=\scriptsize,
    %                     /tikz/every even column/.append style={column sep=5pt}
    %                 },
    %                 legend columns=-1,
    %             ]
                
    %             \node[align=center, opacity=0] {
    %             \addlegendimage{buildcolor,mark=diamond,thick}
    %             \addlegendentry{Build};
    %             \addlegendimage{costcolor,mark=x,thick}
    %             \addlegendentry{Cost};
    %             };
    %         \end{axis}
    %     \end{tikzpicture}
    % \end{subfigure}
    \begin{subfigure}[b]{\subplotwidthThree}
        \centering
        \begin{tikzpicture}
            \begin{axis}[
                accfigLeft,
                ymin=0.0,
                ymax=150.0,
                ytick={0,50,100,150},
            ]
                \addplot[mark=diamond, color={buildcolor}, thick, mark repeat=4]
                table[x=k,y=build_s] {\paramKData}; \label{fig:VII_topk_build}
            \end{axis}
            
            \begin{axis}[
                accfigRight,
                ymin=780.0,
                ymax=840.0,
                ytick={780,800,820,840},
            ]
                \addlegendimage{/pgfplots/refstyle=fig:VII_topk_build}\addlegendentry{Build Time}
            
                % \addplot[mark=o, color={wakeprecisioncolor}, mark repeat=20]
                % table[x=time_avg_s,y=precision_p] {\powerHundredqIX};
                % % \addlegendentry{Precision}
            
                \addplot[mark=x, color={costcolor}, thick,  mark repeat=4]
                table[x=k,y=cost_us] {\paramKData};
                \addlegendentry{Cost $\calL$}
                \label{fig:VII_topk_cost}
                
                \addplot[thick, samples=50, smooth,domain=0:6,color={Acolor}] coordinates {(5,0)(5,840)};
                \addlegendentry{$k=5$}
            \end{axis}
        \end{tikzpicture}
        % \caption{Q9}
        % \label{fig:II_accuracy_rp_curve_1_q9}
    \end{subfigure}
    \hfill
    \vspace{-3mm}
    \caption{Effects of hyperparameter $k$ in selecting top-$k$ candidates to overall \system's building time and cost $\calL$. We use \texttt{books} dataset and SSD profile.}
    \label{fig:VII_topk}
    % \vspace{-3mm}
\end{figure}

Across all experiments, we set $k = 5$ as an arbitrary constant greater than one and less than the number of node builders ($|\calF| = 45$). In this experiment, we vary this hyperparameter $k$ to verify our understanding: as $k$ increases, build time should increase in $L$-degree polynomial ($L = 2$ in this setting) while the optimized cost should monotonically decrease. \cref{fig:VII_topk} reaffirms this hypothesis, but also shows that the available parallelism (192 CPUs) is able to hide the polynomial build time more than we had expected (taking around 50 seconds up until $k = 20$), implying that we could have selected a higher $k$ to get a faster index at no additional build time cost.

\end{document}